\pdfoutput=1
\documentclass[PAPER, atlasdraft=false, UKenglish, cernpreprint, texmf, orcidlogo]{atlasdoc}
\usepackage[]{atlaspackage}
\usepackage{atlasbiblatex}

\usepackage{atlasphysics}
\graphicspath{{logos/}{figures/}}

\usepackage{ANA-STDM-2020-16-PAPER-defs}

\usepackage{multirow}

\AtlasTitle{Measurements of $W^+W^-$ production cross-sections in $pp$ collisions at $\sqrt{s}=13$\,TeV with the ATLAS detector}

\AtlasAbstract{%
Measurements of  $W^+W^-\rightarrow e^\pm \nu \mu^\mp \nu$ production cross-sections are presented, providing a test of the predictions of perturbative quantum chromodynamics and the electroweak theory.
The measurements are based on data from $pp$ collisions at $\sqrt{s}=13$\,TeV recorded by the ATLAS detector at the Large Hadron Collider in 2015--2018, corresponding to an integrated luminosity of 140\,fb$^{-1}$.
The number of events due to top-quark pair production, the largest background, is reduced by rejecting events containing jets with $b$-hadron decays.
An improved methodology for estimating the remaining top-quark background enables a precise measurement of $W^+W^-$ cross-sections with no additional requirements on jets.
The fiducial  $W^+W^-$ cross-section is determined in a maximum-likelihood fit with an uncertainty of 3.1\%.
The measurement is extrapolated to the full phase space, resulting in a total $W^+W^-$ cross-section of $127\pm4$\,pb.
Differential cross-sections are measured as a function of twelve observables that comprehensively describe the kinematics of $W^+W^-$ events.
The measurements are compared with state-of-the-art theory calculations and excellent agreement with predictions is observed.
A charge asymmetry in the lepton rapidity is observed as a function of the dilepton invariant mass, in agreement with the Standard Model expectation. A CP-odd observable is measured to be consistent with no CP violation.
Limits on Standard Model effective field theory Wilson coefficients in the Warsaw basis are obtained from the differential cross-sections.
}

\AtlasRefCode{STDM-2020-16}
\PreprintIdNumber{CERN-EP-2025-091}

\AtlasJournal{JHEP}
\AtlasJournalRef{\JHEP 08 (2025) 142}
\AtlasDOI{10.1007/JHEP08(2025)142}

\hypersetup{pdftitle={ATLAS document},pdfauthor={The ATLAS Collaboration}}

\begin{document}

\maketitle

%
%

%

\section{Introduction}
\label{sec:intro}

\newcommand{\AtlasCoordFootnote}{%
ATLAS uses a right-handed coordinate system with its origin at the nominal interaction point (IP)
in the centre of the detector and the \(z\)-axis along the beam pipe.
The \(x\)-axis points from the IP to the centre of the LHC ring,
and the \(y\)-axis points upwards.
Cylindrical coordinates \((r,\phi)\) are used in the transverse plane,
\(\phi\) being the azimuthal angle around the \(z\)-axis.
The rapidity is defined as $y=(1/2)\ln[(E+p_z)/(E-p_z)]$ while the pseudorapidity is defined in terms of the polar angle \(\theta\) as \(\eta = -\ln \tan(\theta/2)\).
Angular distance is measured in units of \(\Delta R \equiv \sqrt{(\Delta\eta)^{2} + (\Delta\phi)^{2}}\).
}

Measurements of $W$-boson pair (\WW) production cross-sections provide an important test of the Standard Model (SM). The \WW production at hadron colliders is sensitive to the properties of electroweak-boson self-interactions and allows predictions of perturbative quantum chromodynamics (QCD) and electroweak (EW) theory to be tested.
It also constitutes a large background in measurements of Higgs boson production as well as in some searches for physics beyond the SM. Inclusive and fiducial \WW production cross-sections have been measured in proton--proton ($pp$) collisions at $\sqrt{s}=5\,\TeV$~\cite{CMS-SMP-20-012}, $7\,\TeV$~\cite{STDM-2012-01,CMS-SMP-12-005}, $8\,\TeV$~\cite{STDM-2013-07,CMS-SMP-14-016,STDM-2015-24} 13\,\TeV~\cite{STDM-2015-20,STDM-2017-24,CMS-SMP-18-004,STDM-2018-34,SUSY-2021-06}, and 13.6\,\TeV~\cite{CMS-SMP-24-001}, as well as in $e^+e^-$ collisions at LEP~\cite{schael:2013ita} and in $p\bar{p}$ collisions at the Tevatron~\cite{Abe:1996dw,Aaltonen:2009aa,Abazov:2009ys}.

\begin{figure}[htbp]
\centering
\subfloat[\label{fig:feyn-a}]{\includegraphics[width=.3\linewidth]{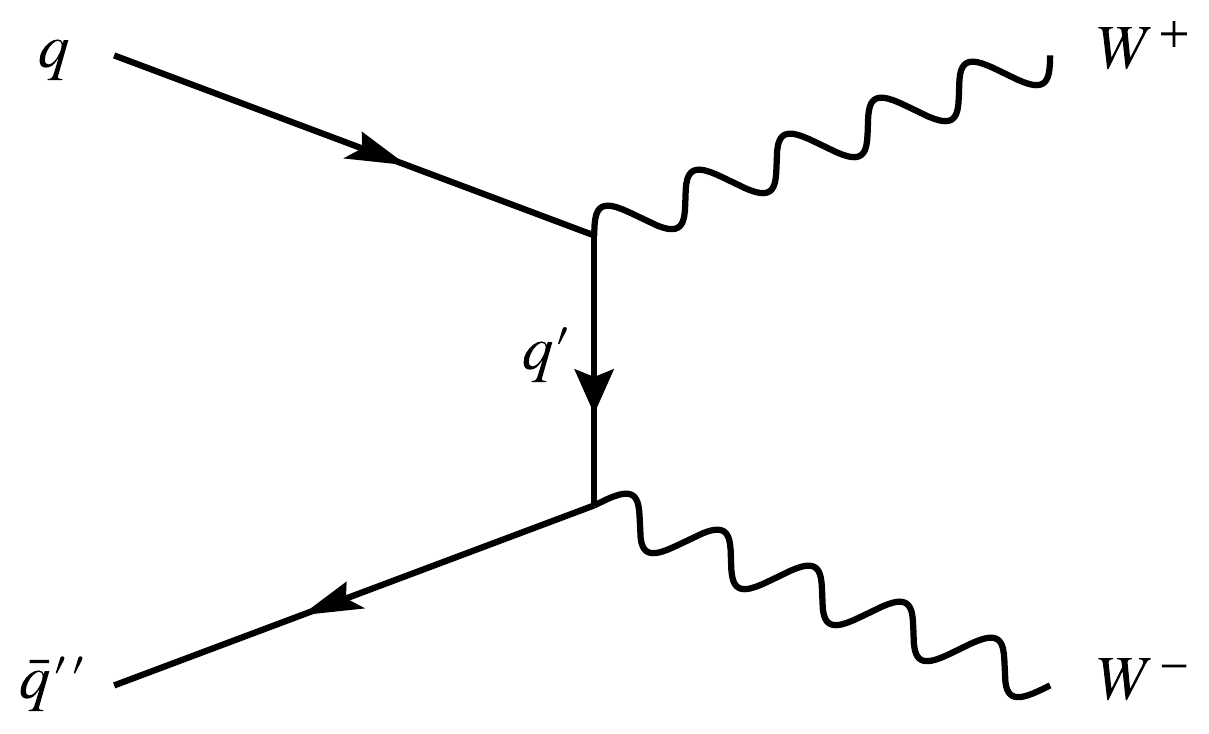}}\hfill
\subfloat[\label{fig:feyn-b}]{\includegraphics[width=.3\linewidth]{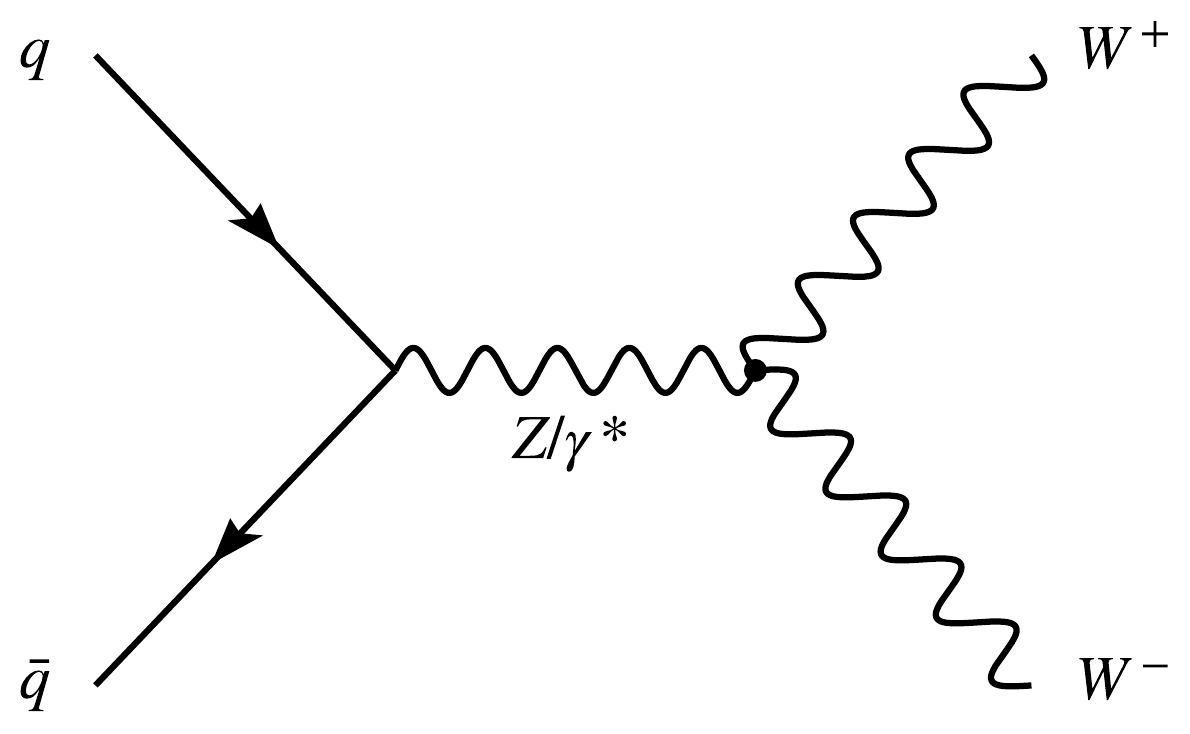}}\hfill
\subfloat[\label{fig:feyn-c}]{\includegraphics[width=.3\linewidth]{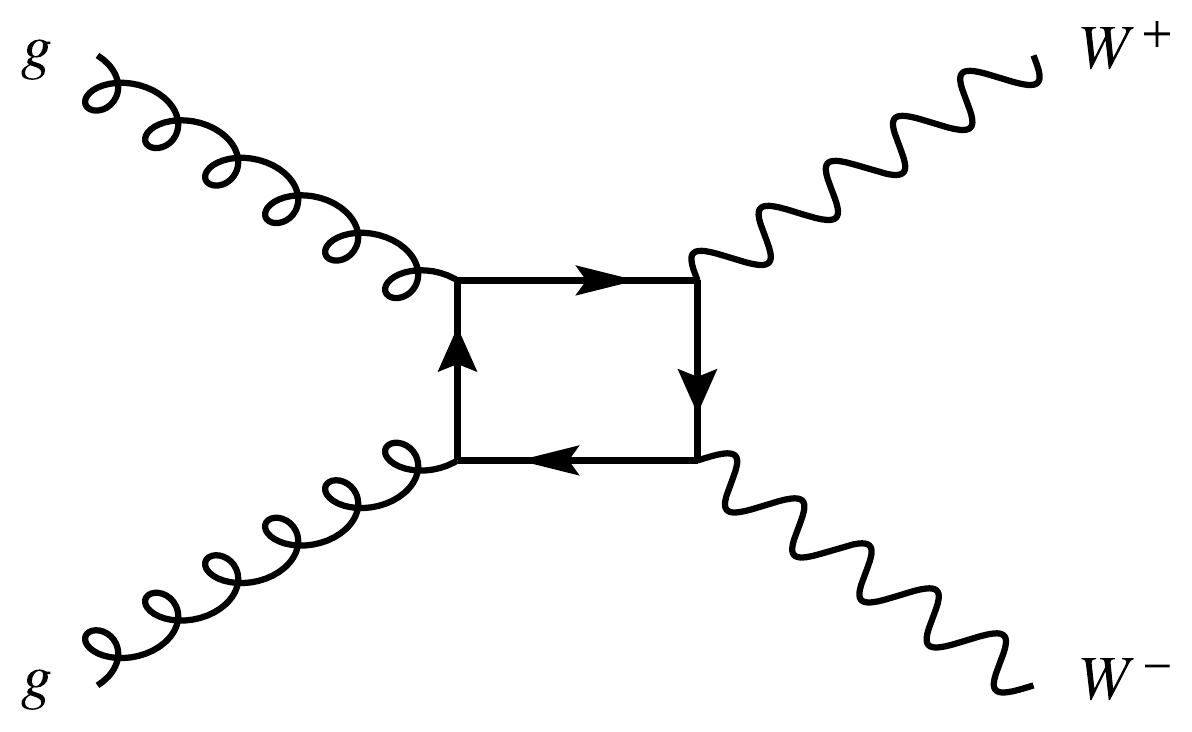}}
\caption{Illustrative examples of Feynman diagrams for the (a) $t$-channel and (b) $s$-channel \qqWW production, as well as the (c) \ggWW production.}
\label{fig:feyn}
\end{figure}

Illustrative Feynman diagrams for \WW production are shown in Figure~\ref{fig:feyn}.
In $pp$ collisions, 95\% of the \WW production rate is due to $t$-channel and $s$-channel $q\bar{q}$-induced $WW$ production (\qqWW), shown in Figures~\ref{fig:feyn-a} and \ref{fig:feyn-b} respectively. Loop-induced gluon--gluon fusion, \ggWW\ (Figure~\ref{fig:feyn-c}), represents 5\% of the total \WW production cross-section, despite formally being only a next-to-next-to-leading-order (NNLO) QCD correction to \WW production.
Beyond leading order in perturbation theory, additional partonic initial states can contribute to the processes labelled \qqWW and \ggWW.
Resonant \ggHWW production is included in the signal definition and simulation, although it is strongly suppressed by the selection requirement on the electron--muon invariant mass.

This paper reports measurements of integrated and differential $\WW\rightarrow e^\pm \nu \mu^\mp \nu$ production cross-sections at $\sqrt{s}=13\,\TeV$ using $pp$ collision data recorded by the ATLAS experiment in 2015--2018, corresponding to an integrated luminosity of 140\,\ifb~\cite{DAPR-2021-01}.
The number of events due to top-quark pair (\ttbar) production, the largest background for this measurement, is reduced by rejecting events containing jets with $b$-hadron decays (\bjets).
The dilepton invariant mass is required to be greater than $85\,\GeV$, to reduce the background due to Drell--Yan production of leptonically decaying $\tau$-lepton pairs.

Measurements of \WW production are challenging because a large top-quark pair production background remains even after rejecting events containing \bjets.
In contrast to previous ATLAS measurements~\cite{STDM-2012-01,STDM-2013-07,STDM-2015-20,STDM-2017-24,SUSY-2021-06}, no jet veto or other selection requirements are employed here to reduce this background.
Instead the background contribution is precisely estimated in situ using two control regions.
The control regions provide a way to the reduce the uncertainties due to the theoretical modelling of top-quark pair production and the efficiency of identifying \bjets.
This jet-inclusive measurement is advantageous not only from an experimental perspective, but also because it allows a comparison with precise predictions of perturbative QCD that are not subject to the large logarithmic corrections that jet vetoes entail~\cite{Meade:2014fca}.

Backgrounds from $W$+jets events with a misidentified or non-prompt lepton, another major source of uncertainty in measurements of \WW cross-sections, are estimated with a data-driven method. The remaining backgrounds are estimated with simulated event samples.

The integrated cross-section is determined from a fit to the data with a statistical model representing background estimates as well as a state-of-the-art NNLO signal prediction matched to the parton shower with the \MiNNLOPS method~\cite{Lombardi:2021rvg,Monni:2019whf}.

Fiducial differential cross-sections for $\WW$ production are measured after background subtraction, using an iterative Bayesian unfolding method~\cite{DAgostini:1994fjx,BayesUnfold2}.
Cross-sections are measured differentially as a function of
\begin{itemize}
\item the transverse momentum of the leading lepton, \ptleadlep,
\item the transverse momentum of the sub-leading lepton, \ptsublep,
\item the transverse momentum of the dilepton system, \ptll,
\item the absolute rapidity of the dilepton system, $|\yll|$,
\item the invariant mass of the lepton pair, \mll,
\item the azimuthal separation of the two leptons, \dphill,
\item $\costhetastar = | \tanh(\detall/2)|$, which is sensitive to the spin structure of the $W$-boson pair~\cite{Barr:2005dz},\footnote{\AtlasCoordFootnote}
\item the magnitude \met of the missing transverse momentum \ptmiss, defined as the negative vectorial sum of the transverse momenta of all visible particles,
\item the scalar sum of  \met and the lepton transverse momenta, \HTlep,
\item the transverse mass of the dilepton system and the missing transverse momentum,\footnote{The transverse mass is defined as $\mtll = \sqrt{ (E_{\mathrm{T},e\mu} + \met)^2 - (\vec{p}_{\mathrm{T},e\mu} + \ptmiss )^2}$, where $E_{\mathrm{T},e\mu} = \sqrt{|\vec{p}_{\mathrm{T},e\mu}|^2 + m_{e\mu}^2}$.} \mtll,
\item the scalar sum of all jet and lepton transverse momenta, \ST, and
\item the jet multiplicity.
\end{itemize}

The $|\yll|$ and \ptll observables are also measured in three regions of \mll. These two observables are proxies for the longitudinal and transverse boost of the \WW system, and their measurement at different invariant masses can be used to constrain parton distribution functions (PDFs).

Two cross-section asymmetries are measured, each defined as
\begin{equation*}
A = \frac{\sigma^{x>0}_\text{fid}-\sigma^{x<0}_\text{fid}}{\sigma_\text{fid}},
\end{equation*}
where $\sigma_\text{fid}$ is the fiducial cross-section and $\sigma^{x>0}_\text{fid}$ ($\sigma^{x<0}_\text{fid}$) is the fiducial cross-section for the phase space for which the relevant observable $x$ is greater than (less than) zero:
\begin{itemize}
\item $A_\text{C}$, the asymmetry in the distribution of the difference of the absolute pseudorapidities of positively and negatively charged leptons,  \dabsetadef, which is predicted to be non-zero due to the charge asymmetry of the partonic initial state~\cite{Re:2018vac} and is useful in the study of PDFs~\cite{Yang:2022bxv}.
\item $A_\text{CP}$, the asymmetry in the CP-odd observable $O_z=\Ozdef$~\cite{Han:2009ra,Degrande:2021zpv}, constructed as the cross product of the lepton three-momenta projected onto the beam direction, with the sign fixed, by convention, by the difference of the longitudinal lepton momenta. The expectation value of the asymmetry is zero and it is sensitive to additional sources of CP violation not predicted by the SM.
\end{itemize}

Finally, limits on boson--quark couplings and boson self-couplings are obtained in the framework of the Standard Model Effective Field Theory (SMEFT)~\cite{Buchmuller:1985jz,Grzadkowski:2010es} by analysing the unfolded \mtll distribution.

This paper is organized as follows. Section~\ref{sec:atlas} gives a brief overview of the ATLAS detector. The recorded dataset and the simulated event samples used are listed in Section~\ref{sec:data_mc_samples}. The criteria used to define the object and event selection in the signal region are given in Section~\ref{sec:event-selection}. Section~\ref{sec:background} presents estimates of the top-quark, non-prompt-lepton, Drell--Yan and other backgrounds, as well as an overview of the selected \WW candidate events. The statistical analysis used to determine the fiducial and differential cross-sections is presented in Section~\ref{sec:unfolding}, while Section~\ref{sec:systematics} considers the systematic uncertainties and their treatment. State-of-the-art predictions of fiducial cross-sections (both integrated and differential) are introduced in Section~\ref{sec:predictions}. Section~\ref{sec:results} reports the measured fiducial and total cross-sections, along with asymmetries, compares them to theoretical predictions, and provides an interpretation in the SMEFT framework.
The paper ends with the conclusions in Section~\ref{sec:conclusions}.


\section{The ATLAS detector}
\label{sec:atlas}

The ATLAS experiment~\cite{PERF-2007-01} at the LHC~\cite{Evans:2008zzb} is a multipurpose particle detector
with a forward--backward symmetric cylindrical geometry and nearly \(4\pi\) coverage in
solid angle.
It consists of an inner tracking detector surrounded by a thin superconducting solenoid
providing a \SI{2}{\tesla} axial magnetic field, electromagnetic and hadron calorimeters, and a muon spectrometer with three large superconducting toroidal magnets with eight coils each.

The inner tracking detector (ID) covers the pseudorapidity range \(|\eta| < 2.5\). It contains a high-granularity silicon pixel detector, including the insertable B-layer installed before Run~2~\cite{ATLAS-TDR-2010-19,PIX-2018-001}, followed by a silicon microstrip tracker. The silicon detectors are complemented by a transition radiation tracking detector, enabling extended track reconstruction within $|\eta| < 2.0$.

Lead/liquid-argon (LAr) sampling calorimeters provide electromagnetic (EM) energy measurements
with high granularity.
A steel/scintillator-tile hadron calorimeter covers the central pseudorapidity range (\(|\eta| < 1.7\)).
The endcap and forward regions are instrumented with copper/LAr and tungsten/LAr calorimeters for EM and hadronic energy measurements up to \(|\eta| = 4.9\).

The muon spectrometer surrounds the calorimeters and is based on
three large air-core toroidal superconducting magnets with eight coils each.
The field integral of the toroids ranges between \num{2.0} and \num{6.0}\,Tm across most of the spectrometer.
The muon spectrometer includes a system of precision chambers for tracking and fast detectors for triggering.

Events are selected using a two-level trigger system.
The first-level trigger is implemented in hardware and uses a subset of the detector information
to accept events at a rate of about \SI{100}{\kHz}.
It is followed by a software-based trigger that
reduces the accepted event rate to about \SI{1.25}{\kHz}.

An extensive software suite~\cite{SOFT-2022-02} is used in data simulation, in the reconstruction
and analysis of real and simulated data, in detector operations, and in the trigger and data acquisition
systems of the experiment.


%

\section{Data and Monte Carlo samples}
\label{sec:data_mc_samples}

The analysis uses data collected from proton--proton collisions at a centre-of-mass energy of 13\,\TeV from 2015 to 2018. After applying data quality criteria~\cite{DAPR-2018-01}, the dataset corresponds to an integrated luminosity of 140\,\ifb, with an uncertainty of  0.83\%~\cite{DAPR-2021-01}, determined using the LUCID-2 detector~\cite{LUCID2} for the primary luminosity measurements.

Monte Carlo (MC) simulated event samples are used to correct the signal yield for detector effects and to estimate background contributions.
All samples were passed through a full simulation of the ATLAS detector~\cite{SOFT-2010-01}, based on $\GEANT$~\cite{Agostinelli:2002hh}. Table~\ref{tab:nominal-mc-samples} lists the configurations for the nominal MC simulations used in the analysis.

\begin{table}[htbp]
\caption{Summary of the nominal Monte Carlo simulated events samples used in the analysis. The $gg\rightarrow WW$ simulation includes Higgs boson contributions. The last two columns give the order in the strong coupling constant, \alphas, of the matrix-element calculation and the overall cross-section normalization. All nominal Monte Carlo samples use the \NNPDF[3.0] PDF set. The samples generated with \SHERPA use its default set of tuned parton-shower parameters, while for the \POWHEGBOX and \textsc{MiNNLO} samples the A14 set of tuned parameters and the \NNPDF[2.3lo] PDF set are used for the parton shower.}
\label{tab:nominal-mc-samples}
\centering
\begin{tabular}{lllll}
\toprule
Process                  & Generator       & Parton shower & Matrix element $\mathcal{O}(\alphas)$ & Normalization                                                                                                             \\ \midrule
$q\bar{q}\rightarrow WW$ & \textsc{MiNNLO} & \PYTHIA[8.2]    & NNLO                                          & Generator                                                                                                                 \\
$gg\rightarrow WW$       & \SHERPA[2.2.2]  & \SHERPA       & LO (0--1 jet)                                 & NLO                                                                                                                       \\
$q\bar{q}\rightarrow WWjj$      & \SHERPA[2.2.2]  & \SHERPA       & LO (EWK only)                                 & Generator \\
\ttbar                   & \POWHEGBOX[v2]  & \PYTHIA[8.2]    & NLO                                           & NNLO+NNLL                                                                                                                 \\
$Wt$                     & \POWHEGBOX[v2]  & \PYTHIA[8.2]    & NLO                                           & NLO+NNLL                                                                                                                  \\
$Z$+jets                 & \SHERPA[2.2.1]  & \SHERPA       & NLO (0--2 jets), LO (3--4 jets)               & NNLO                                                                                                                      \\
$WZ, ZZ$                 & \SHERPA[2.2.2]  & \SHERPA       & NLO (0--1 jet), LO (2--3 jets)                & Generator$^\dagger$                                                                                                       \\
$W\gamma,Z\gamma$        & \SHERPA[2.2.8]  & \SHERPA       & NLO (0--1 jet), LO (2--3 jets)                & Generator$^\dagger$                                                                                                       \\
\bottomrule
\multicolumn{5}{l}{\footnotesize$\dagger$: The cross-section calculated by \SHERPA is found to be in good agreement with the NNLO result~\cite{Grazzini:2016swo,Grazzini:2017ckn,Grazzini:2015hta,Cascioli:2014yka,Grazzini:2016ctr}.} \\
\end{tabular}
\end{table}

Signal \qqWW events were modelled using \PowMiNNLO~\cite{Lombardi:2021rvg,Monni:2019whf}, which is accurate at NNLO in QCD for inclusive observables.
This signal sample and all other \WW production processes described in the following were generated using the \NNPDF[3.0nnlo] PDF set~\cite{Ball:2014uwa}.
The events were interfaced to \PYTHIA[8.245]~\cite{Sjostrand:2014zea} for the modelling of the parton shower, hadronization, and underlying
event, with parameter values set according to the A14 set of tuned parameters (tune)~\cite{ATL-PHYS-PUB-2014-021}.
The matrix-element calculation of $gg\rightarrow WW$ production, which includes off-shell effects and Higgs boson contributions, incorporates up to one additional parton emission at LO. This contribution was simulated with the \SHERPA[2.2.2]~\cite{Bothmann:2019yzt,Cascioli:2013gfa} generator, and the matrix element was matched
and merged with the \SHERPA parton shower based on the Catani--Seymour
dipole factorization~\cite{Gleisberg:2008fv,Schumann:2007mg} using the MEPS@NLO
prescription~\cite{Hoeche:2011fd,Hoeche:2012yf,Catani:2001cc,Hoeche:2009rj}.
The virtual QCD corrections were provided by the
\OPENLOOPS\ library~\cite{Cascioli:2011va,Denner:2016kdg}.
The \SHERPA[2.2.2]~\cite{Bothmann:2019yzt} generator was also used to simulate
electroweak production of a diboson in association with two jets.
The LO-accurate matrix elements were matched to a parton
shower based on Catani--Seymour dipole factorization using the MEPS@LO
prescription~\cite{Hoeche:2011fd,Hoeche:2012yf,Catani:2001cc,Hoeche:2009rj}.
A dedicated set of tuned parton-shower parameters developed by the \SHERPA authors was used for all \SHERPA samples.

The production of \ttbar\ and single-top $Wt$ events was modelled using the
\POWHEGBOX[v2]~\cite{Frixione:2007nw,Nason:2004rx,Frixione:2007vw,Alioli:2010xd,Re:2010bp}
generator at NLO with the \NNPDF[3.0nlo]~\cite{Ball:2014uwa} PDFs. The events were interfaced
to \PYTHIA[8.230]~\cite{Sjostrand:2014zea} to model the parton shower,
hadronization, and underlying event, using the A14 tune and the \NNPDF[2.3lo]
set of PDFs~\cite{Ball:2012cx}. For \ttbar event generation, the \hdamp\ parameter\footnote{The
\hdamp\ parameter is a resummation damping factor and one of the
parameters that control the matching of \POWHEG matrix elements to
the parton shower and thus effectively regulates the
high-\pt\ radiation against which the \ttbar\ system recoils.} was set
to 1.5 times the top quark mass~\cite{ATL-PHYS-PUB-2016-020}. The diagram-removal scheme~\cite{Frixione:2008yi}
was employed to handle the interference between the $Wt$ and \ttbar\ production processes~\cite{ATL-PHYS-PUB-2016-020}.
Alternative samples were generated to assess the uncertainties in the top-background modelling.
The uncertainty due to initial-state radiation and higher-order QCD effects was estimated by
simultaneous variations of the \hdamp\ parameter and the renormalization and
factorization scales, and by choosing the VAR3c up/down variants of the A14 tune.
The VAR3c variations correspond to alternative values of \alphas in the parton shower, as described in Refs.~\cite{ATL-PHYS-PUB-2014-021,ATL-PHYS-PUB-2017-007}.
The impact of
final-state radiation was evaluated with weights that account for
the effect of raising or lowering the renormalization scale for final-state
parton-shower emissions by a factor of two.
To assess the measurement's dependence on the \ttbar--$Wt$ overlap removal scheme, the diagram-subtraction scheme~\cite{Frixione:2008yi}
was employed as an alternative to the diagram-removal scheme.
The uncertainty due to the parton shower and hadronization model was
evaluated by comparing the nominal sample of events with an event sample
generated by \POWHEGBOX[v2] and interfaced to
\HERWIG[7.04]~\cite{Bahr:2008pv,Bellm:2015jjp}, using the H7UE set
of tuned parameters~\cite{Bellm:2015jjp} and the \MMHT[lo] PDF set~\cite{Harland-Lang:2014zoa}.
To estimate the uncertainty from the matching of NLO matrix elements to the
parton shower, the nominal sample was compared with a sample generated
by \MGNLO[2.6.2]~\cite{Alwall:2014hca} at NLO in QCD using the five-flavour
scheme and the \NNPDF[2.3nlo] PDF set. The events were
interfaced to \PYTHIA[8.230], as done for the nominal sample. The \ttbar\ sample was normalized to the cross-section prediction at NNLO in QCD including the resummation of next-to-next-to-leading logarithmic (NNLL) soft-gluon terms calculated using
\TOPpp[2.0]~\cite{Beneke:2011mq,Cacciari:2011hy,Baernreuther:2012ws,Czakon:2012zr,Czakon:2012pz,Czakon:2013goa,Czakon:2011xx}. The inclusive cross-section for single-top $Wt$ production was corrected to the theory prediction
calculated at NLO in QCD with NNLL soft-gluon corrections~\cite{Kidonakis:2010ux,Kidonakis:2013zqa}.

The background due to $Z/\gamma^*$+jets production was simulated with the
\SHERPA[2.2.1] generator using NLO-accurate matrix elements for up to two associated jets, and LO-accurate matrix elements for three or four jets, calculated with the \textsc{Comix} and \OPENLOOPS libraries.
They were matched with the \SHERPA parton shower using the MEPS@NLO
prescription
and the set of tuned parameters developed by the \SHERPA authors.
The \NNPDF[3.0nnlo] set of PDFs was used, and the samples
were normalized to an NNLO prediction~\cite{Anastasiou:2003ds}.

The production of diboson final states was simulated with the
\SHERPA[2.2.2] ($VZ$ with $V=W,Z$) and \SHERPA[2.2.8] ($V\gamma$) generators using \OPENLOOPS\ at NLO QCD accuracy for up to one additional parton and LO accuracy for two to three additional parton emissions, matched and merged with
the \SHERPA parton shower. The $VZ$ simulation includes $V\gamma^*$ contributions for $m(\ell\ell) > 4\,\GeV$. Samples were generated using the  \NNPDF[3.0nnlo] PDF set and normalized to the cross-section calculated by the event generator.

Samples generated with \POWHEGBOX or \MGNLO used the \EVTGEN program~\cite{Lange:2001uf} to model the decay of bottom and charm hadrons.
The effect of multiple interactions in the same and neighbouring bunch crossings (pile-up) was modelled by overlaying the hard-scattering event with simulated inelastic $pp$ events generated with \PYTHIA[8.186]~\cite{Sjostrand:2007gs} using the \NNPDF[2.3lo] set of PDFs and the A3 set of tuned parameters~\cite{ATL-PHYS-PUB-2016-017}.


\section{Event reconstruction and selection}
\label{sec:event-selection}

Candidate \WW events are selected by requiring exactly one electron and one muon with opposite electric charges. These leptons are required to be isolated, i.e.\ there should be little hadronic activity in the vicinity of the lepton, to suppress backgrounds due to misidentified leptons or leptons from hadron decays. Events with two isolated leptons of the same flavour are not considered in the analysis, due to the higher background from Drell--Yan events.

Events were selected online by either single-electron or single-muon triggers. During the data-taking period, the lowest \pt threshold was raised from 24\,\GeV to 26\,\GeV for electrons and from 20\,\GeV to 26\,\GeV for muons, both requiring `loose' to `medium' isolation criteria~\cite{TRIG-2018-05,TRIG-2018-01}. Triggers with higher \pt thresholds and looser isolation requirements are used to increase the efficiency. The trigger selection efficiency is more than 99\% for signal events fulfilling all other selection requirements detailed in this section.

Trajectories of charged particles are reconstructed as tracks in the ID, whose common vertices are used to identify interaction vertex candidates. Candidate events are required to have at least one vertex having at least two associated tracks with $\pt > 500\,\MeV$. The vertex with the highest $\sum \pt^2$ of the associated tracks is taken as the primary vertex.

Electrons are reconstructed from energy deposits in the calorimeter that are matched to tracks~\cite{EGAM-2018-01}. Electron candidates are required to meet the \texttt{TightLH} likelihood-based identification criteria as defined in Ref.~\cite{EGAM-2018-01}. Furthermore, they are required to have $\pt > 27\,\GeV$ and $|\eta|<2.47$, excluding the transition region between the barrel and endcap calorimeter regions, $1.37 < |\eta| < 1.52$.

Muon candidates are reconstructed by combining a track in the inner detector with a track in the muon spectrometer~\cite{MUON-2018-03}. Muons are required to have $\pt > 27\,\GeV$ and $|\eta| < 2.5$ and to satisfy the \texttt{Medium} identification selection, as defined in Ref.~\cite{MUON-2018-03}.

Leptons are required to be compatible with the primary vertex by imposing constraints on the impact parameters of their tracks. The transverse impact parameter's significance, $d_0 / \sigma_{d_0}$, is required to satisfy $|d_0 / \sigma_{d_0}|  < 5\,(3)$ for electrons (muons). The longitudinal impact parameter $z_0$ must satisfy $|z_0\cdot\sin\theta| < \SI{0.5}{\milli\meter}$, where $\theta$ is the polar angle of the track. Additionally, leptons are required to be isolated using information from the ID tracks and calorimeter energy clusters in a cone around the lepton. The \texttt{Gradient} working point is used for electrons~\cite{EGAM-2018-01}, while for muons the \texttt{Tight\_FixedRad} working point is used. The latter is similar to the \texttt{Tight} selection defined in Ref.~\cite{PERF-2015-10} but with a fixed cone size for muons with $\pt > 50\,\GeV$ in order to increase the background rejection. The electron or muon trigger object is required to match the respective reconstructed lepton.

Jets are reconstructed from particle-flow objects~\cite{PERF-2015-09} by using the anti-$k_t$ algorithm~\cite{Cacciari:2008gp,Fastjet} with a radius parameter of $R=0.4$. They are required to have $\pt > 30\,\GeV$ and $|\eta| < 4.5$.  To suppress jets that originate from pile-up, a  jet-vertex tagger~\cite{PERF-2014-03}  is applied to jets with $\pt < 60\,\GeV$ and $|\eta|<2.4$. The jet energy scale is calibrated using an $\eta$- and \pt-dependent correction~\cite{JETM-2018-05}.
Jets with $\pt > 20\,\GeV$ and $|\eta|<2.5$ containing decay products of a $b$-hadron are identified using the DL1r \btagging algorithm~\cite{FTAG-2019-07} at the 85\% efficiency working point.
The lower \pt threshold of $20\,\GeV$ when selecting \bjets allows a larger fraction of the \ttbar background to be rejected.

To resolve the overlap between particles reconstructed as multiple physics objects in the detector, an overlap removal procedure is applied.
Non-\btagged jets are removed if they overlap, within $\Delta R < 0.2$, with an electron, or with a muon if the jet has less than three associated tracks with $\pt > 500\,\MeV$ and satisfies $\pt^\mu / \pt^{\mathrm{jet}} > 0.5$, and the ratio of the muon \pt to the sum of the track \pt associated with the jet is greater than 0.7.
Subsequently, electrons or muons overlapping within $\Delta R < 0.4$  with any jet, including \btagged jets,  are removed.

The missing transverse momentum, \met, computed as the negative vectorial sum of the transverse momenta of calibrated objects in the detector~\cite{JETM-2020-03}, measures the momentum imbalance produced by weakly interacting particles, such as neutrinos in the $W^+W^-\rightarrow e^\pm \nu \mu^\mp \nu$  decays. It includes a track-based term to account for the energy from particles that are not associated with a reconstructed lepton or jet.

Events with one or more \btagged jets are vetoed to reduce the dominant background from top-quark pair production. To reduce the Drell--Yan backgrounds, dominated by $Z$+jets events with $Z\rightarrow\tau^+\tau^-$ decays, the invariant mass of the electron--muon pair is required to be $m_{e\mu} > 85\,\GeV$. This requirement also reduces the contribution from resonant $gg \rightarrow H \rightarrow WW$ production.

Events with additional leptons that have $\pt > 10\,\GeV$ and satisfy \texttt{Loose} isolation and \texttt{LooseLH} (\texttt{Loose}) identification requirements for electrons (muons) are vetoed to reduce backgrounds due to $WZ$ and $ZZ$ production.

Table~\ref{tab:selection} gives a summary of the lepton, jet, and event selection criteria used to define the signal region.
The expected $W^+W^-\rightarrow e^\pm \nu \mu^\mp \nu$ acceptance and efficiency are approximately 25\% and 50\%, respectively, with a predicted signal yield of $56900 \pm 1100$ events.

\begin{table}
\caption{Summary of the object and event selection criteria. }
\label{tab:selection}
\centering
\begin{tabular}{lll}
\toprule
& Requirement       & Criteria                                                                     \\
\midrule
\multirow{5}{*}{Electron} & \pt             & $>27\,\GeV$                                                                  \\
& $|\eta|$          & $\in[0\,,\,1.37)\cup(1.52\,,\,2.47)$                                         \\
& Identification    & \texttt{TightLH} working point                                                          \\
& Isolation         & \texttt{Gradient}                                                            \\
& Impact parameters & $|d_0 / \sigma_{d_0}| < 5$; $|z_0 \cdot \sin\theta| <0.5\,\si{\milli\meter}$ \\
\hline
\multirow{5}{*}{Muon}     & $\pt$             & $>27\,\GeV$                                                                  \\
& $|\eta|$          & $< 2.5$                                                                      \\
& Identification    & \texttt{Medium} working point                                                           \\
& Isolation         & \texttt{Tight\_FixedRad}                                                     \\
& Impact parameters & $|d_0 / \sigma_{d_0}| < 3$; $|z_0 \cdot \sin\theta| <0.5\,\si{\milli\meter}$ \\
\hline
\multirow{4}{*}{\bjets}   & \pt               & $ > 20\,\GeV$                                                                \\
& $\abseta$         & $<2.5$                                                                       \\
& Pile-up suppression    & jet-vertex tagger~\cite{PERF-2014-03}  for $\pt < 60\,\GeV$ and $|\eta|<2.4$ \\
& \btagging         & DL1r, 85\% efficiency working point                                                          \\
\hline
\multirow{3}{*}{Jets}     & $\pt$             & $ > 30\,\GeV$                                                                \\
& $\abseta$         & $<4.5$  \\
& Pile-up suppression    & jet-vertex tagger~\cite{PERF-2014-03}  for $\pt < 60\,\GeV$ and $|\eta|<2.4$ \\
\hline
\multirow{5}{*}{Event}    &                   & 1 electron and 1 muon of opposite electric charge;                           \\
& Leptons           & no additional lepton with $\pt > 10\,\GeV$, \texttt{Loose} isolation,        \\
&                   & and \texttt{LooseLH} (electron) / \texttt{Loose} (muon) identification       \\
& Number of \bjets  & 0                                                                            \\
& \mll              & $> 85\,\GeV$                                                                 \\
\bottomrule
\end{tabular}
\end{table}


%
\newcommand{\RecoFigCaption}{Data are shown as black markers together with histograms of the predictions for signal and background processes. The rightmost bin contains overflow events. The lower panels show the ratio of the data to the total prediction. Top-quark and fake-lepton backgrounds are determined using data-driven methods. The uncertainty bands shown include statistical and systematic uncertainties, excluding theory uncertainties in the signal, which largely cancel out in the measurement of \WW cross-sections.
}

\section{Background estimate}
\label{sec:background}

About 60\% of the events passing the event selection are background events.
The top-quark background, from either \ttbar or single-top $Wt$ production, is the largest background and comprises about 80\% of the total background.
Additional backgrounds considered come from $Z$+jets production, events with non-prompt or misidentified leptons, and diboson production ($WZ$, $W\gamma$, $ZZ$, and $Z\gamma$).

\subsection{Top-quark background}
\label{sec:topquarkbkg}
An estimate of the \ttbar background is obtained using a data-driven technique referred to as the \btag counting method, which is applied separately in each bin of the differential measurements.
Following the procedure established in a measurement of the \ttbar production cross-section~\cite{TOPQ-2018-17} and in the measurement of \WWj production~\cite{STDM-2018-34}, two control regions are defined, requiring exactly one or exactly two \btagged jets respectively. All other selection criteria are the same as in the signal region. The contribution of non-\ttbar events in the 1-\bjet and 2-\bjet control regions is 13\% and 4\% of the expected events, respectively, of which more than 80\% can be attributed to single-top $Wt$ production.

After subtracting the estimated non-\ttbar backgrounds, the \ttbar event yields in the two control regions and the signal region are parameterized by three quantities. The first, $N^{\ttbar}$, is the total number of \ttbar events without requirements on the \bjet multiplicity.
The second, $\varepsilon_{b}$, is the efficiency of identifying and selecting a \bjet in a \ttbar event, reflecting both the efficiency of the \btagging algorithm and the acceptance for \bjets.
The third parameter is the \bjet correlation factor $C_b$, which accounts for the fact that the probability of identifying both \bjets in a \ttbar event is not exactly $\varepsilon_{b}^2$ but rather $C_b\varepsilon_{b}^2$, due to correlation effects that depend on the interplay of event selection and \ttbar kinematics, as well as the presence of additional light-flavour jets and \bjets.
The number of \ttbar events with exactly $i$ \btagged jets, $N^{\ttbar}_{ib}$, is given by
\begin{align}
N_{2b}^{\ttbar} & = N^{\ttbar} \cdot C_b \varepsilon_{b}^2 \ , \nonumber                        \\
N_{1b}^{\ttbar} & = N^{\ttbar} \cdot 2 \varepsilon_{b} \left(1 - C_b \varepsilon_{b} \right) \ , \nonumber    \\
N_{0b}^{\ttbar} & = N^{\ttbar} \cdot \left(1 - 2 \varepsilon_{b} +  C_b \varepsilon_{b}^2 \right) \ . \nonumber
\end{align}

Using these equations, the number of \ttbar events in the signal region can be expressed as
\begin{equation*}
N_{0b}^{\ttbar} = \frac{C_b}{4} \frac{\left(N_{1b}^{\ttbar} + 2 N_{2b}^{\ttbar}\right)^2}{N_{2b}^{\ttbar}} - N_{1b}^{\ttbar} - N_{2b}^{\ttbar}\ ,
\end{equation*}
where $\varepsilon_{b}$ and $N^{\ttbar}$ are expressed in terms of the observed yields in the two control regions, $N_{1b}^{\ttbar}$ and $N_{2b}^{\ttbar}$, which are obtained after the subtraction of minor backgrounds.
Consequently, the only input from \ttbar simulation is the correction factor $C_b$, which is typically very close to unity -- between 0.99 and 1.01 across jet-inclusive distributions such as \mll (Figure~\ref{fig:corrfact-a}).
Since uncertainties in the modelling of top-quark processes and \btagging primarily affect $\varepsilon_{b}$ and $N^{\ttbar}$ but not $C_b$, the uncertainty in $C_b$ is less than 1\% in most analysis bins.

In certain configurations, for example in events with exactly one jet with $\pt>30\,\GeV$, $C_b$ can be as low as 0.8 (Figure~\ref{fig:corrfact-b}).
Although identifying two \bjets is possible in events with only one jet with $\pt>30\,\GeV$, given that the \bjet \pt threshold is 20\,\GeV, the probability of tagging both \bjets is substantially lower than $\varepsilon_{b}^2$, as reflected by $C_b<1$.
This is due to the fact that, in such a configuration, the second jet is often outside the tracker acceptance or has a transverse momentum between 20\,\GeV and 30\,\GeV, resulting in a smaller \btagging efficiency.
The disparity in kinematics also leads to a lower correlation between the uncertainties associated with finding the first and second jet, which in turn increases the uncertainty on $C_b$.

\begin{figure}
\centering
\subfloat[\label{fig:corrfact-a}]{\includegraphics[width=.45\linewidth]{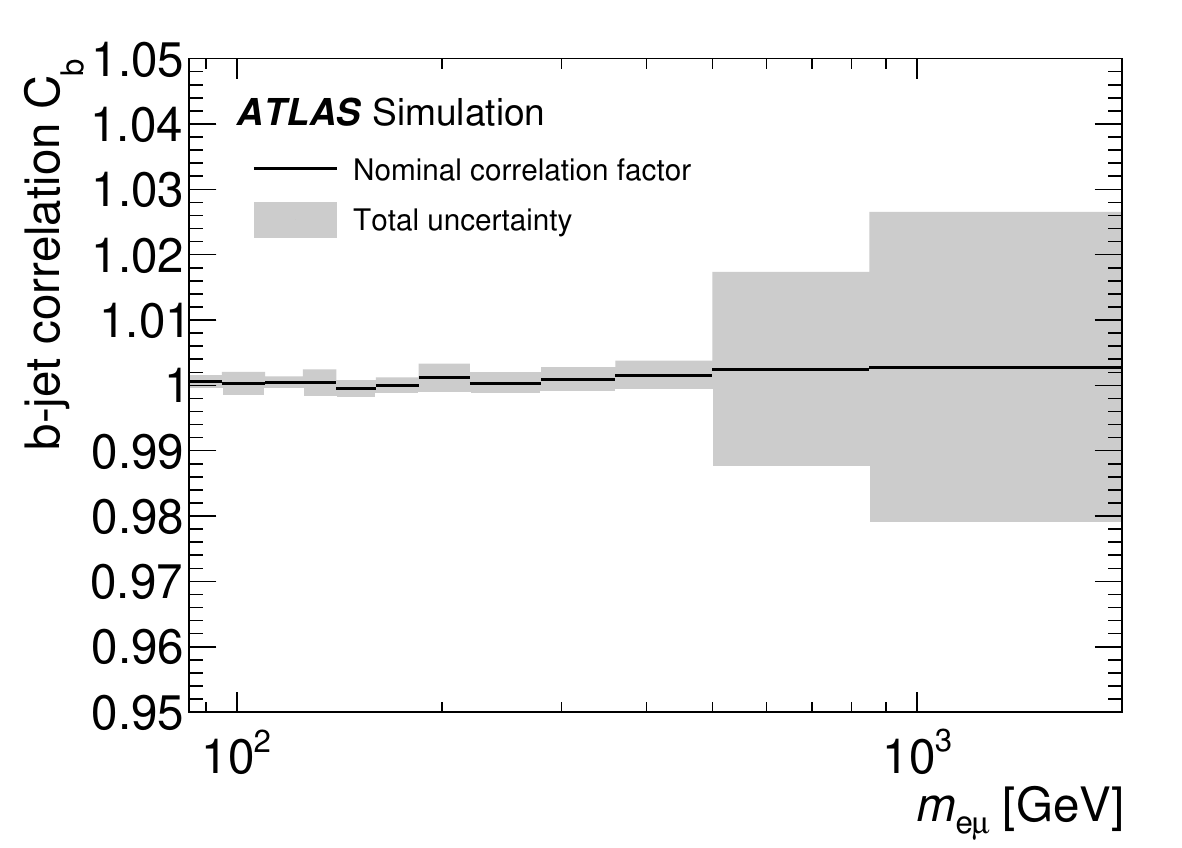}}%
\subfloat[\label{fig:corrfact-b}]{\includegraphics[width=.45\linewidth]{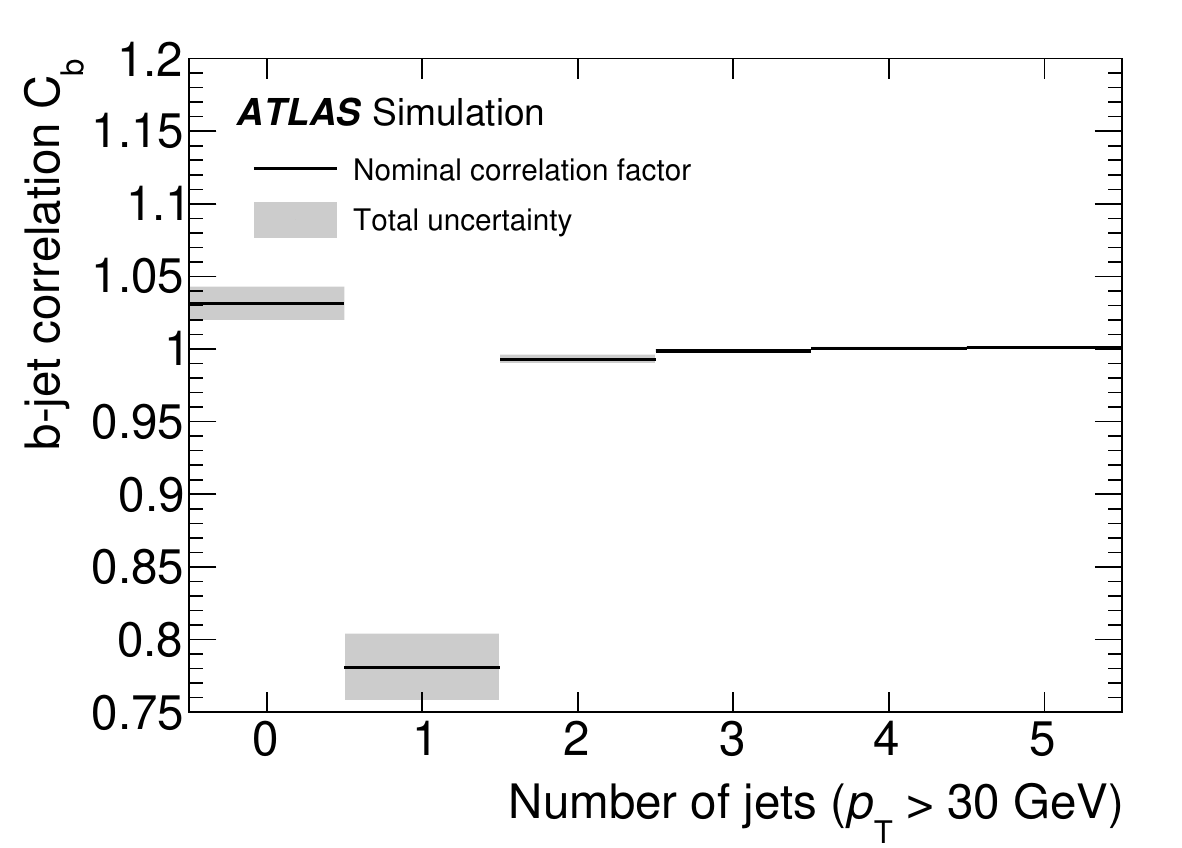}}\\
\caption{The \bjet correlation correction factor $C_b$ as evaluated from simulation as a function of (a) \mll and (b) the number of jets. The error bands show the total uncertainty, including systematic effects.}
\end{figure}

The \bjet counting method significantly reduces experimental and theoretical uncertainties in the \ttbar background estimate, thereby lowering its total uncertainty to about 3\%, approximately a factor of five less than an estimate based purely on MC simulation.

The single-top $Wt$ background is estimated using simulation.
The subtraction of $Wt$ backgrounds in the \ttbar control regions introduces an anti-correlation between the estimated $Wt$ and \ttbar yields in the signal region so that modelling uncertainties in $Wt$ have a reduced impact on the measurement.

In a few analysis bins, the 2-\bjet control region contains only a limited number of events, leading to a large statistical uncertainty in the \ttbar background estimate.
To mitigate this, the top-quark background is instead estimated using a transfer-factor method in these bins.
The \ttbar yield from the corresponding bins of a control region, constructed by requiring at least one \bjet, with the other selection criteria being the same as in the signal region, is extrapolated into the signal region using a transfer factor calculated from \ttbar simulation.
Although this transfer factor has a systematic uncertainty of about 10\% (larger than that of the \btag counting method), it still reduces the impact of experimental and theoretical uncertainties relative to an estimate based purely on simulation.

The top-quark background estimate is validated in a top-enriched region that overlaps with the signal region, by requiring at least one jet and $m_{\ell j} < 140\,\GeV$, as well as $\dphill < \pi/2$ in addition to the normal event selection. Here $m_{\ell j}$ is the invariant mass of the leading jet and the closest lepton. This region is approximately 70\% pure in top-quark events and shows good agreement between the data and the combined signal and background prediction.
\cref{fig:vr-datamc-1-a,fig:vr-datamc-1-b} show the distributions of \ptleadlep and the jet multiplicity, confirming that lepton and jet-related properties are accurately modelled in events without \bjets.

\subsection{Backgrounds with non-prompt or misidentified leptons}
\label{sec:fakebackground_main}

Reducible backgrounds arising from events with non-prompt or misidentified leptons -- referred to as fake-lepton backgrounds or \enquote{fakes} -- originate from heavy-flavour hadron decays and jets misidentified as electrons.
These fake-lepton events, which stem mostly from $W$+jets production, constitute about 4\% of the selected events.

The fake-lepton background is estimated using a data-driven technique. A control region is defined by selecting events in which one of the two lepton candidates fails the nominal selection but meets a looser set of requirements. This strategy enhances the contribution of fake-lepton events while ensuring that the fake lepton's characteristics remain similar to those in the signal region.
Specifically, the lepton that fails the nominal selection is required to fulfil a loose (instead of a tight) isolation requirement and either the medium (instead of tight) identification criteria if it is an electron or, if it is a muon, an impact parameter significance requirement of $|d_0 / \sigma_{d_0}|  < 10$ (instead of $|d_0 / \sigma_{d_0}|  < 3$).
Although this control region is enriched in fake leptons, approximately 75\% of its events still contain two prompt leptons (including electrons or muons from prompt $\tau$-lepton decays).
These prompt-lepton contributions are modelled using simulated event samples, with a dedicated correction factor applied to adjust the probability with which leptons pass the loose, but not the tight, isolation and identification cuts.
Signal contamination in the control region is estimated using the \WW simulation, corrected in each bin by the ratio of background-subtracted data events to the \WW simulation yield in the signal region.
An uncertainty of 5\% is assigned to this correction to account for uncertainties in the signal subtraction.
Although the method can, in principle, be applied iteratively to refine the background subtraction involved, it is found to converge after just one iteration.

The fake-lepton background in the signal region is then obtained by scaling the number of data events in the control region -- after subtracting processes with two prompt leptons -- by an extrapolation factor.
This factor, determined from a data sample dominated by fake leptons, depends on the \pt, $|\eta|$, and flavour of the lepton. The sample is selected by requiring a dijet-like configuration where one lepton candidate recoils against a jet, with $\Delta \phi_{\ell j}>2.8$. To further suppress contamination from $W$+jets events, the sum of the \met and the transverse mass of the lepton-and-\met system is required to be smaller than 50~\GeV.
The extrapolation factor is defined as the ratio of the number of events in which a lepton satisfies the nominal selection to the number in which it meets only the looser criteria.
Systematic uncertainties associated with the contributions from different sources of fake leptons are estimated by varying the selection of the data sample in which the extrapolation factors are determined. The variations include selecting events with a \bjet recoiling against the lepton candidate, selecting only events without a \bjet, and changing the \met requirements to increase the fake-lepton contributions. The total uncertainty in the fake-lepton background is about 25\%.

To validate the fake-lepton background estimate, the opposite-charge requirement in the signal region selection is inverted, and events containing same-charge electron--muon pairs are selected instead.
This selection increases the contribution of $W$+jets events to about $25\%$ by suppressing prompt-lepton processes, which are mostly charge asymmetric.
The modelling of the fake-lepton background can be validated despite the relatively low purity
since the dominant $WZ$ background in this region is known with a precision of about 10\%.
Good agreement between the prediction and the data is observed, as is shown for the \ptsublep distribution in Figure~\ref{fig:vr-datamc-1-c}.

\subsection{Drell--Yan background}
The Drell--Yan $Z$+jets background is estimated using MC simulation. The $\mll>85~\GeV$ requirement strongly suppresses this background.
The contribution of this background to the selected events in the signal region is about 5\%, almost entirely due to $Z/\gamma^*(\rightarrow\tau^+\tau^-)$+jets events.

The $Z$+jets estimate is checked in a validation region requiring a dilepton invariant mass between 45\,\GeV and 80\,\GeV, and either $\ptll< 30\,\GeV$ or $\met < 20\,\GeV$, in addition to the \bjet veto.
The $Z$+jets purity of this region is about 85\% and the data is observed to be well modelled. Figure~\ref{fig:vr-datamc-1-d} shows the distribution of the dilepton invariant mass \mll in the validation region, which features the resonant $Z\rightarrow\tau\tau$ distribution over a rising background of top-quark events.

In addition to the theoretical uncertainty of 5\% in the $Z$+jets cross-section~\cite{STDM-2015-03}, uncertainties in the $Z$+jets background are estimated by varying the renormalization and factorization scales used in the matrix-element calculation by factors of two and one-half, avoiding variations in opposite directions.

\begin{figure}
\centering
\subfloat[\label{fig:vr-datamc-1-a}]{\includegraphics[width=.45\linewidth]{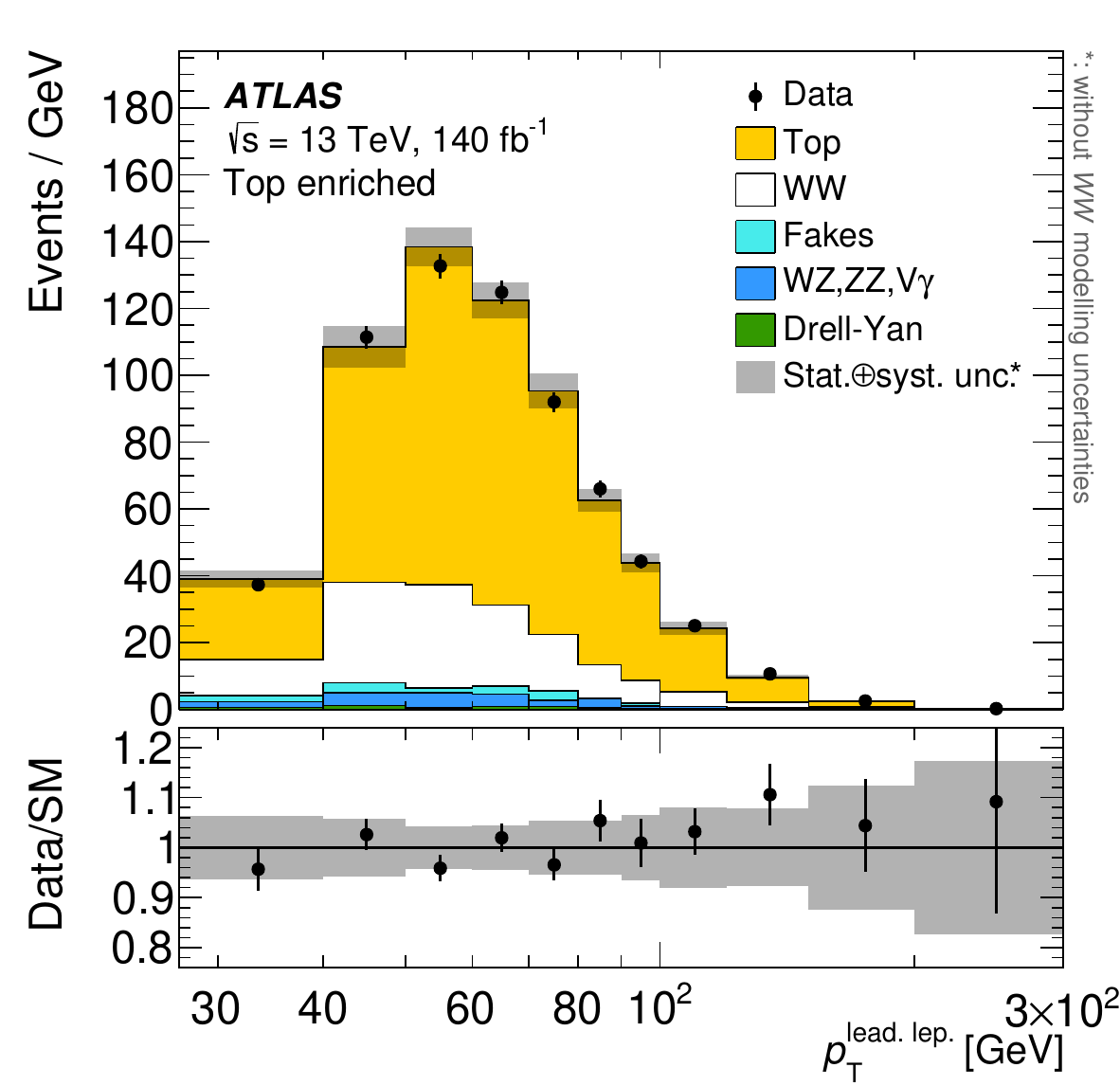}}
\subfloat[\label{fig:vr-datamc-1-b}]{\includegraphics[width=.45\linewidth]{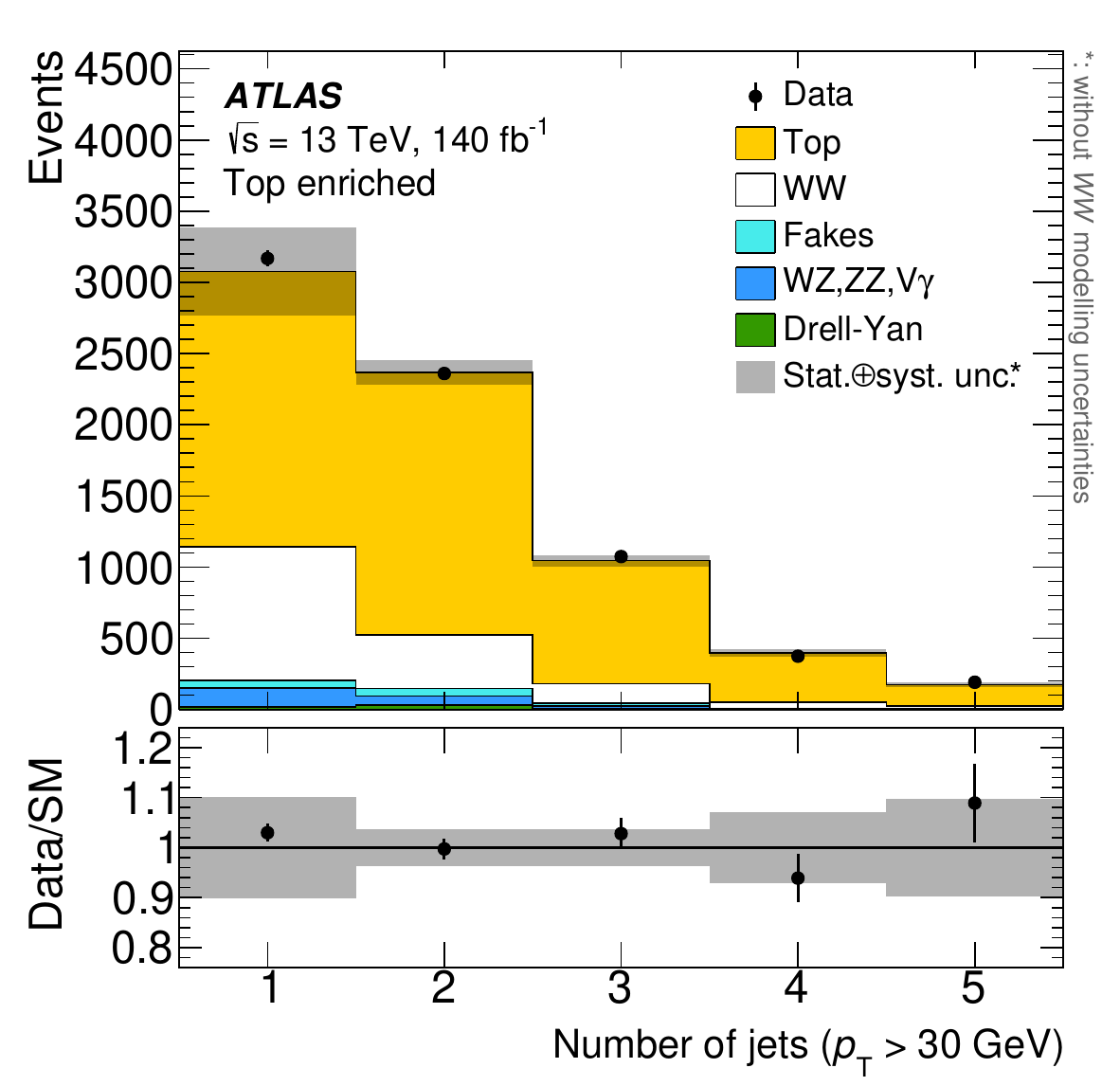}}\\
\subfloat[\label{fig:vr-datamc-1-c}]{\includegraphics[width=.45\linewidth]{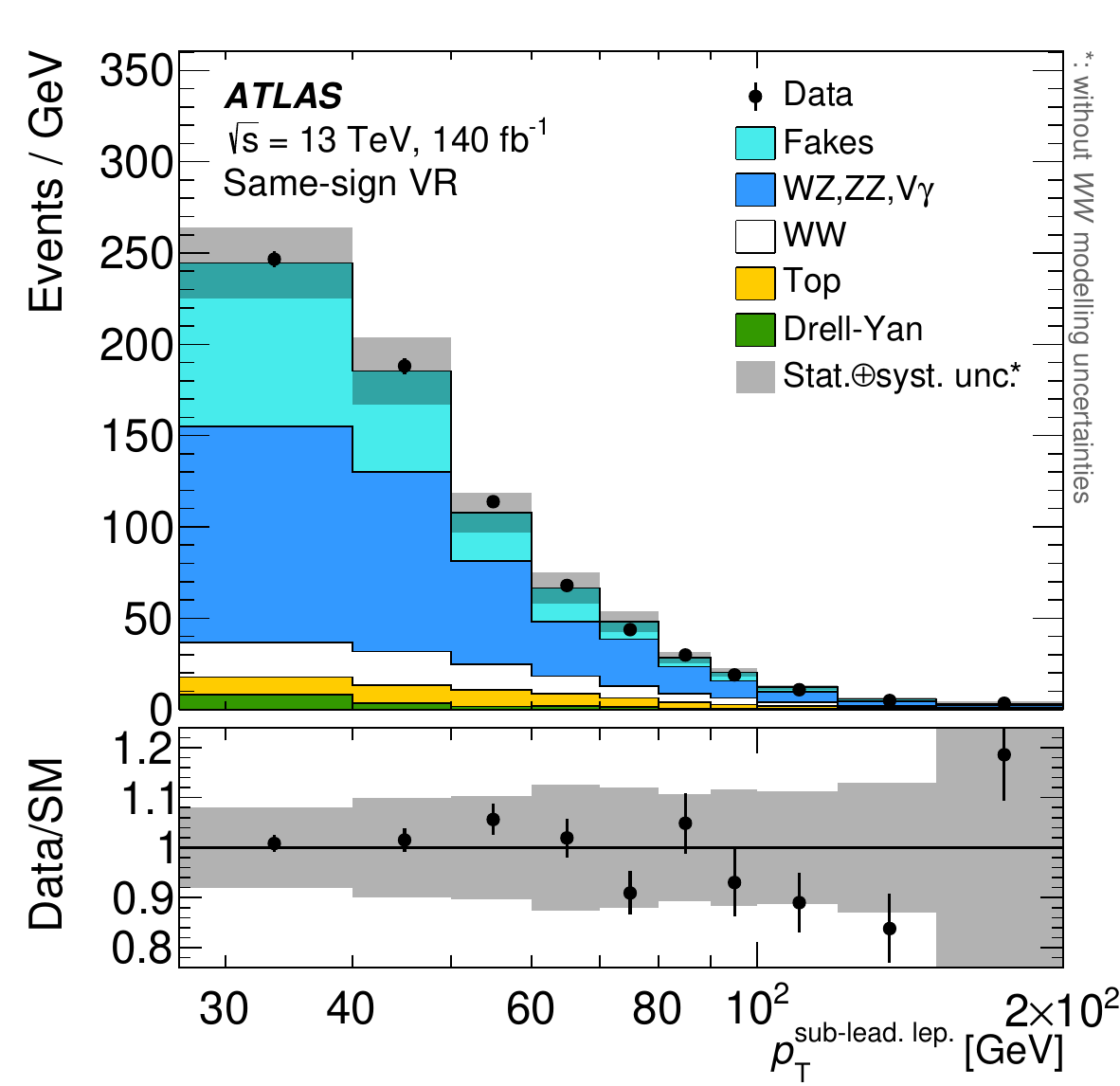}}
\subfloat[\label{fig:vr-datamc-1-d}]{\includegraphics[width=.45\linewidth]{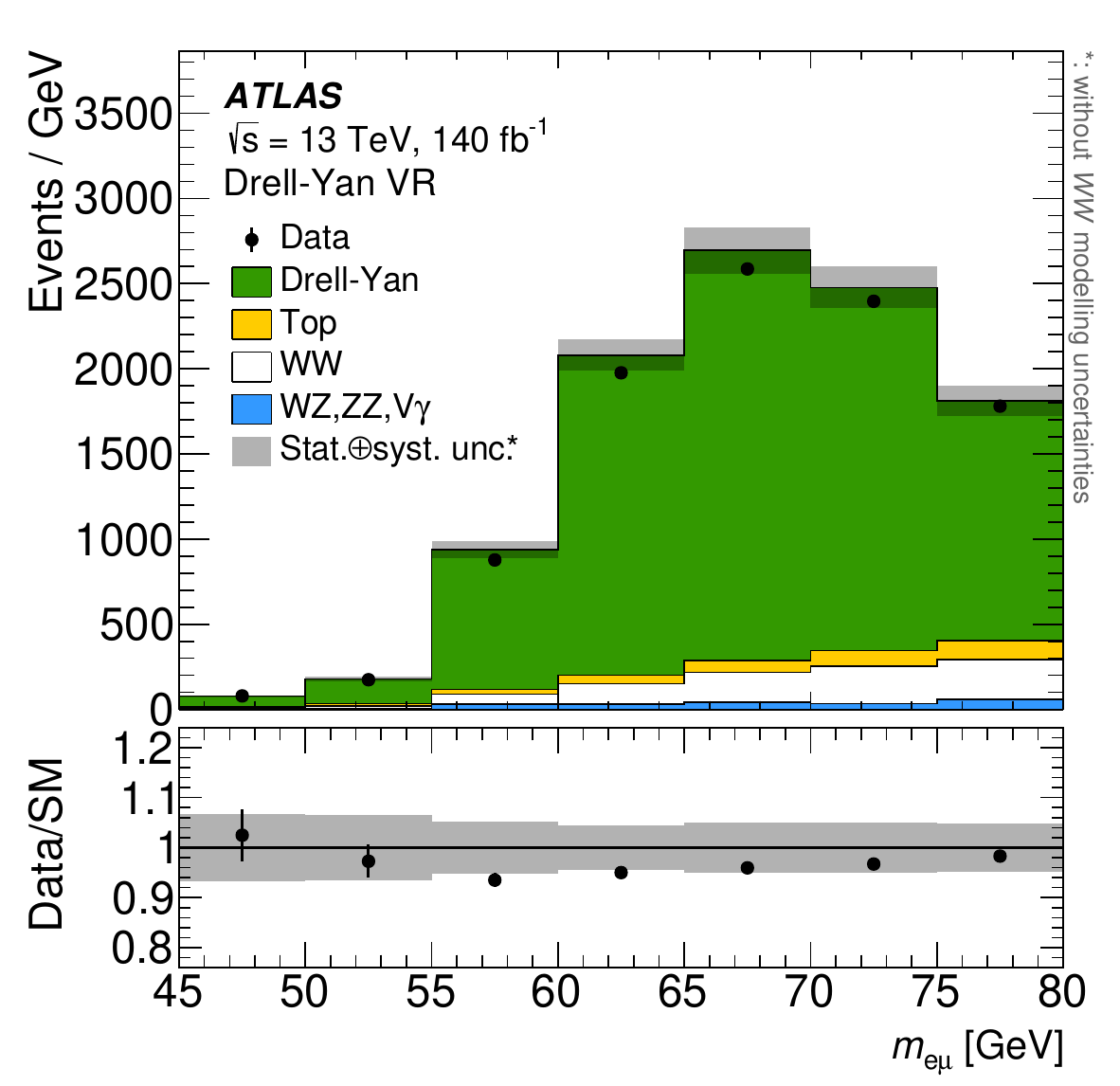}}
\hspace{.4\linewidth}
\caption{Detector-level distributions of (a) \ptleadlep and (b) the jet multiplicity in the top-enriched region, (c) \ptsublep in the same-sign validation region (VR), and (d) \mll in the Drell--Yan VR. The rightmost bin contains overflow events.
Data are shown as black markers, together with histograms of the predictions for signal and background processes. The top-quark background in the top-enriched region is estimated using the data-driven method explained in Section~\ref{sec:topquarkbkg}; in the other two regions its contribution is small and the nominal MC prediction is used instead.
The lower panels show the ratio of the data to the total prediction.
The uncertainty bands shown include statistical and systematic uncertainties. Theory uncertainties in the signal are negligible and not shown.
}
\end{figure}

\subsection{Other backgrounds}
Backgrounds from $WZ$, $ZZ$, $W\gamma$ and $Z\gamma$ production are estimated from simulation, and are found to contribute about 4\% of the selected events. The dominant contribution comes from $WZ$ events, which are observed to be well described by the nominal \SHERPA simulation in Ref.~\cite{STDM-2018-03}.
Uncertainties due to missing higher-order QCD corrections are derived by varying the factorization and renormalization scales used in the matrix-element calculations by factors of two, avoiding variations in opposite directions. Additionally, an uncertainty of 10\% is assigned to the diboson production cross-section~\cite{ATL-PHYS-PUB-2016-002,Grazzini:2015nwa}.

The $VZ$ ($WZ$ and $ZZ$) prediction is validated in events containing a third lepton that has $\pt \ge 10$\,\GeV and satisfies loosened identification criteria. The invariant mass of the resulting same-flavour opposite-charge pair of leptons is required to be between 80\,\GeV and 100\,\GeV, close to the $Z$-boson mass, while the remaining selection criteria are identical to those in the signal region. These selections give a very pure sample of diboson events, and the prediction is in good agreement with the data, as seen in Figure~\ref{fig:vr-datamc-2-a}, where the \met distribution in the $VZ$ validation region shows separation between $ZZ$ and $WZ$ events.

Some $V\gamma$ ($W\gamma$ and $Z\gamma$) events enter the signal region as backgrounds when the photon is reconstructed and selected as an electron candidate. To validate estimates of these backgrounds, the electron identification requirements are loosened and electron track requirements changed such that contributions from photon conversions increase. As the electron candidates reconstructed from photon conversion are charge symmetric,
both electron-candidate charge states are selected, regardless of the selected muon's charge.
Figure~\ref{fig:vr-datamc-2-b} shows the \pt distribution of the electron candidates in the $V\gamma$ validation region with opposite-charge $e{-}\mu$ pairs. It is dominated by electrons from photon conversion. Excellent agreement with the data is found in both validation regions.

\begin{figure}
\centering
\subfloat[\label{fig:vr-datamc-2-a}]{\includegraphics[width=.45\linewidth]{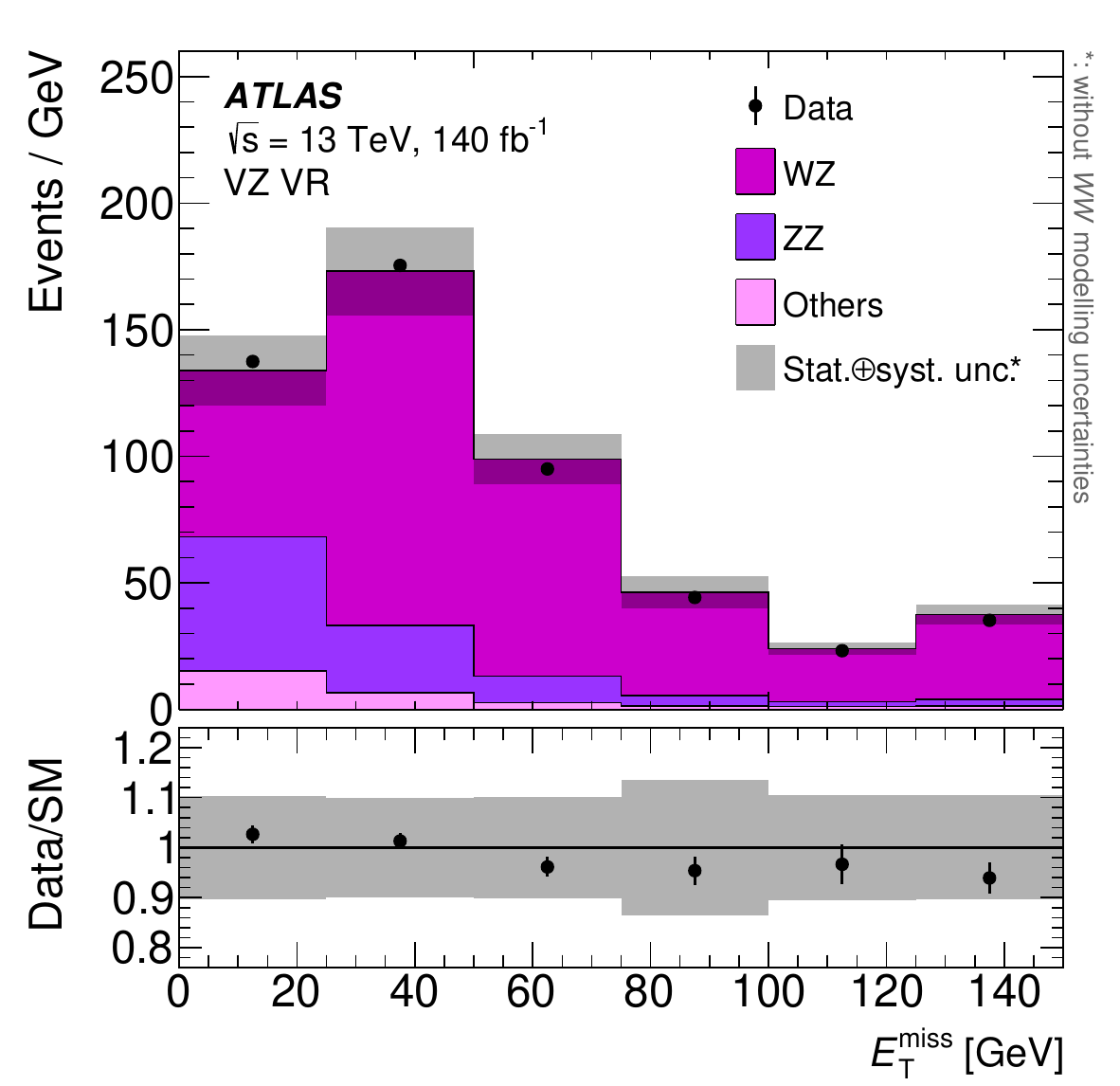}}
\subfloat[\label{fig:vr-datamc-2-b}]{\includegraphics[width=.45\linewidth]{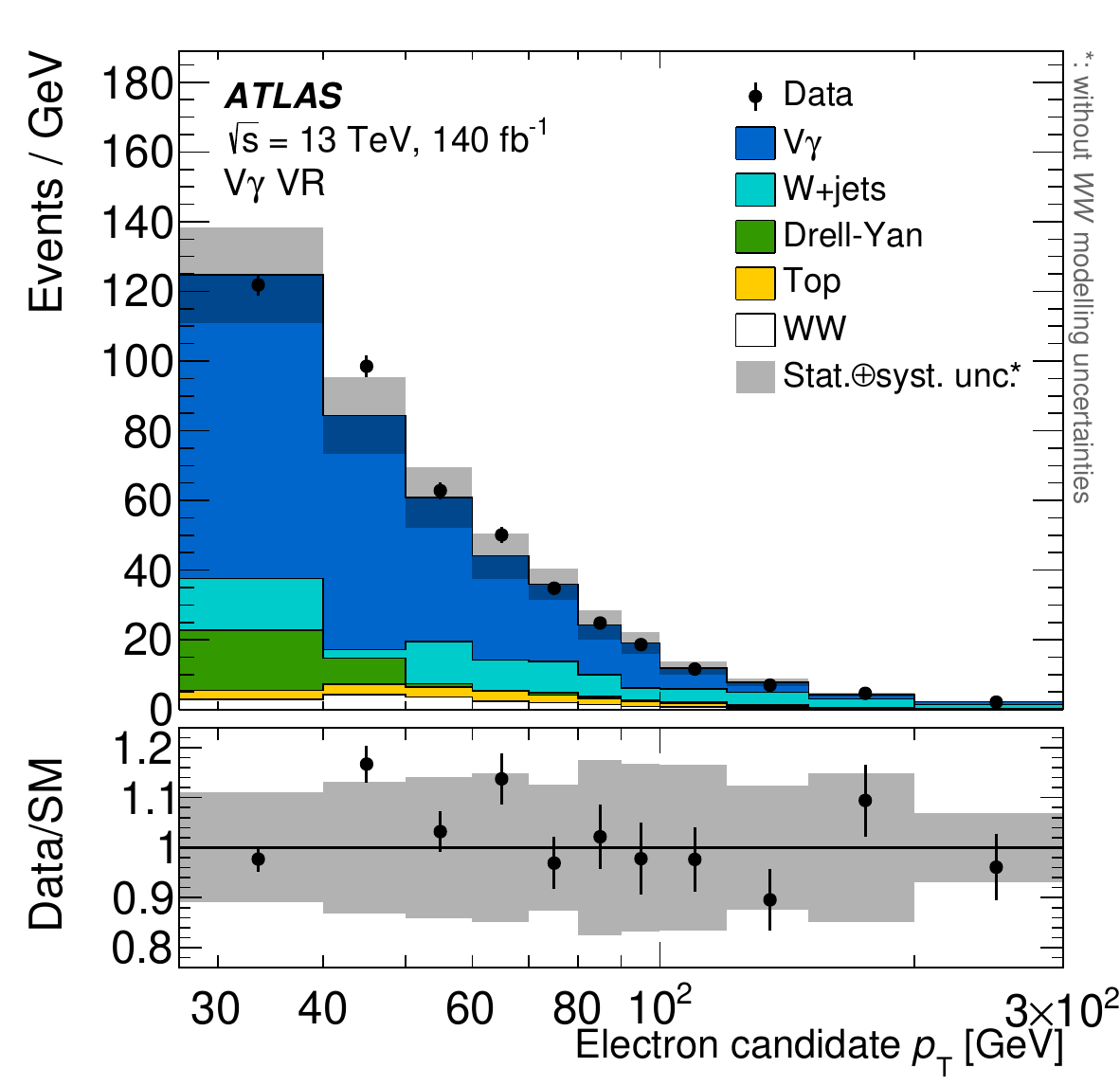}}
\hspace{.4\linewidth}
\caption{Detector-level distributions of (a) \met in the $VZ$ validation region (VR) and (b) the electron candidate \pt in the $V\gamma$ VR with opposite-charge leptons. The rightmost bin contains overflow events.
Data are shown as black markers, together with histograms of the predictions for signal and background processes. The nominal MC prediction is used for the top-quark background in both regions.
The lower panels show the ratio of the data to the total prediction.
The uncertainty bands shown include statistical and systematic uncertainties.  Theory uncertainties in the signal are negligible and not shown.
}
\end{figure}

\subsection{Selected \WW candidate events}
\begin{table}
\caption{Selected \WW candidate events in data, together with the signal prediction and the background estimates. The uncertainties include statistical and systematic contributions, excluding theory uncertainties in the signal. The fractions in percent give the relative contribution to the total SM prediction. The individual uncertainties are correlated, and do not add up in quadrature to the total uncertainty. }
\label{tab:sr-yields}
\centering
\begin{tabular}{@{}lr@{$\,\pm\,$}rrrr@{$\,\pm\,$}rr@{}}
\toprule
Category            & \multicolumn{2}{r}{Event yield} & Fraction        \\
\midrule
Data                & \multicolumn{2}{r}{144221}      & \rule{0.5em}{0pt}        \\
\midrule
Total SM            & 139500                          & 2400                     \\
\midrule
$WW$                & 56900                           & 1100              & 41\%  \\
Total background          & 82600                           & 2100              & 59\%  \\
\midrule
Top                 & 66500                           & 1900              & 48\%  \\
Drell--Yan           & 6500                            & 400               & 5\%   \\
Fakes               & 5000                            & 1300              & 4\%   \\
$WZ$, $ZZ$, $V\gamma$ & 4500                            & 600               & 3\%   \\
\bottomrule
\end{tabular}
\end{table}

\begin{figure}
\centering
\subfloat[]{\includegraphics[width=.45\linewidth]{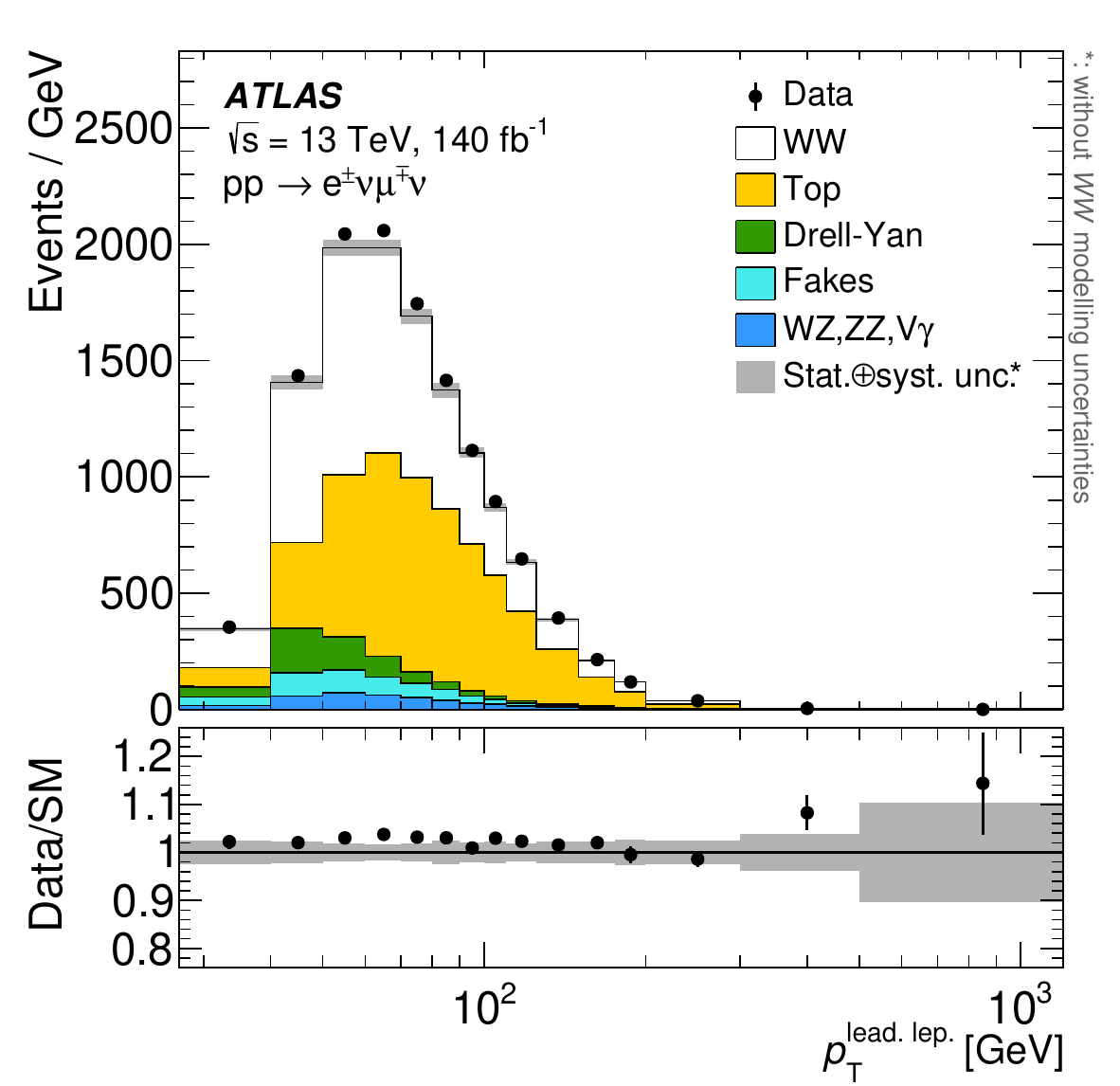}}
\subfloat[]{\includegraphics[width=.45\linewidth]{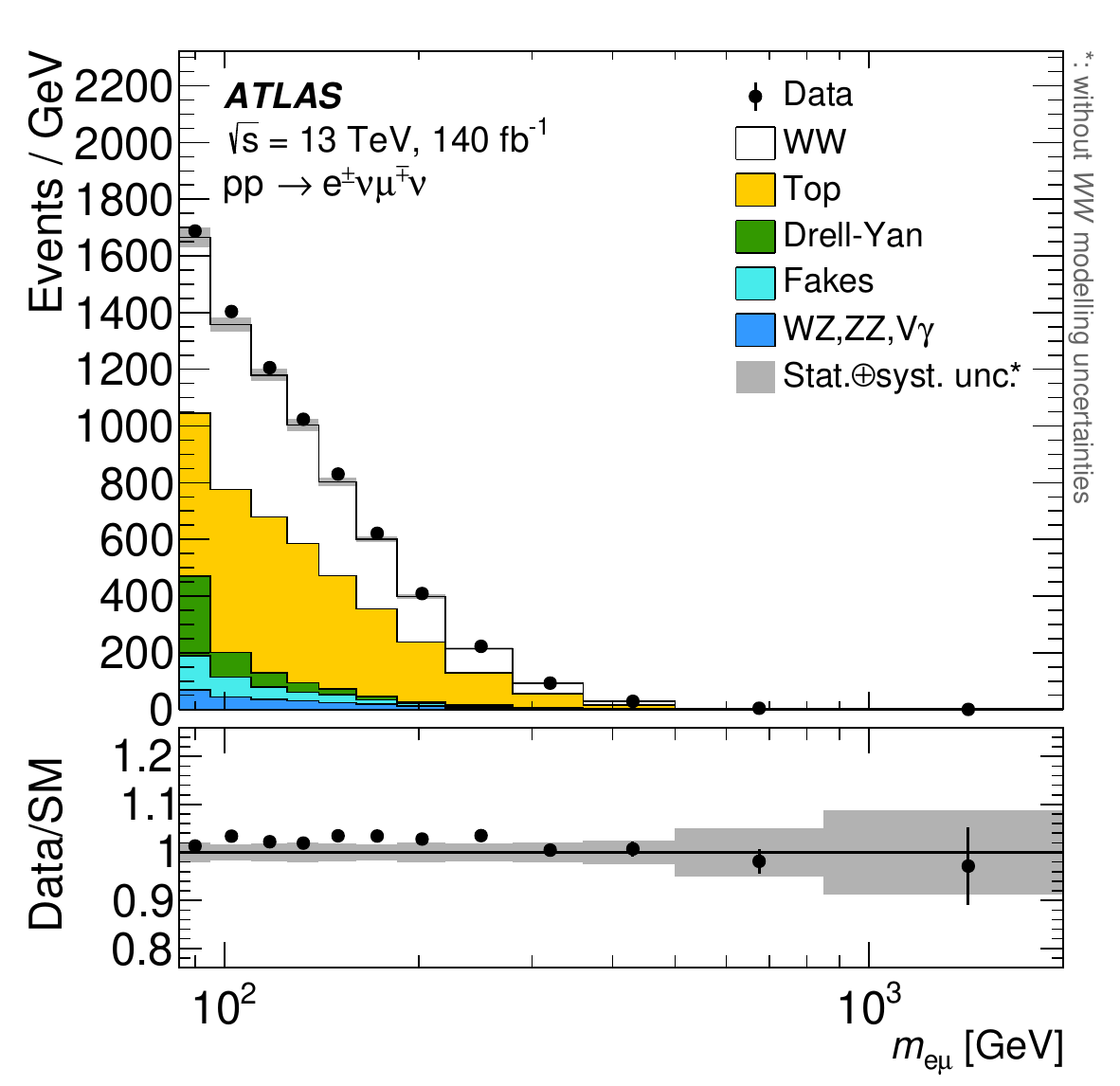}}\\[-2.8ex]
\subfloat[]{\includegraphics[width=.45\linewidth]{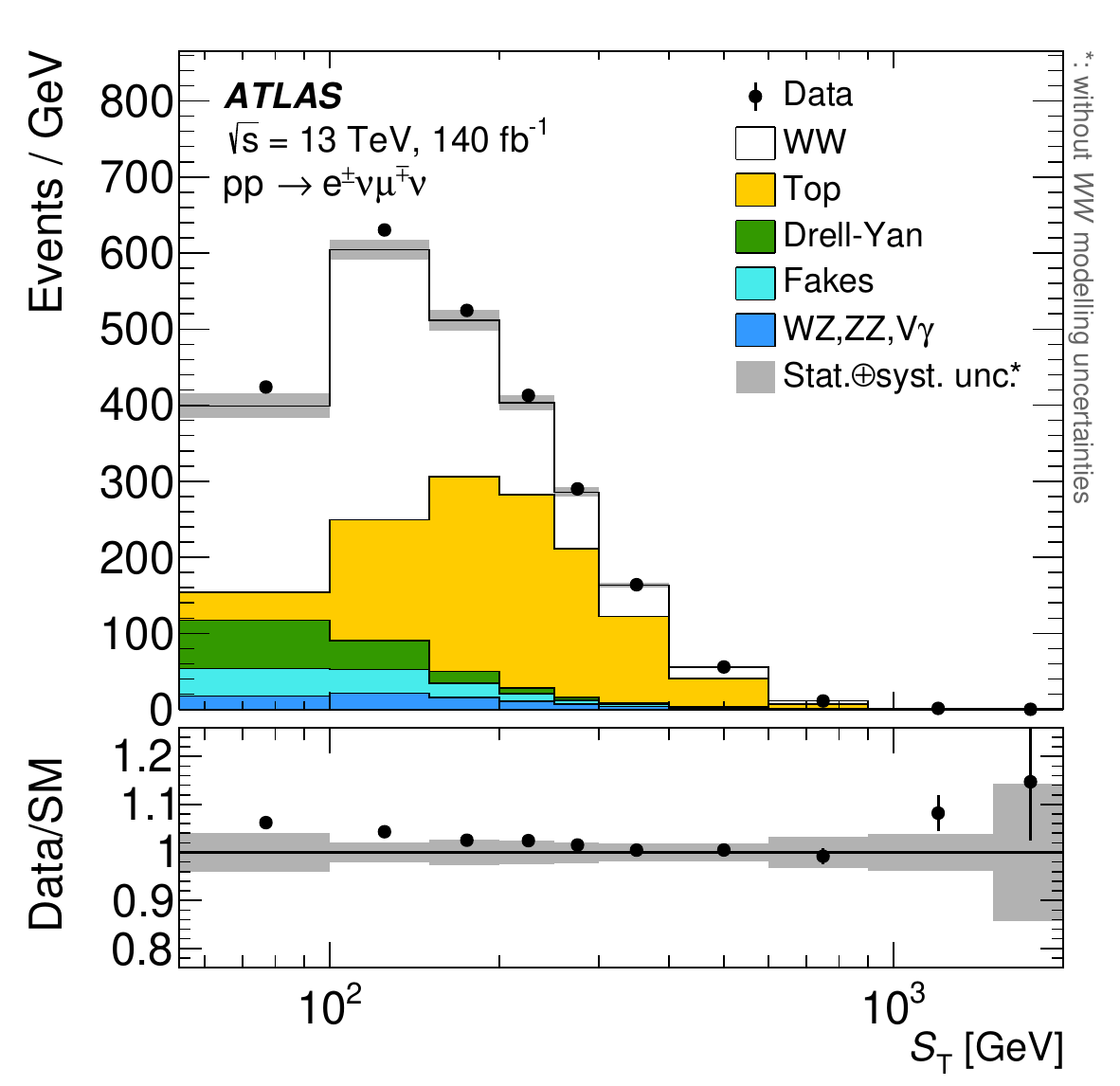}}
\subfloat[]{\includegraphics[width=.45\linewidth]{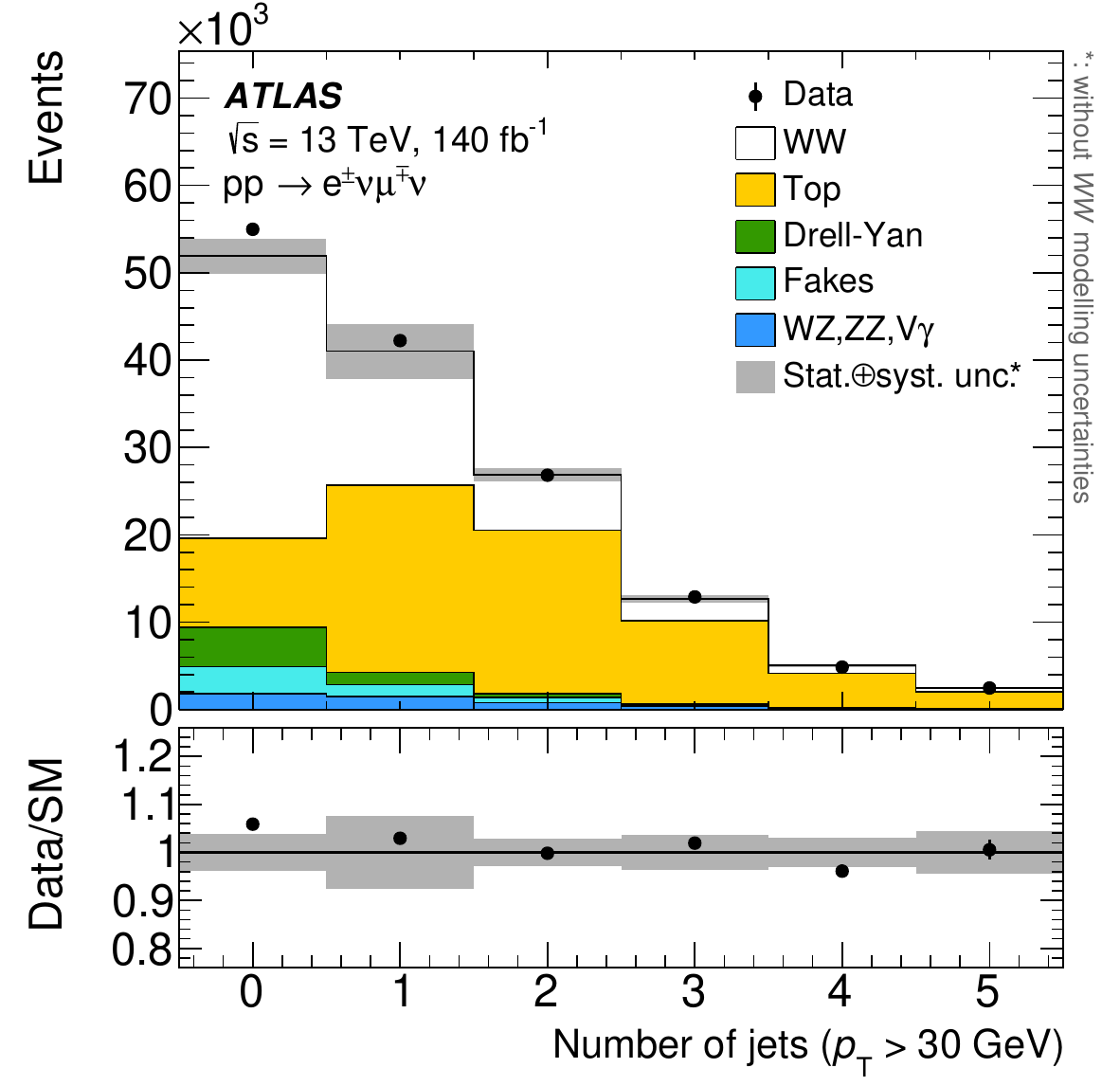}}\\%
\caption{Detector-level distributions of (a) \ptleadlep, (b) \mll, (c) \ST, and (d) the jet multiplicity.
Data are shown as black markers together with histograms of the predictions for signal and background processes. The rightmost bin contains overflow events. The lower panels show the ratio of the data to the total prediction. Top-quark and fake-lepton backgrounds are determined using data-driven methods. The uncertainty bands shown include statistical and systematic uncertainties, excluding theory uncertainties in the signal, which largely cancel out in the measurement of $WW$ cross-sections.}
\label{fig:reco-level-dist}
\end{figure}

Table~\ref{tab:sr-yields} lists the number of selected \WW candidate events, as well as the breakdown of the background predictions. Details of the systematic uncertainties are given in Section~\ref{sec:systematics}.
Figure~\ref{fig:reco-level-dist} shows distributions at detector level and compares the observed data with the signal prediction and the background estimate.
A small excess of data events is observed, in particular at low jet-activity values. In these bins, \WW production is expected to make the dominant contribution to the event yield.


%
\section{Fiducial and differential cross-section determination}
\label{sec:unfolding}

The \WW cross-section is evaluated in the fiducial volume defined in Table~\ref{tab:fiducialtruthlevelcuts} at particle level, i.e.\ without detector effects and reconstruction inefficiencies.
Exactly one prompt electron and one prompt muon of opposite electric charge are required.
Prompt leptons are defined as leptons that do not originate from $\tau$-lepton or hadron decays.
The momenta of photons not originating from hadron decays but emitted in a cone of size $\Delta R = 0.1$ around the lepton direction are added to the lepton momentum to form infrared-safe \enquote{dressed} leptons.
Kinematic requirements on leptons reflect the analysis requirements of Table~\ref{tab:selection}, and the invariant mass of the two leptons, \mll, is required to be greater than 85\,\GeV. Events with additional prompt leptons fulfilling a looser \pt requirement are vetoed.
Stable final-state particles,\footnote{Particles are considered stable if their decay length $c\tau$ is greater than 1\,cm.} excluding prompt neutrinos as well as prompt charged leptons and the associated photons, are clustered into particle-level jets using the anti-$k_t$ algorithm with radius parameter $R = 0.4$.
The nominal definition of the fiducial phase space includes a veto on particle-level \bjets with $\pT > 20$\,\GeV that are defined by ghost-association~\cite{Cacciari:2007fd} of $b$-hadrons.
The missing transverse momentum is defined at particle level as the negative vectorial sum of the transverse momenta of visible particles.

\begin{table}[htbp]
\caption{Definition of the particle-level objects and fiducial phase space.
Jets containing $b$-hadrons are labelled \emph{$b$-jet} while other jets are labelled \emph{jet}.}
\begin{center}

\begin{tabular}{lll}
\toprule
Object                         & Requirement      & Criteria                            \\
\midrule
\multirow{2}{*}{Prompt lepton} & $\pt$            & $>27\,\GeV$                         \\
& $|\eta|$         & $< 2.5$                             \\
\hline
\multirow{2}{*}{Loose lepton}  & $\pt$            & $>10\,\GeV$                         \\
& $|\eta|$         & $< 2.5$                             \\
\hline
\multirow{2}{*}{\bjets}        & $\pt$            & $> 20\,\GeV$                        \\
& $\abseta$        & $<2.5$                              \\
\hline
\multirow{2}{*}{Jets}          & $\pt$            & $> 30\,\GeV$                        \\
& $\abseta$        & $<4.5$                              \\
\hline
\multirow{5}{*}{Event}         &                  & 1 prompt electron and 1 prompt muon \\
& Leptons          & of opposite electric charge.        \\
&                  & No additional loose leptons         \\
& Number of \bjets & 0                                   \\
& \mll             & $> 85\,\GeV$                        \\
\bottomrule
\end{tabular}
\end{center}
\label{tab:fiducialtruthlevelcuts}
\end{table}

The integrated fiducial cross-section is determined using a profile likelihood fit of the \ST distribution.
The fit of the \ST distribution reduces uncertainties associated with the top-quark background, which dominates at high \ST, and, to a lesser extent, uncertainties in Drell--Yan and fake-lepton backgrounds, which are most important at low \ST.
A model based on the sum of background estimates, which for top-pair production and fake-lepton events are determined using the data-driven methods presented in Sections~\ref{sec:topquarkbkg} and~\ref{sec:fakebackground_main}, is fitted to the data, along with the signal prediction, which includes \qqWW and \ggWW production as well as electroweak \WWjj production.
The number of events in each bin is modelled as the predicted number of signal events scaled by a signal-strength modifier $\mu$, plus the number of background events. Systematic uncertainties in the signal and background predictions are parameterized by nuisance parameters.
The likelihood is modelled using RooFit~\cite{RooFit} as a product of Poissonian distributions multiplied by Gaussian constraints on the nuisance parameters.
The fiducial cross-section is obtained by multiplying the unconditional maximum-likelihood estimate of $\mu$ by the particle-level fiducial cross-section prediction of the signal model.
Since the fit relies on a correct prediction of the signal shape from MC simulation, it is subject to signal modelling uncertainties and only valid for a \WW signal like that in the SM.

The differential cross-sections are determined using an iterative Bayesian unfolding method~\cite{DAgostini:1994fjx,BayesUnfold2}.
In contrast to the integrated cross-section measurement, these results depend only weakly on the signal model, which is only used to estimate detector resolution and efficiency, and remain approximately valid in the presence of physics beyond the SM.
Background events are subtracted in accord with the estimates discussed in Section~\ref{sec:background}.
A fiducial correction is applied to take into account events that are reconstructed in the signal region, but originate from outside the fiducial region at particle level (about 13\% of selected events, mostly events with leptons from $\tau$ decays).
The unfolding procedure corrects for migrations between bins in the distributions during the reconstruction of the events.
Finally, efficiency corrections take into account events inside the fiducial region that are not reconstructed in the signal region due to detector inefficiencies (about 44\% of events).
To reduce the bias from using the chosen theory prediction to give the \enquote{true} distribution, the method can be applied with multiple iterations, at the cost of increased statistical uncertainty.
Due to the good modelling of the data by simulation and the relatively small migration effects, the method converges quickly and only two iterations are required for most observables, the only exception being the jet multiplicity distribution, which is subject to larger modelling uncertainties and is unfolded using three iterations.
The bias from using the simulation distributions as priors in the unfolding is estimated by reweighting the simulation with a smooth function such that it closely resembles the background-subtracted data.
This reweighted detector-level prediction is unfolded using the nominal unfolding setup.
The unfolding procedure is able to very accurately recover the generator-level distribution, indicating that any potential bias introduced by the unfolding method can be neglected.


\newcommand{\SystFigCaption}{ \enquote{Jet calibration} uncertainties encompass jet energy scale and resolution uncertainties, \enquote{Top modelling} is the uncertainty in the theoretical modelling of the top-quark background, and \enquote{Fake lepton backgr.} is the uncertainty in the estimate of the fake-lepton background. All systematic uncertainties related to minor prompt-lepton backgrounds, flavour-tagging efficiencies and mis-tag rates, the luminosity, lepton calibration, pile-up reweighting, and signal modelling, are included in \enquote{Other systematics}. \enquote{Statistical uncertainty} combines statistical uncertainties arising in the signal region and the control regions used for the data-driven top-quark and fake-lepton estimates, and those in backgrounds estimated using MC simulations.}

\section{Uncertainties}
\label{sec:systematics}
Systematic uncertainties in the \WW cross-section measurements arise from experimental sources, the background determination, the procedures used to correct for detector effects, and theoretical uncertainties in the signal modelling.
Systematic uncertainties in the differential cross-sections are evaluated by repeating the unfolding procedure with simulations making varied assumptions in the signal, background, and detector model. The resulting uncertainties are symmetrized and treated as being uncorrelated.
In the likelihood that is used to determine the integrated fiducial cross-section, the same uncertainty sources are instead modelled by nuisance parameters that are profiled in the fit.

The dominant experimental systematic uncertainties arise in the determination of the \btagging efficiency and mis-tag rates~\cite{FTAG-2018-01}, the calibration of the jet energy scale and resolution~\cite{JETM-2018-05}, and the luminosity measurement~\cite{DAPR-2021-01}.
Experimental uncertainties also encompass uncertainties in the correction of lepton trigger~\cite{TRIG-2018-05,TRIG-2018-01}, reconstruction, identification and isolation efficiencies~\cite{EGAM-2018-01,MUON-2018-03}, the calibration of the lepton momentum or energy scale and resolution~\cite{EGAM-2018-01,MUON-2022-01}, and the modelling of pile-up.
All experimental uncertainties are evaluated by varying the respective calibrations, and propagating their effects through the analysis, affecting both the background estimates and the unfolding of detector effects. Both the effect on the total rate and the effect on the shape of distributions are taken into account for all sources of systematic uncertainty.

The estimate of the top-quark background is affected by the statistical uncertainty of the number of events in the control region, and by uncertainties in the modelling of \ttbar and single-top $Wt$ events, such as the uncertainty in the matrix-element calculation, the parton-shower modelling, the QCD scales, the  amount of initial- and final-state radiation, and the modelling of  interference between \ttbar and single-top $Wt$ events~\cite{ATL-PHYS-PUB-2020-023}.
These  are e\-val\-uat\-ed by using the alternative simulations described in Section~\ref{sec:data_mc_samples} and propagating the results through the top-quark background estimation.

Systematic uncertainties in the estimate of fake leptons are derived by changing the selection used to estimate the extrapolation weights, in order to change the contributions from different sources of fake leptons. Additionally, the subtraction of the prompt-lepton sources in the control region is varied, and the statistical uncertainties of the weights are propagated. More details of the uncertainties affecting the fake-lepton estimate can be found in Section~\ref{sec:background}.

The uncertainty in additional backgrounds is estimated by varying each of their cross-sections within its uncertainty and by using scale variations to account for missing higher-order QCD corrections and parton-shower modelling uncertainties.
The impact of these uncertainties on the cross-section measurements is small compared to the uncertainties associated with the fake-lepton and top-quark background.

The uncertainty due to missing higher-order QCD corrections in the signal simulation is evaluated by varying the renormalization and factorization scales by factors of two and one-half, avoiding variations in opposite directions. The renormalization scales are varied independently in the matrix element and the parton shower. Additional uncertainties from the parton shower are evaluated by varying the parameters of the A14 tune within their uncertainties, with the largest uncertainty resulting from the variation VAR3c~\cite{ATL-PHYS-PUB-2014-021}, affecting the modelling of initial-state radiation.
An uncertainty due to the modelling of heavy-flavour jets is estimated by varying the fraction of events containing at least one jet originating from a $b$-quark or a $c$-quark by 30\%, which covers the difference between predictions from \PYTHIA[8.230] and \SHERPA[2.2.2].

Statistical uncertainties in the unfolded distributions are evaluated by creating pseudo-data samples by varying the data within their Poisson uncertainties in each bin and then propagating these varied samples through the unfolding.
The statistical uncertainties in the background estimates, which include statistical uncertainties in MC predictions and others due to the control regions used in estimating the top-quark and fake-lepton backgrounds, are evaluated using the same method.

Table~\ref{tab:syst-breakdown} gives a breakdown of the uncertainties in the fiducial cross-section measured in the profile likelihood fit.
Figures~\ref{fig:DiffErrs-mll}~and~\ref{fig:DiffErrs-st} display the uncertainties as a function of the unfolded \mll and \ST distributions, respectively.

\begin{table}
\caption{Effects of uncertainties on the integrated fiducial cross-section measurement with the profile likelihood fit. They are evaluated by fixing each nuisance parameter individually to the boundaries of its post-fit 68\% confidence interval, repeating the fit, and adding in quadrature the effects for each uncertainty group.
\enquote{Top modelling} and \enquote{Signal modelling} are uncertainties in the theoretical modelling of the respective processes, \enquote{Fake-lepton background} is the uncertainty in the fake-lepton estimate while \enquote{Other background} is the uncertainty due to minor prompt-lepton backgrounds, \enquote{Flavour tagging} is all uncertainties in the flavour-tagging efficiency and mis-tag rate,  \enquote{Jet calibration} uncertainties encompass jet energy scale and resolution uncertainties, and \enquote{Luminosity} is the uncertainty in the luminosity measurement. All systematic uncertainties belonging to none of the above categories are included in \enquote{Other systematic uncertainties}. Statistical uncertainties arise in both the signal region and the control region used for the data-driven top-quark and fake-lepton estimates, and from backgrounds estimated using MC simulations.
}
\label{tab:syst-breakdown}
\centering

\begin{tabular}{@{}lr@{}}
\toprule
Uncertainty source             & Effect \\
\midrule
Total uncertainty              & 3.1\%  \\
Statistical uncertainty              & 1.1\%  \\
\midrule
Top modelling                  & 1.6\%  \\
Fake-lepton background         & 1.5\%  \\
Flavour tagging                & 0.7\%  \\
Other background               & 0.9\%  \\
Signal modelling               & 1.0\%  \\
Jet calibration                & 0.6\%  \\
Luminosity                     & 0.8\%  \\
Other systematic uncertainties & 0.9\%  \\
\bottomrule
\end{tabular}
\end{table}

\begin{figure}
\centering
\subfloat[\label{fig:DiffErrs-mll}]{\includegraphics[width=.45\linewidth]{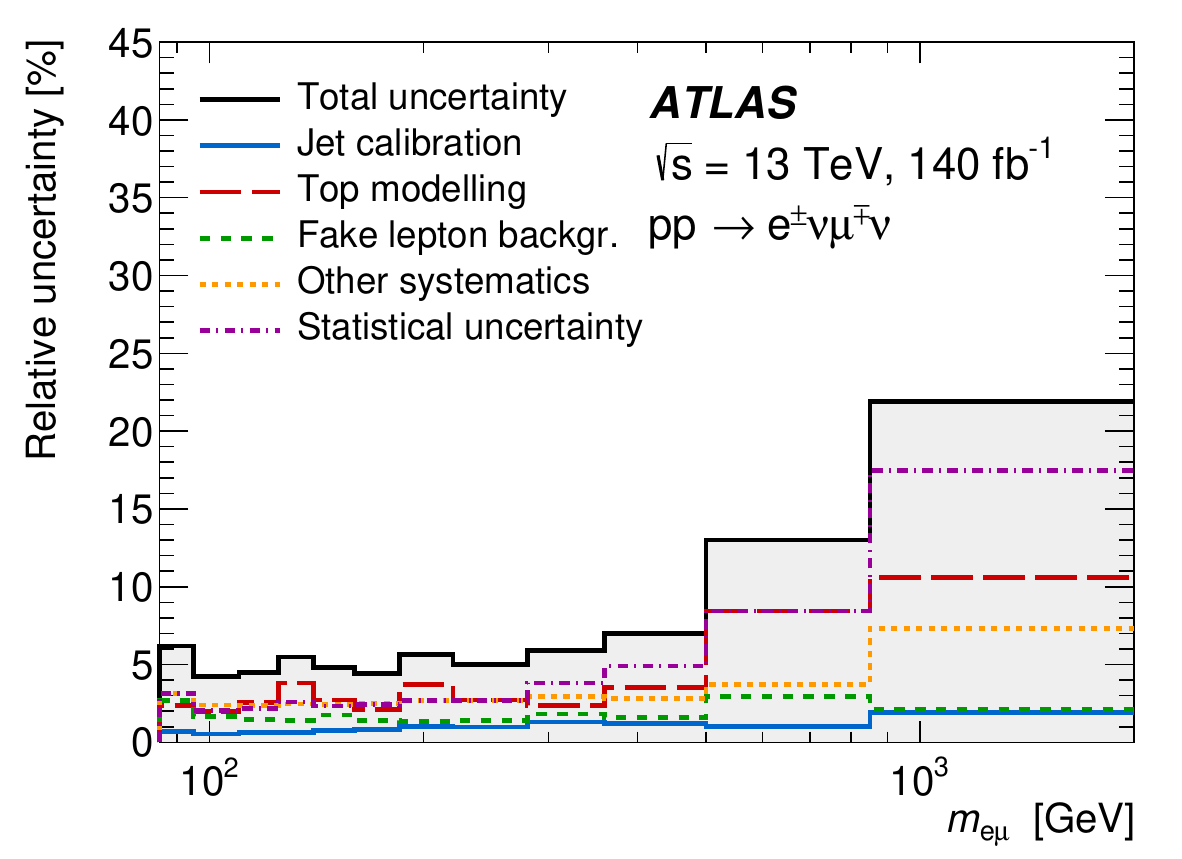}}
\subfloat[\label{fig:DiffErrs-st}]{\includegraphics[width=.45\linewidth]{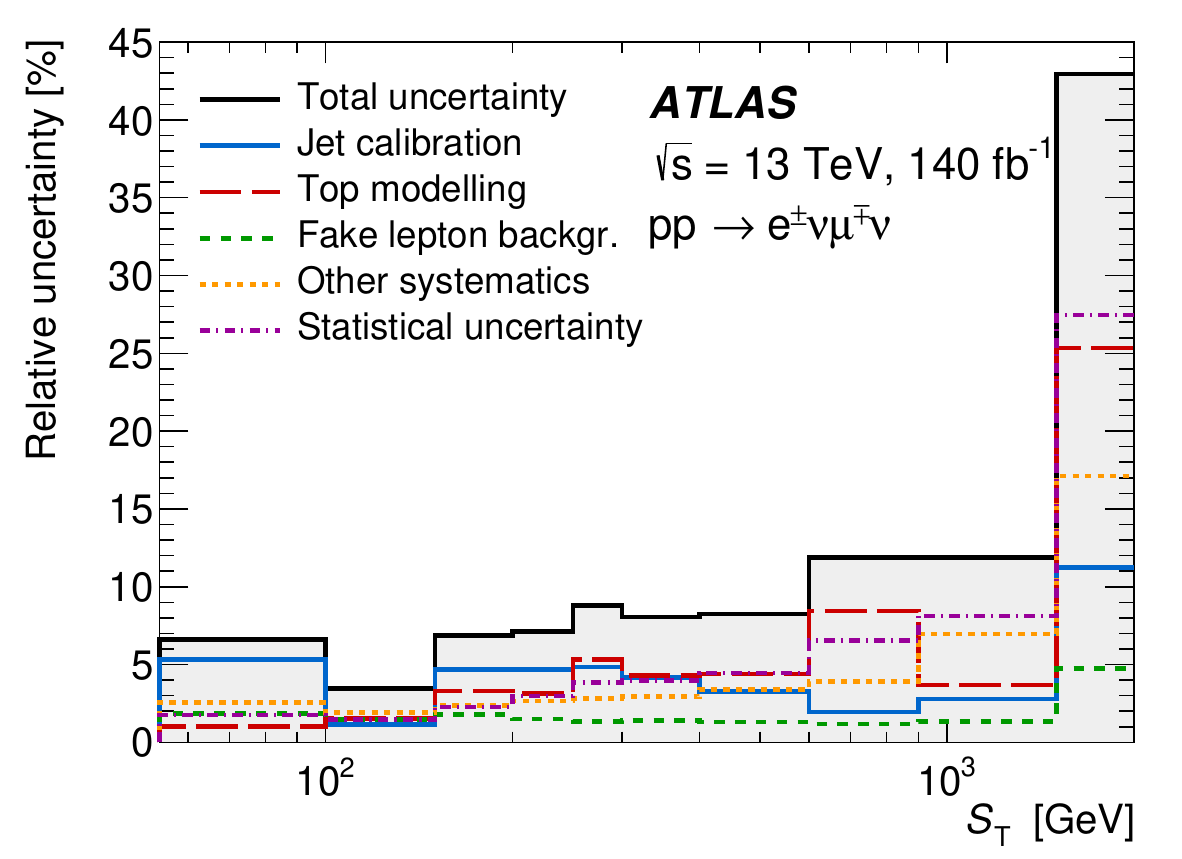}}
\caption{Uncertainties in the unfolded (a) \mll and (b) \ST distributions. \SystFigCaption}
\end{figure}


\section{Theoretical predictions}
\label{sec:predictions}
Fiducial differential cross-section measurements are compared with fixed-order predictions as well as three predictions where the matrix element is matched to a parton shower.
Fixed-order predictions are calculated using \textsc{Matrix\,2.1}~\cite{Grazzini:2019jkl,Grazzini:2020stb,Grazzini:2016ctr,Gehrmann:2014fva,Grazzini:2017mhc,vonManteuffel:2015msa,Gehrmann:2015ora,Denner:2016kdg,Cascioli:2011va,Buccioni:2019sur,Buccioni:2017yxi,Catani:2012qa,Catani:2007vq}, which includes the NNLO QCD correction to \qqWW production, and the NLO QCD correction to \ggWW production. The latter correction constitutes part of the N$^3$LO correction to \WW production and the combined prediction is thus labelled nNNLO.
The prediction also accounts for photon-induced contributions and NLO electroweak corrections by using \OPENLOOPS. The electroweak correction is combined, by default, multiplicatively with the QCD correction to \qqWW.
An additive application of the NLO EW correction is preferred for the \ptll, \met, and jet multiplicity distributions because their tails are populated by events with hard radiation that arise only beyond leading order. Since the LO prediction is zero or close to zero while the NLO prediction is finite, multiplicative NLO corrections are ill-defined.
The \NNPDF[3.1nnlo]~\cite{NNPDF:2017mvq} set of PDFs was used, with a prediction of  the photon PDF with the \textsc{luxQED} method~\cite{Bertone:2017bme}. The cross-section uncertainty associated with the \textsc{NNPDF3.1nnlo\,luxQED} set is estimated as the standard deviation of replicas, yielding a value of 1.2\%.
The factorization and renormalization scales were set to half the sum of the $W$-boson transverse masses~\cite{Grazzini:2019jkl}. Theoretical uncertainties due to missing higher-order QCD corrections are evaluated by varying the factorization and renormalization scales by factors of two and one-half, avoiding variations in opposite directions.
This results in an uncertainty estimate of 2.1\%.

The nominal parton-shower-matched prediction is derived from the \textsc{MiNNLO}+\PYTHIA[8] \qqWW sample in combination with the \SHERPA[2.2.2] \ggWW sample. These samples are also used to estimate efficiencies and resolution, in order to model the \WW signal in the maximum-likelihood fit, and were introduced in Section~\ref{sec:data_mc_samples}.
The \NNPDF[3.0nnlo] set of PDFs was used in the generation of these samples while the more recent \NNPDF[3.1nnlo] was used for the \textsc{Matrix} prediction introduced above. The cross-section uncertainty associated with the \NNPDF[3.0nnlo] set is 1.4\%.
Compared to the results from \textsc{Matrix}, this prediction lacks photon-induced contributions as well as NLO electroweak corrections beyond the effects of photon radiation included in the \PYTHIA parton shower. On the other hand, the inclusion of the parton-shower effects improves the modelling of jets as well as observables correlated with the transverse momentum of the \WW system.
The \SHERPA[2.2.2] simulation of \ggWW includes an additional jet at matrix-element level but lacks the full NLO correction included in the \textsc{Matrix} prediction.

An alternative NNLO+PS prediction is computed using the \textsc{Geneva} method to produce an MC sample of \qqWW events~\cite{Gavardi:2023aco}.
The matching is achieved using a resummed calculation for the hardest jet transverse momentum, instead of the colour-singlet transverse momentum resummation employed for the \textsc{MiNNLO} prediction.
The event simulation uses the \NNPDF[3.1nnlo] parton distribution functions and is interfaced with the \PYTHIA[8.307] parton shower using the \textsc{Monash} tune~\cite{Skands:2014pea}.
As in the \textsc{MiNNLO} prediction, the expected yields are augmented with the \SHERPA[2.2.2] \ggWW sample.

A third parton-shower-matched prediction is generated using \SHERPA[2.2.12]~\cite{Bothmann:2019yzt}.
The prediction has NLO accuracy in QCD for up to one additional parton, and leading-order accuracy for two to three additional parton emissions for $q\bar q$ initial states.
The matrix-element calculations are matched
and merged with the \SHERPA parton shower based on Catani--Seymour
dipole factorization using the MEPS@NLO
prescription.
The virtual QCD corrections are provided by the
\OPENLOOPS library~\cite{Buccioni:2019sur,Cascioli:2011va,Denner:2016kdg}. The
\NNPDF[3.0nnlo] set of PDFs is used, along with the
dedicated set of tuned parton-shower parameters developed by the
\SHERPA authors.
This generator combination lacks the full NNLO QCD corrections but does include an extra parton emission at LO, which improves the modelling of high-multiplicity events.

In comparisons with differential distributions, the parton-shower-matched predictions are augmented by a simulation of electroweak production of a diboson pair in association with two jets (\VVjj), which was generated with \SHERPA[2.2.12].
The contribution from this subprocess is negligible, inclusively, but \WWjj production with vector-boson scattering kinematics constitutes a correction of several percent in bins dominated by events with at least two jets.

An overview of fiducial cross-section predictions is given in Table~\ref{tab:predictions}.
The \PowMiNNLO prediction for \qqWW, using \textsc{NNPDF3.0nnlo}, is 5\% smaller than the fixed-order NNLO prediction using \textsc{NNPDF3.1nnlo}. The different PDF choice results in a 3\% change in cross-section, while parton-shower effects, particularly final-state photon radiation, reduce the lepton momenta and thus the signal acceptance by 2\%.
The latter effect is also modelled by the \textsc{Geneva} prediction, but this uses \textsc{NNPDF3.1nnlo}, resulting in a larger cross-section.
\SHERPA[2.2.12] predicts a fiducial cross-section that is compatible with those from the NNLO generators, albeit with a larger scale uncertainty.
The \SHERPA[2.2.2] \ggWW prediction at LO is significantly smaller than the \textsc{Matrix} NLO QCD prediction.
A $K$-factor of 1.7, calculated as the ratio of the NLO to \SHERPA cross-section in the total phase space, is applied to the \ggWW sample and only partially compensates for the difference, due to the different acceptances predicted by the two models.
The NLO electroweak correction to \qqWW, given by the ratio of the NLO EW prediction to the LO prediction, decreases the \qqWW cross-section by 4\% while doubling the photon-induced contribution.

\begin{table}
\caption{Comparison of theoretical predictions of fiducial cross-sections for various subprocesses contributing to \WW production, with the corresponding scale uncertainties. The specified process, for example in \qqWW, corresponds to the initial state at leading order. Other partons contribute to each sub-process at higher orders in perturbation theory. The names of the \textsc{NNPDF3.0nnlo} and \textsc{NNPDF3.1nnlo\,luxQED} PDF sets are shortened in the table.}
\label{tab:predictions}
\centering
\begin{tabular}{llllrrl}
\toprule
Process                                                         & Code                         & PDF               & Perturbative order      & \multicolumn{3}{r@{}}{Fid. cross-section}                          \\
\midrule
\qqWW                                                           & \textsc{Matrix\,2.1}         & \textsc{NNPDF3.1} & NNLO QCD                &                                           & 672 fb & $\pm 1.8\,\%$ \\
\qqWW                                                           & \textsc{Geneva}+\PYTHIA[8.3] & \textsc{NNPDF3.1} & NNLO QCD + PS           &                                           & 659 fb & $\pm 0.6\,\%$ \\
\qqWW                                                           & \textsc{MiNNLO}+\PYTHIA[8.2] & \textsc{NNPDF3.0} & NNLO QCD + PS           &                                           & 642 fb & $\pm 1.1\,\%$ \\
\qqWW                                                           & \SHERPA[2.2.12]              & \textsc{NNPDF3.0} & NLO QCD + PS $^\dagger$ &                                           & 630 fb & $\pm 7.2\,\%$ \\
\midrule
\ggWW                                                           & \textsc{Matrix\,2.1}         & \textsc{NNPDF3.1} & NLO QCD                 &                                           & 32 fb  & $\pm 13\,\%$  \\
\ggWW                                                           & \SHERPA[2.2.2]               & \textsc{NNPDF3.0} & LO QCD + PS $^\dagger$  &                                           & 15 fb  & $\pm 30\,\%$  \\
\midrule
$\gamma\gamma\rightarrow WW$                                    & \textsc{Matrix\,2.1}         & \textsc{NNPDF3.1} & LO                      &                                           & 5 fb   & $\pm 16\,\%$  \\
$\gamma\gamma\rightarrow WW$                                    & \textsc{Matrix\,2.1}         & \textsc{NNPDF3.1} & NLO EW                  &                                           & 11 fb  & $\pm 2.3\,\%$ \\
\midrule
$q\bar q\rightarrow WWjj$ (EW)                                  & \SHERPA[2.2.12]              & \textsc{NNPDF3.0} & LO + PS                 &                                           & 4 fb   & $\pm 7.0\,\%$ \\
\midrule
\multicolumn{5}{@{}l@{}}{For calculation of NLO EW correction:} &                                                                                                                                                 \\
\qqWW                                                           & \textsc{Matrix\,2.1}         & \textsc{NNPDF3.1} & LO                      &                                           & 436 fb & $\pm 5.1\,\%$ \\
\qqWW                                                           & \textsc{Matrix\,2.1}         & \textsc{NNPDF3.1} & NLO EW                  &                                           & 418 fb & $\pm 5.1\,\%$ \\
\bottomrule
\multicolumn{7}{l}{\footnotesize $\dagger$: Includes matrix elements with additional parton emissions, matched and merged with the parton shower, which increases}                                              \\
\multicolumn{7}{l}{\footnotesize the accuracy of the simulation of high jet-multiplicity events but also increases the nominal scale uncertainty.}                                                                \\
\end{tabular}
\end{table}


\section{Results}
\label{sec:results}
\subsection{Cross-sections}
The measured fiducial cross-section for \WW production, with \WWemu, in $pp$ collisions at $\sqrt{s}=13\,\tev$, for the fiducial volume defined in Table~\ref{tab:fiducialtruthlevelcuts} is determined from the profile likelihood fit to be
\[
\sigma_{\mathrm{fid}}=707 \pm 7~\mathrm{(stat.)} \pm 20~\mathrm{(syst.)} \pm 6~\mathrm{(lumin.)} \,\si{\femto\barn},
\]

with a total uncertainty of 3.1\%.
Figures~\ref{fig:postfit-pre} and \ref{fig:postfit-post} present the pre-fit and post-fit distributions of \ST, respectively.
In the fit, nuisance parameters remain very close to their initial values and the background normalizations are changed by less than 2\% from their estimates based on control regions and theoretical calculations.

\begin{figure}[htbp]
\centering
\subfloat[\label{fig:postfit-pre}]{\includegraphics[width=.45\linewidth]{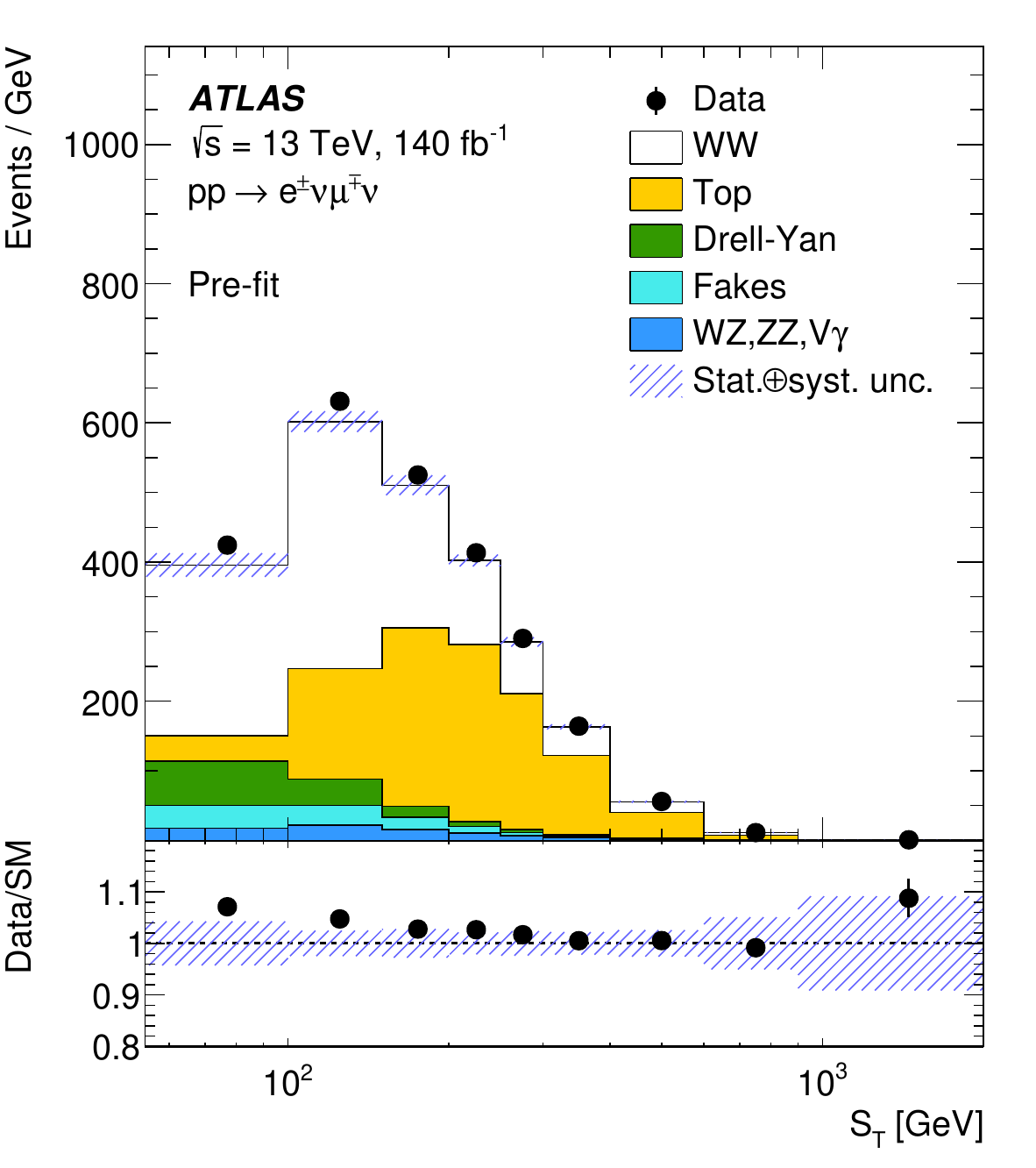}}
\subfloat[\label{fig:postfit-post}]{\includegraphics[width=.45\linewidth]{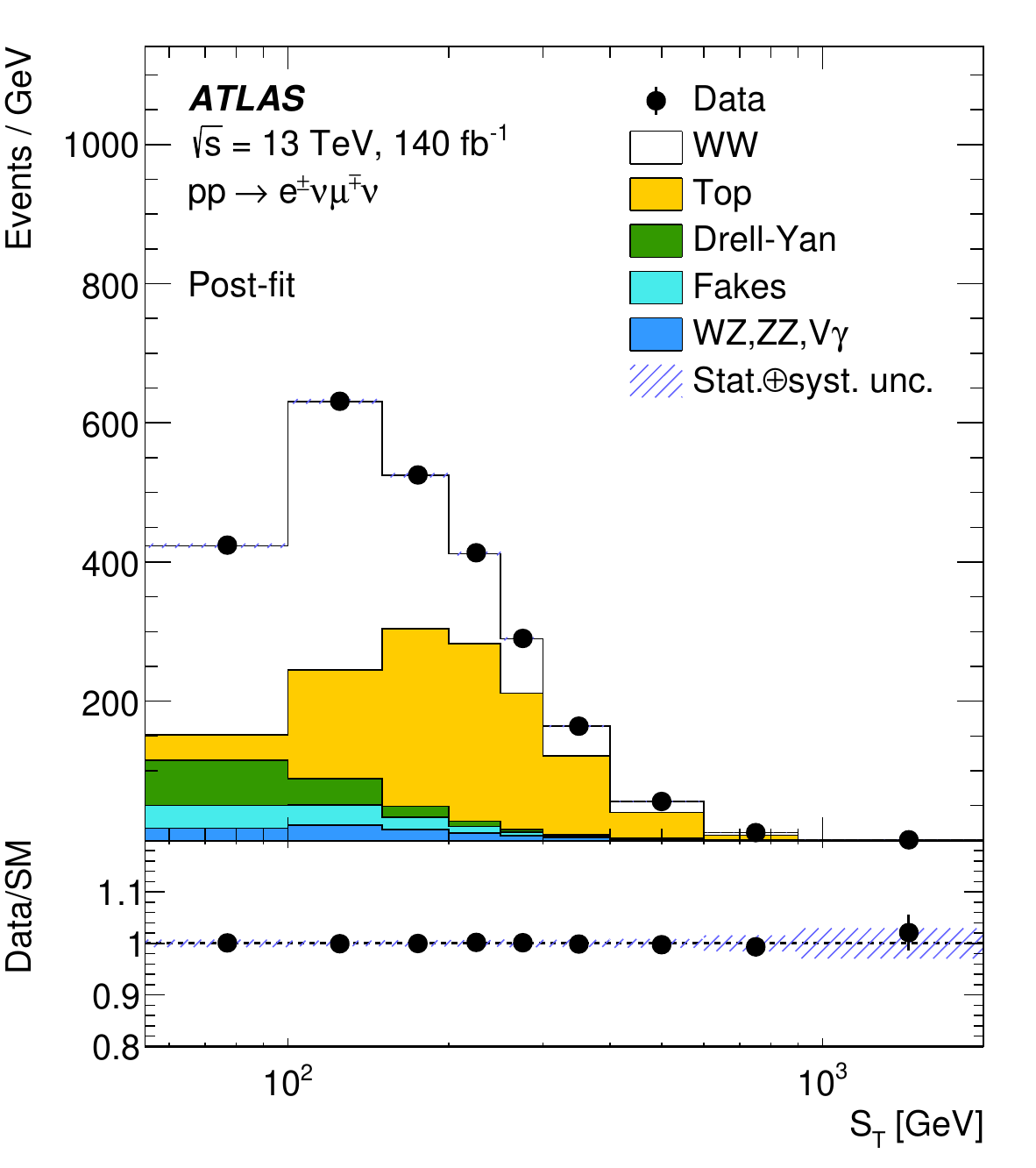}}
\caption{(a) Pre-fit and (b) post-fit  \ST distributions in the profile likelihood fit.
Data are shown as black markers together with the predictions for the signal and background production processes.
The rightmost bin contains overflow events.
The lower panels show the ratio of the data to the total prediction.
Top-quark and fake-lepton backgrounds are determined using data-driven methods.
The uncertainty bands shown include statistical and systematic uncertainties.
Theory uncertainties affecting the shape and selection efficiency of the signal are included.}
\end{figure}

In Figure~\ref{fig:XSResults} the result is compared with several theoretical predictions: the nominal \MiNNLO model used in the analysis, a second NNLO+PS prediction from \textsc{Geneva},  the \SHERPA NLO  prediction, and two pairs of nNNLO prediction from \textsc{Matrix\,2.1}.
One of the \textsc{Matrix} prediction pairs includes NLO electroweak corrections while predictions within each pair differ based on the NNPDF version.
The measured cross-section is about two standard deviations larger than the cross-section predicted by \textsc{MiNNLO}, but is in good agreement with \textsc{Matrix} predictions. The main reasons for the larger cross-section predicted by \textsc{Matrix} are the updated NNPDF PDF version, which increases the cross-section by 28\,fb, and an 11\,fb increase in the cross-section due to photon-induced contributions, as detailed in Section~\ref{sec:predictions}.
The parton-shower-matched \qqWW predictions from \MiNNLO, \textsc{Geneva}, and \SHERPA  are combined with the \SHERPA[2.2.2] prediction of gluon-induced \WW production.
The \textsc{Matrix} prediction without EW corrections is the only prediction not modelling photon FSR; including photon FSR would reduce the predicted cross-section by 2\%.
For all measured distributions, correction factors that provide \enquote{Born-level} cross-sections, which exclude photon FSR effects, are available at the HEPdata entry associated with this paper.

\begin{figure}
\centering
\includegraphics[width=.65\linewidth]{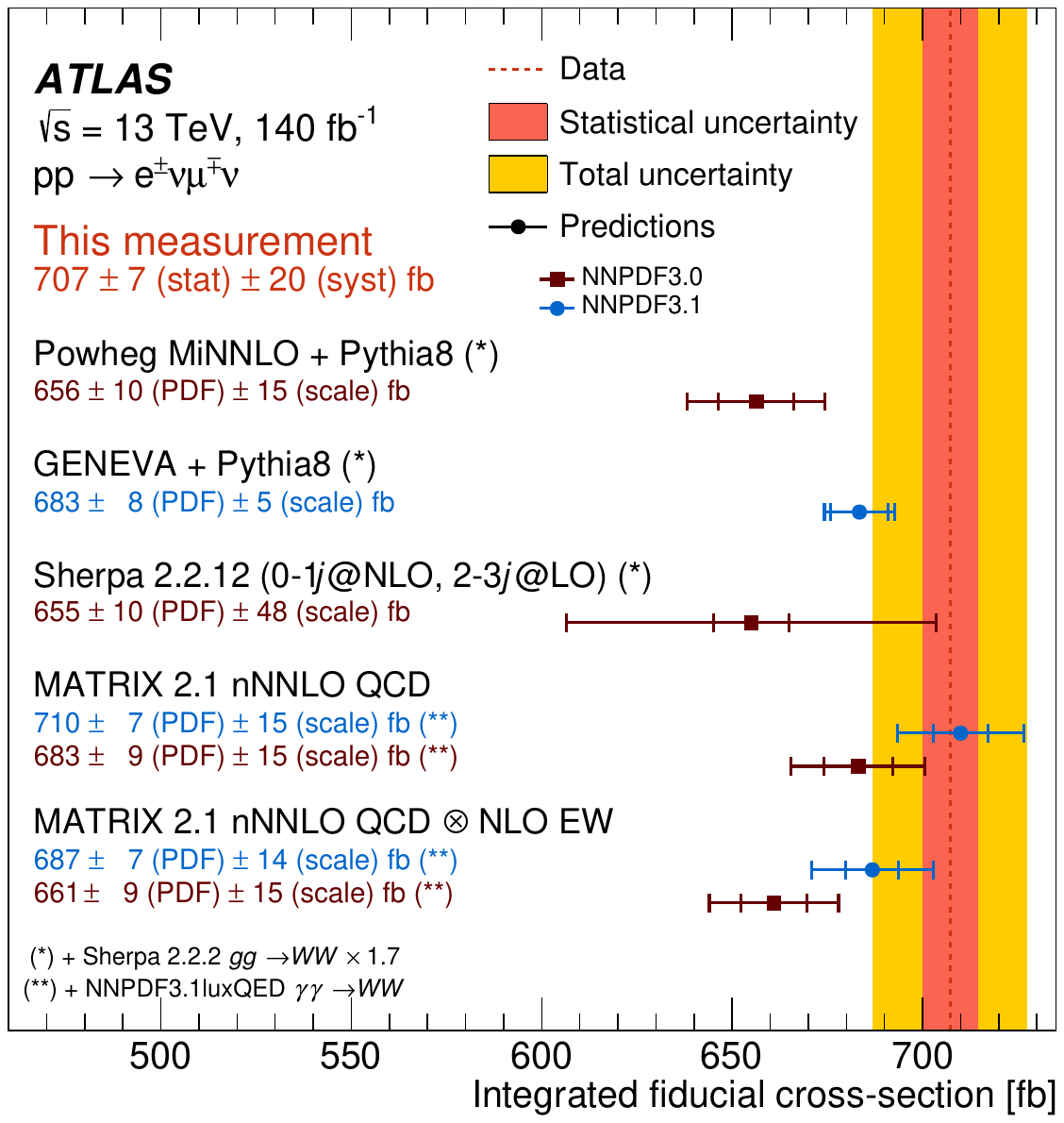}%
\caption{
Measured fiducial cross-section compared with theoretical predictions from \minnlo{}+\PYTHIA[8], \textsc{Geneva}+\PYTHIA[8],  \SHERPA[2.2.12], and \textsc{Matrix\,2.1}.
The predictions are based on the \textsc{NNPDF3.0} (red squares) and \textsc{NNPDF3.1\,luxQED} (blue dots) PDF sets.
The nNNLO predictions include photon-induced contributions (always using \textsc{NNPDF3.1\,luxQED}) and NLO QCD corrections to the gluon-gluon initial state.
The \qqWW predictions from \minnlo, \textsc{Geneva} and \SHERPA[2.2.12] are combined with a \SHERPA[2.2.2] prediction of gluon-induced \WW production, scaled by an inclusive NLO $K$-factor of 1.7.
Inner (outer) error bars on the theory predictions correspond to PDF (the combination of scale and PDF) uncertainties.
The \textsc{Matrix} nNNLO QCD $\otimes$ NLO EW prediction using \textsc{NNPDF3.1\,luxQED}, the best available prediction of the integrated fiducial cross-section, is in good agreement with the measurement.
}
\label{fig:XSResults}
\end{figure}

The measurement is extrapolated to the full phase space of \WW production, accounting for a $W$-boson leptonic decay branching ratio of 10.86\%~\cite{ParticleDataGroup:2024cfk}, and an acceptance of $23.7\%$ for $W^+W^-\rightarrow e^\pm \nu \mu^\mp \nu$ events that was calculated at nNNLO with \textsc{Matrix} and includes NLO electroweak corrections.
The uncertainty in the acceptance is 1.1\%, estimated  by comparing the multiplicative and additive schemes for electroweak corrections, evaluating the PDF uncertainty, and varying the renormalization and factorization scales by factors of two and one-half, avoiding variations in opposite directions.
The choice of combination scheme for electroweak corrections is the dominant source of uncertainty in the signal acceptance.
After the extrapolation, the measured total production cross-section for $W$-boson pairs is found to be
\[
\sigma_\mathrm{total}=127 \pm 1~\mathrm{(stat.)} \pm 4~\mathrm{(syst.)} \pm 1~\mathrm{(lumin.)} \,\si{\pico\barn}.
\]
In Figure~\ref{fig:XSResultsCompared} the total cross-section is compared with measurements by ATLAS~\cite{STDM-2017-24} and CMS~\cite{CMS-SMP-18-004} that are based on datasets of 36~\ifb.
The present measurement is more precise than its predecessor~\cite{STDM-2017-24} because of more precise data-driven top-quark and fake-lepton estimates, an improved luminosity determination~\cite{DAPR-2021-01}, and the use of a jet-inclusive phase space, which reduces jet-related uncertainties as well as theoretical uncertainties in the extrapolation to the full phase space.

\begin{figure}
\centering
\includegraphics[width=.65\linewidth]{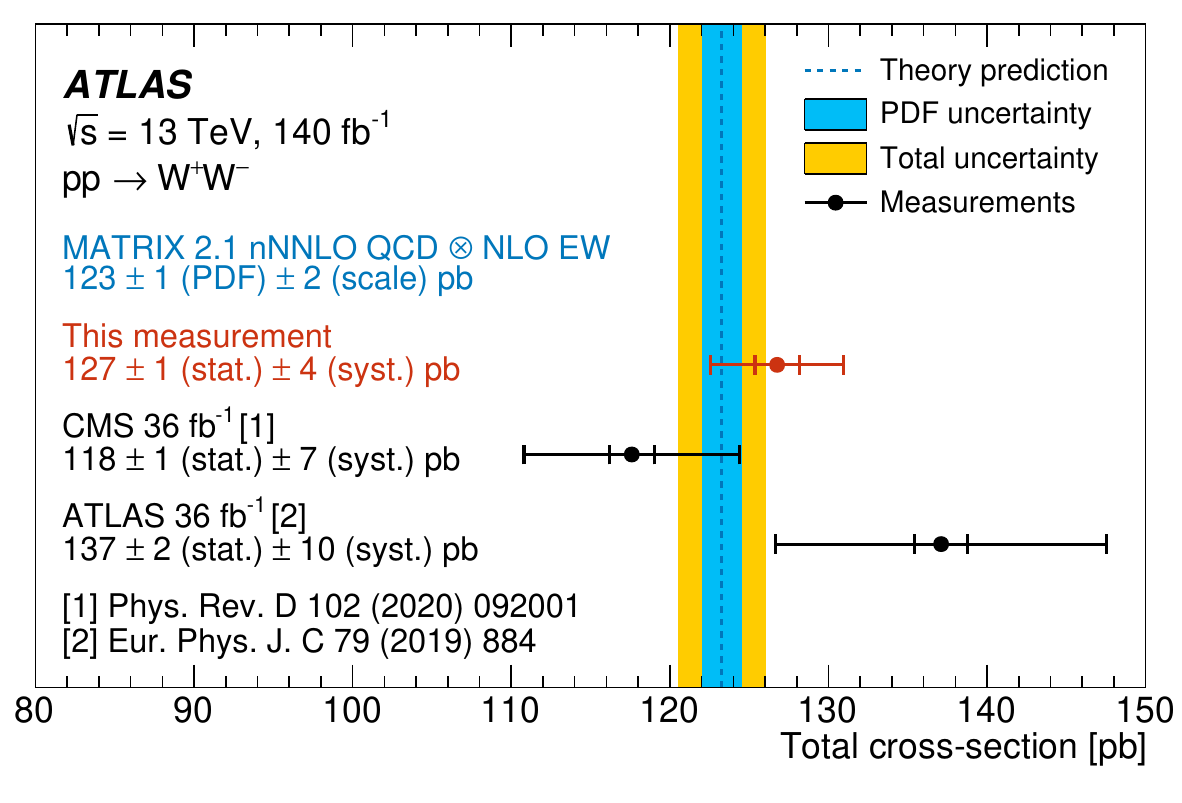}%
\caption{Measured total $WW$ cross-section, compared with a theoretical prediction from \textsc{Matrix} and previous measurements by ATLAS~\cite{STDM-2017-24} and CMS~\cite{CMS-SMP-18-004}. The theoretical prediction uses the \textsc{NNPDF3.1nnlo\,luxQED} set of parton distribution functions, has NNLO accuracy in QCD for \qqWW production and includes NLO QCD corrections to \ggWW production, which constitute part of the N$^3$LO correction. It includes photon-induced contributions and is combined multiplicatively with NLO EW corrections to \qqWW.
Inner (outer) error bars on experimental measurements correspond to statistical (total) uncertainties. The inner (outer) error band  includes PDF uncertainties (PDF and scale uncertainties added in quadrature).}
\label{fig:XSResultsCompared}
\end{figure}

Differential fiducial cross-sections are presented in Figures~\ref{fig:DiffResults-1}--\ref{fig:DiffResults-3}.
Excellent agreement with the fixed-order \textsc{Matrix} prediction is observed for jet-inclusive distributions.
The most notable discrepancy is a local three-standard-deviation difference in the lowest bin of the \HT distribution.  A broader tension with the data is also observed for lower values of \dphill, a region characterized by relatively large systematic uncertainties in the modeling of the \ttbar background, as well as in the signal cross-section prediction, particularly due to the \ggWW contribution.

Electroweak corrections, applied with the multiplicative combination scheme, improve the modelling of high-mass events for some distributions (e.g.\ \mll) but over-correct for other distributions (e.g.\ \ptleadlep).
Over-corrections are expected because the multiplicative combination scheme does not always yield an appropriate estimate of mixed QCD--EW effects, particularly in regions of phase space that are dominated by events with hard QCD radiation, as is the case for high \ptleadlep~\cite{Grazzini:2019jkl}.

The parton-shower-matched predictions also model the data well, except for underestimates of the cross-section by the \SHERPA and \MiNNLO predictions, which can be explained by the different PDF choice used in the generation of these samples. The parton shower allows the simulated events to have more than two jets, and improves the modelling of the \ST distribution in particular.

\begin{figure}
\centering
\subfloat[]{\includegraphics[width=.5\linewidth]{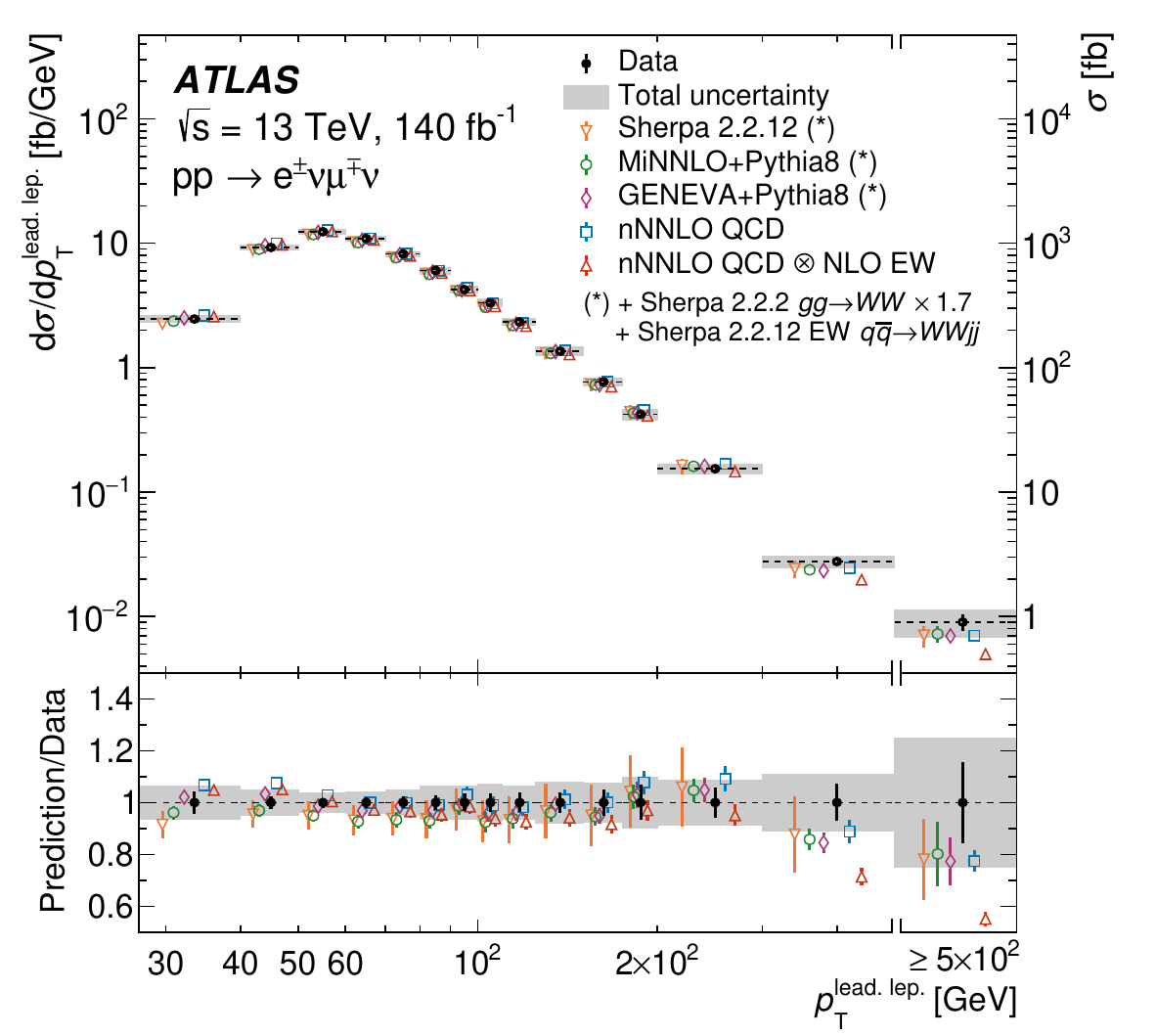}}
\subfloat[]{\includegraphics[width=.5\linewidth]{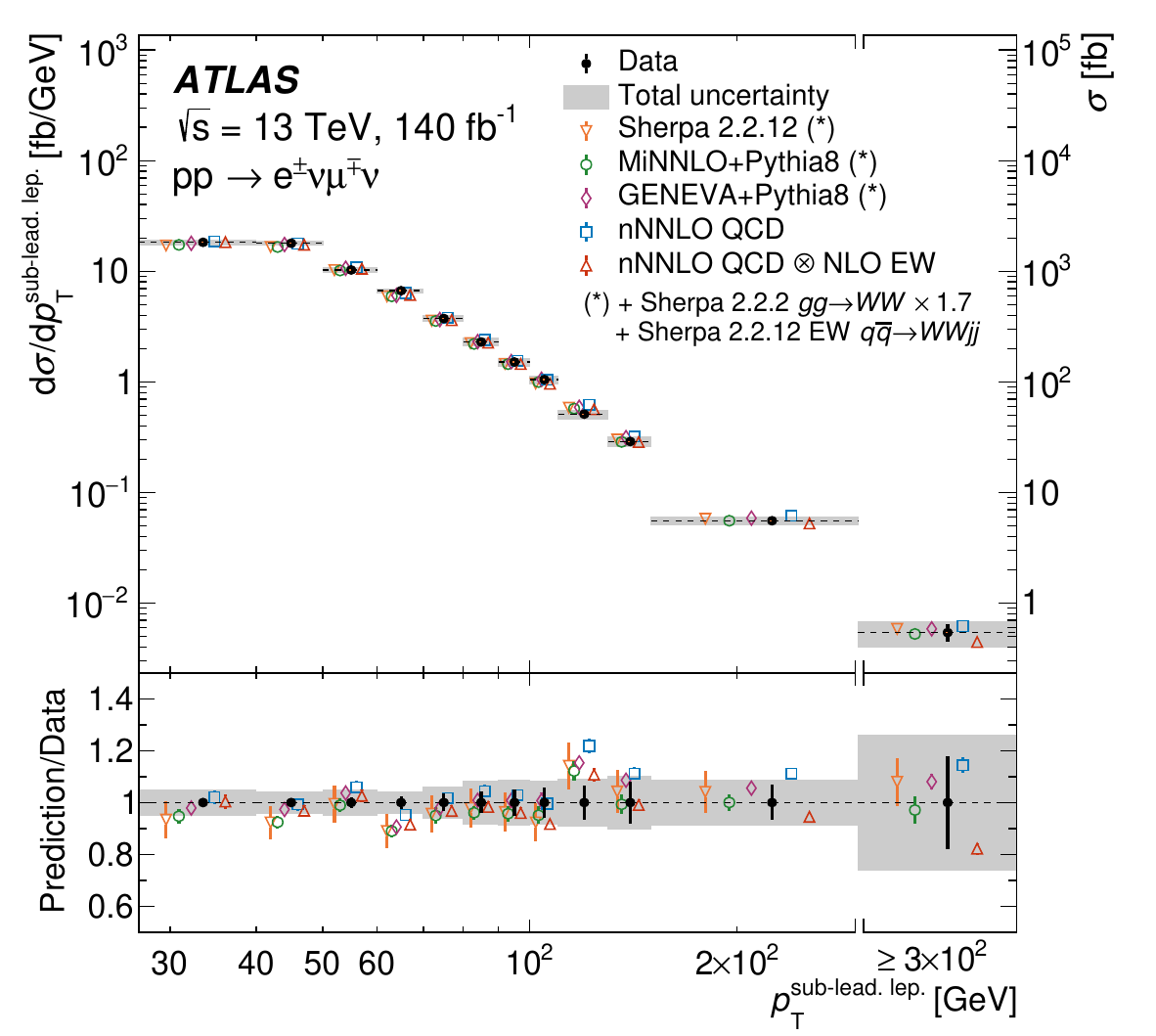}}\\
\subfloat[]{\includegraphics[width=.5\linewidth]{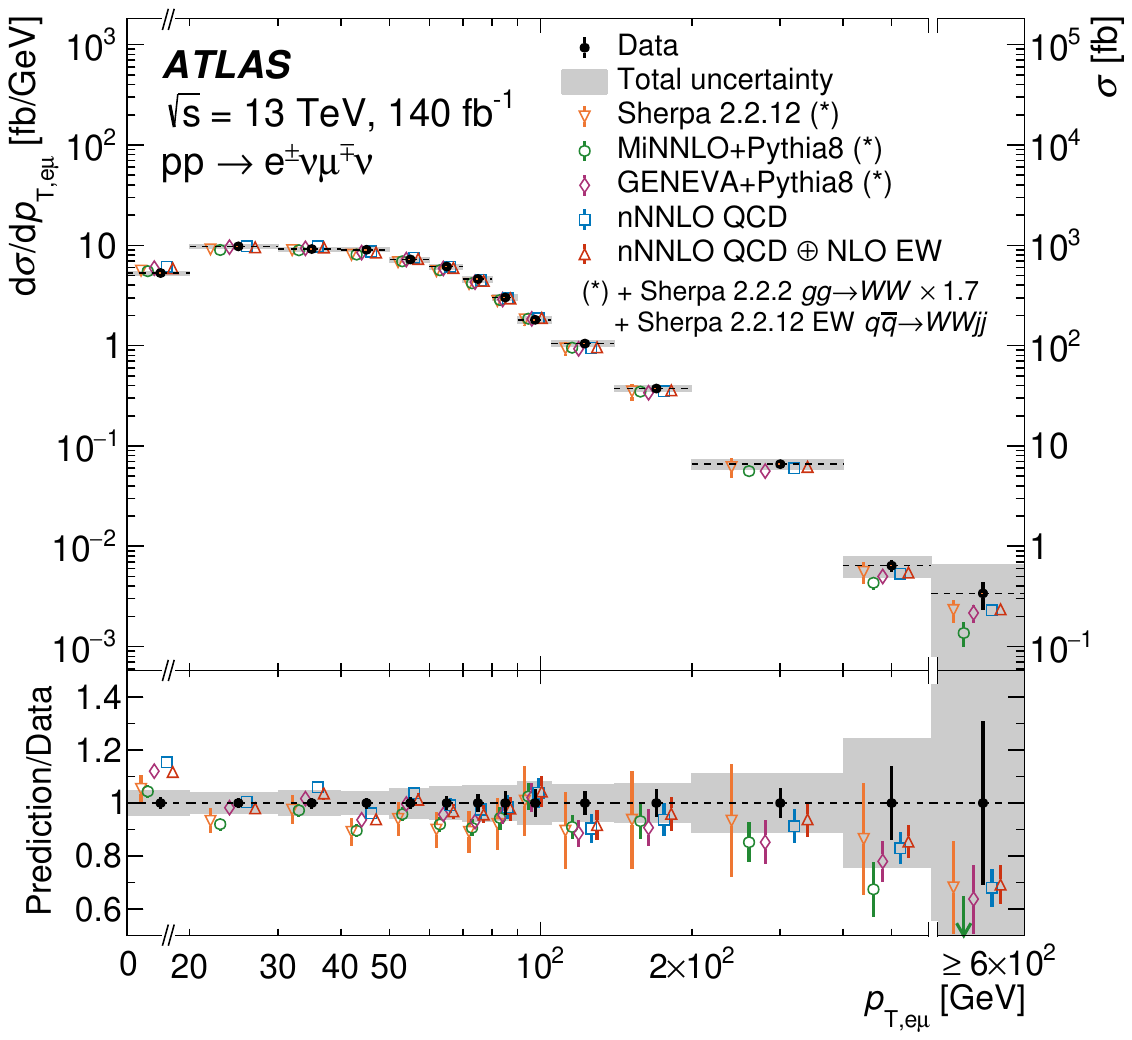}}
\subfloat[]{\includegraphics[width=.5\linewidth]{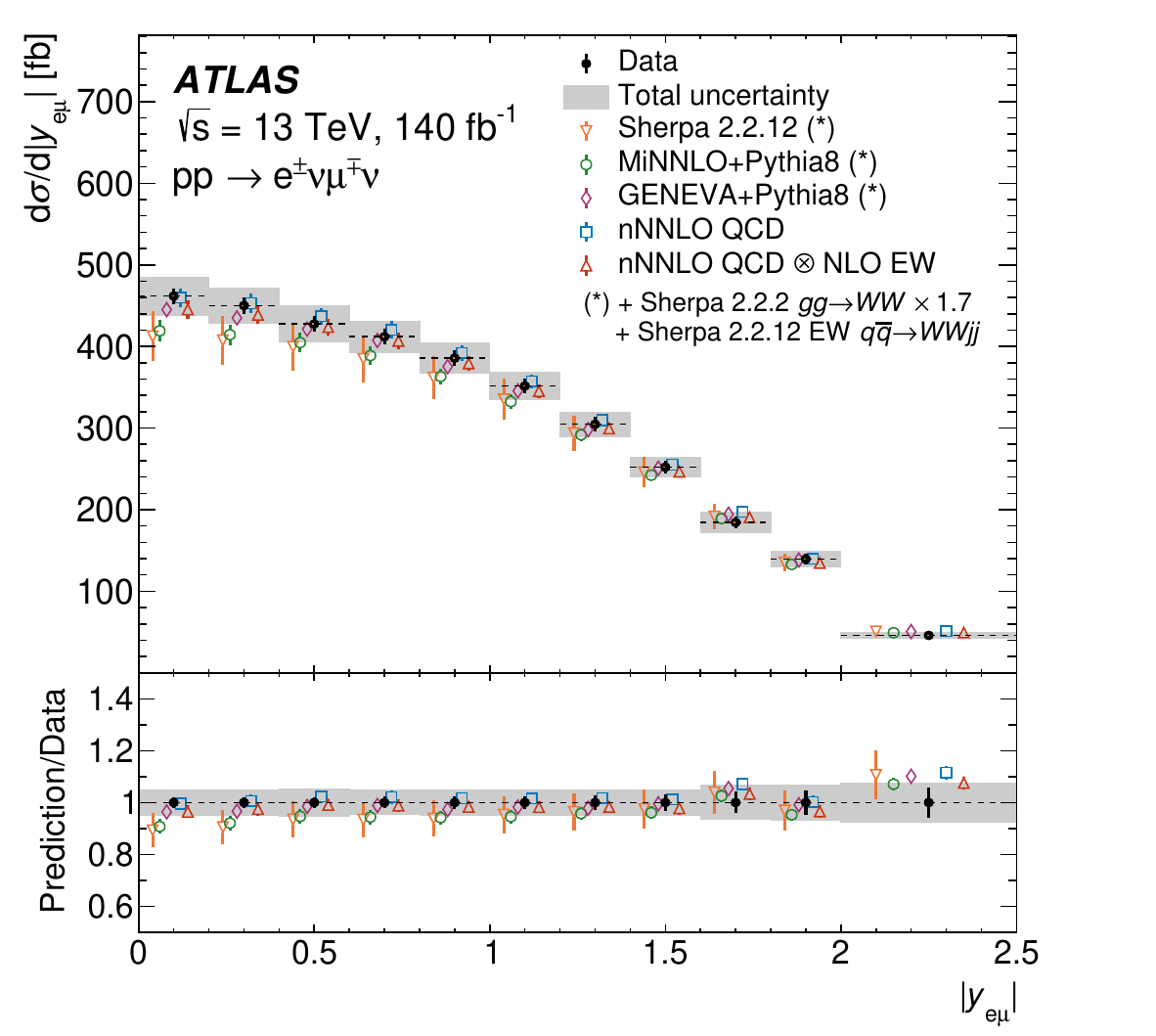}}
\caption{Fiducial differential cross-sections as a function of (a) \ptleadlep, (b) \ptsublep, (c) \ptll, and (d) |\yll{}|.
The measured cross-section values are shown as points with error bars giving the statistical uncertainty and solid bands indicating the size of the total uncertainty. For distributions in which the rightmost bin is inclusive, the right-hand-side axis indicates the integrated cross-section of the rightmost bin. The results are compared with fixed-order nNNLO QCD (+ NLO EW) predictions from \textsc{Matrix\,2.1}, the NNLO+PS predictions from \textsc{Powheg MiNNLO}+\PYTHIA[8] and \textsc{Geneva}+\PYTHIA[8], and the \SHERPA[2.2.12]  NLO+PS predictions. The last three predictions are combined with the ones from \SHERPA[2.2.2] for the $gg$ initial state and \SHERPA[2.2.12] for electroweak $WWjj$ production. These contributions are modelled at LO but an NLO QCD $K$-factor of 1.7 is applied for gluon-induced production. Theoretical predictions are indicated as markers with vertical lines denoting PDF, scale and parton-shower uncertainties. Markers are staggered for better visibility.}
\label{fig:DiffResults-1}
\end{figure}

\begin{figure}
\centering
\subfloat[]{\includegraphics[width=.5\linewidth]{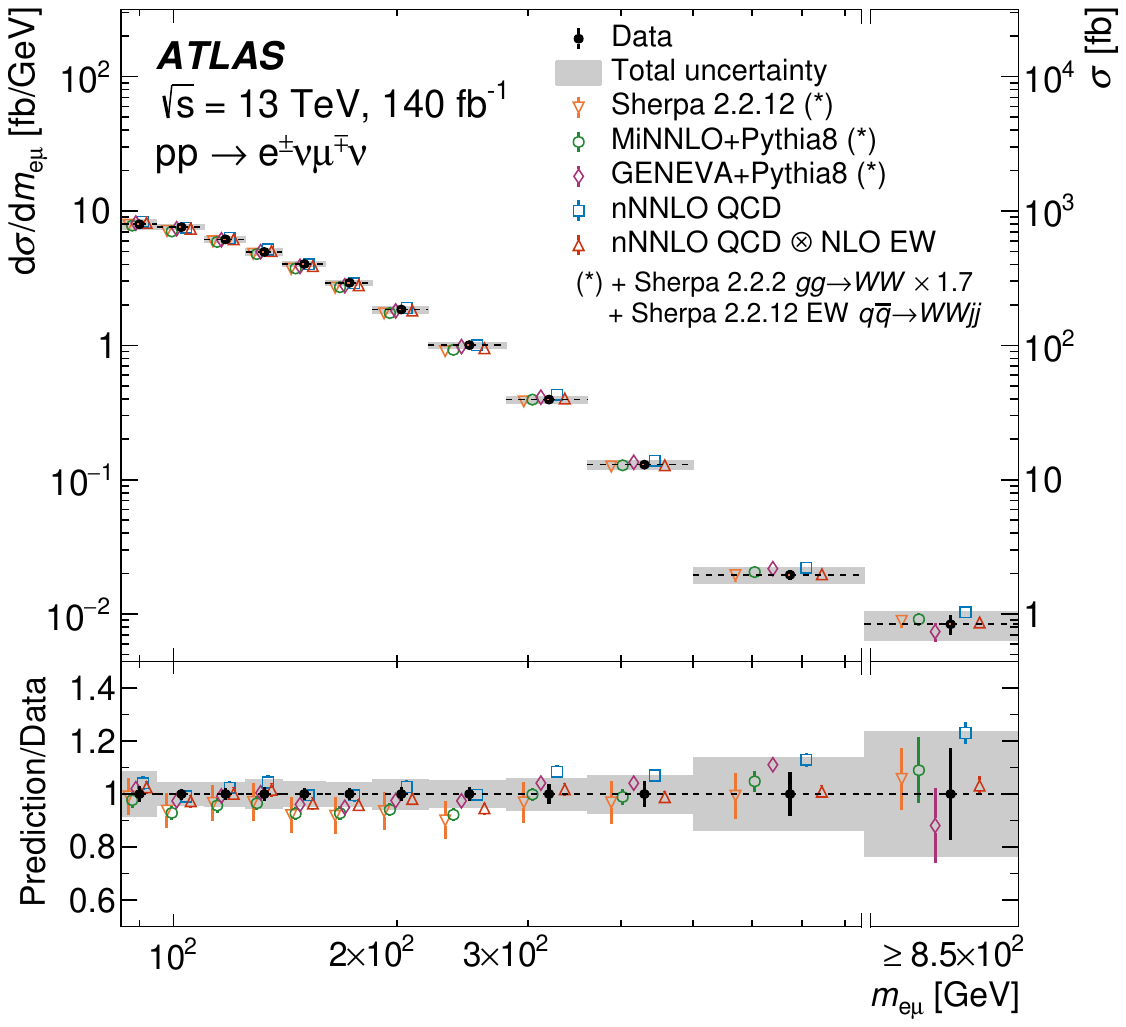}}
\subfloat[]{\includegraphics[width=.5\linewidth]{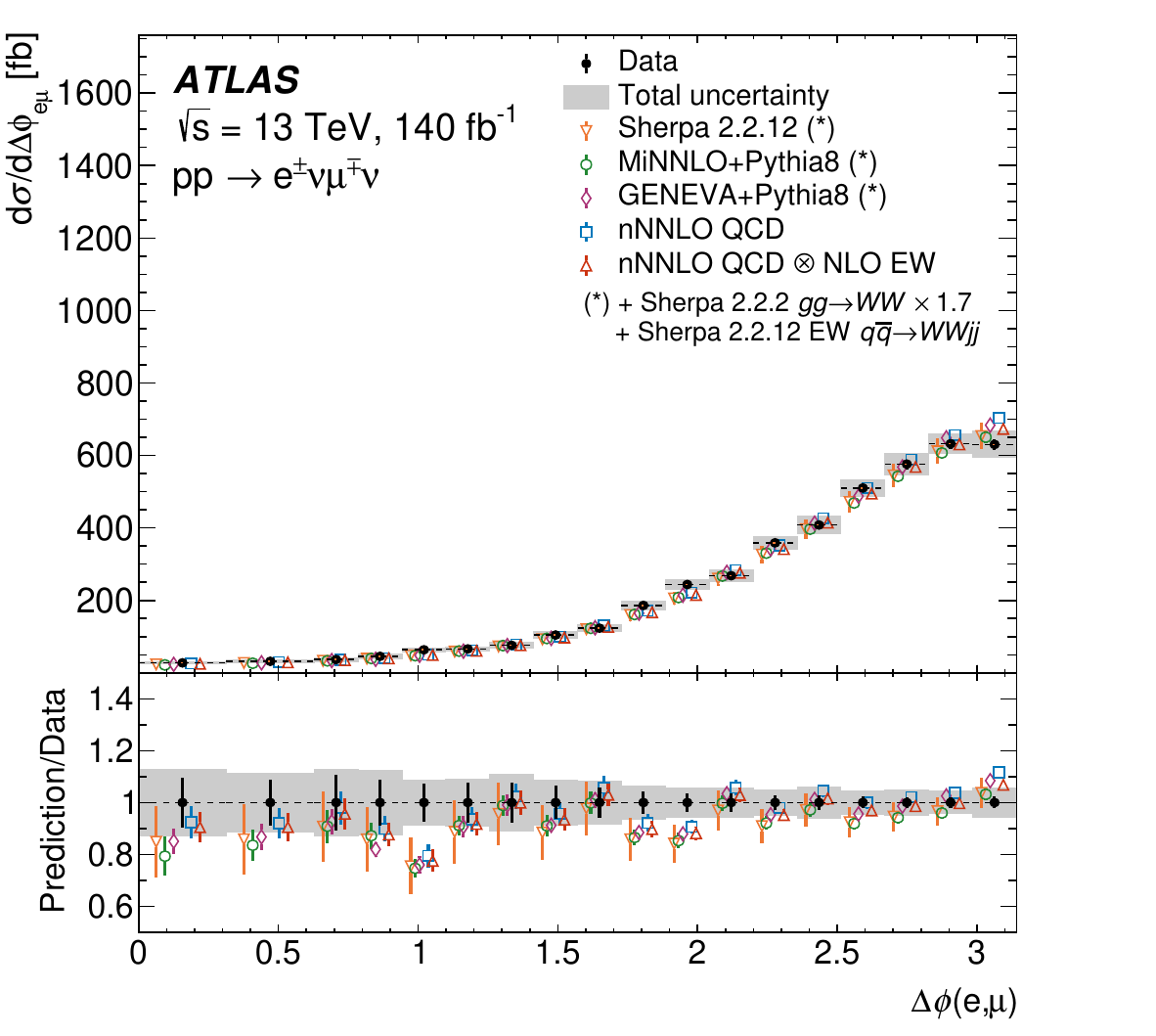}}\\
\subfloat[]{\includegraphics[width=.5\linewidth]{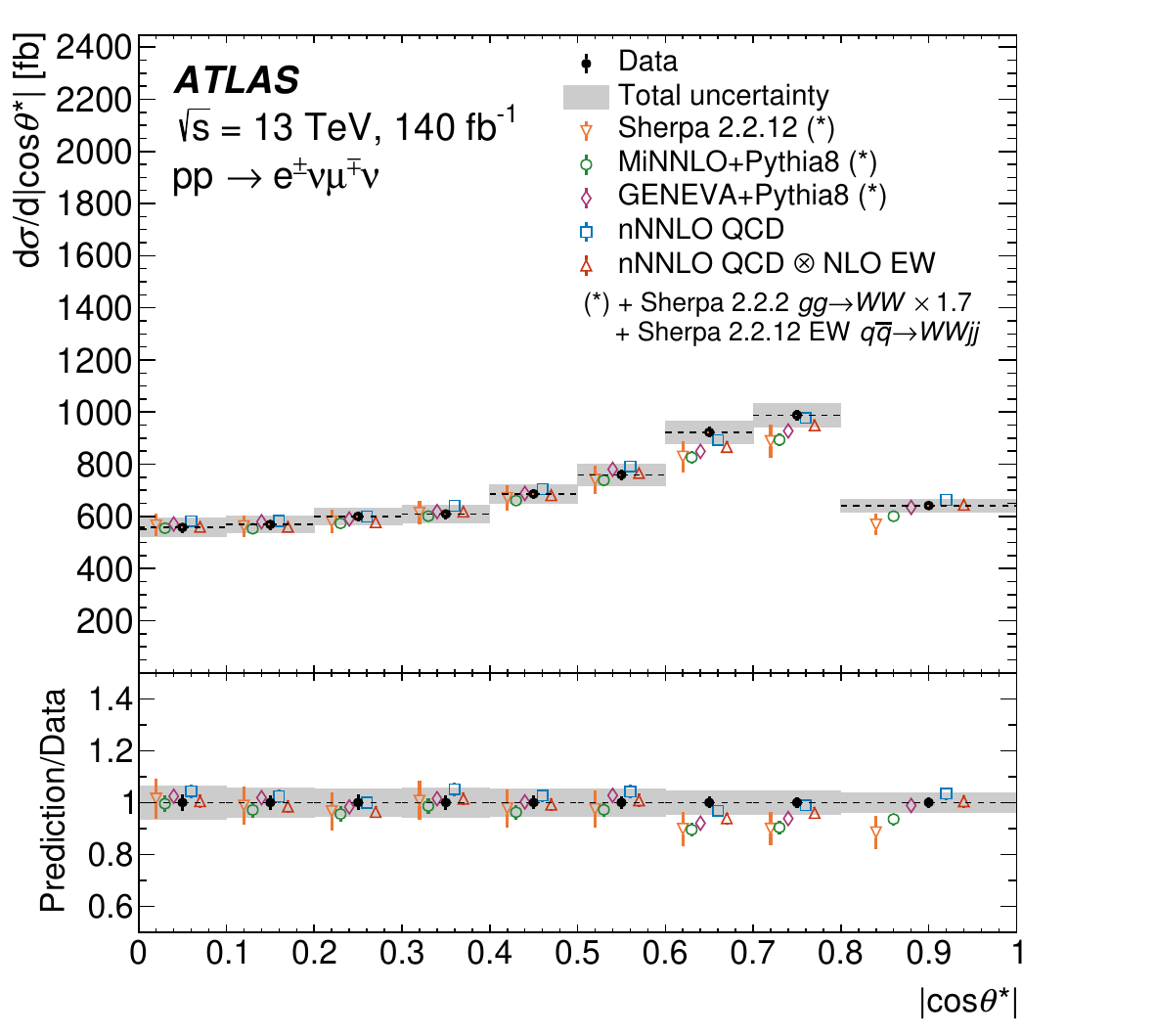}}
\subfloat[]{\includegraphics[width=.5\linewidth]{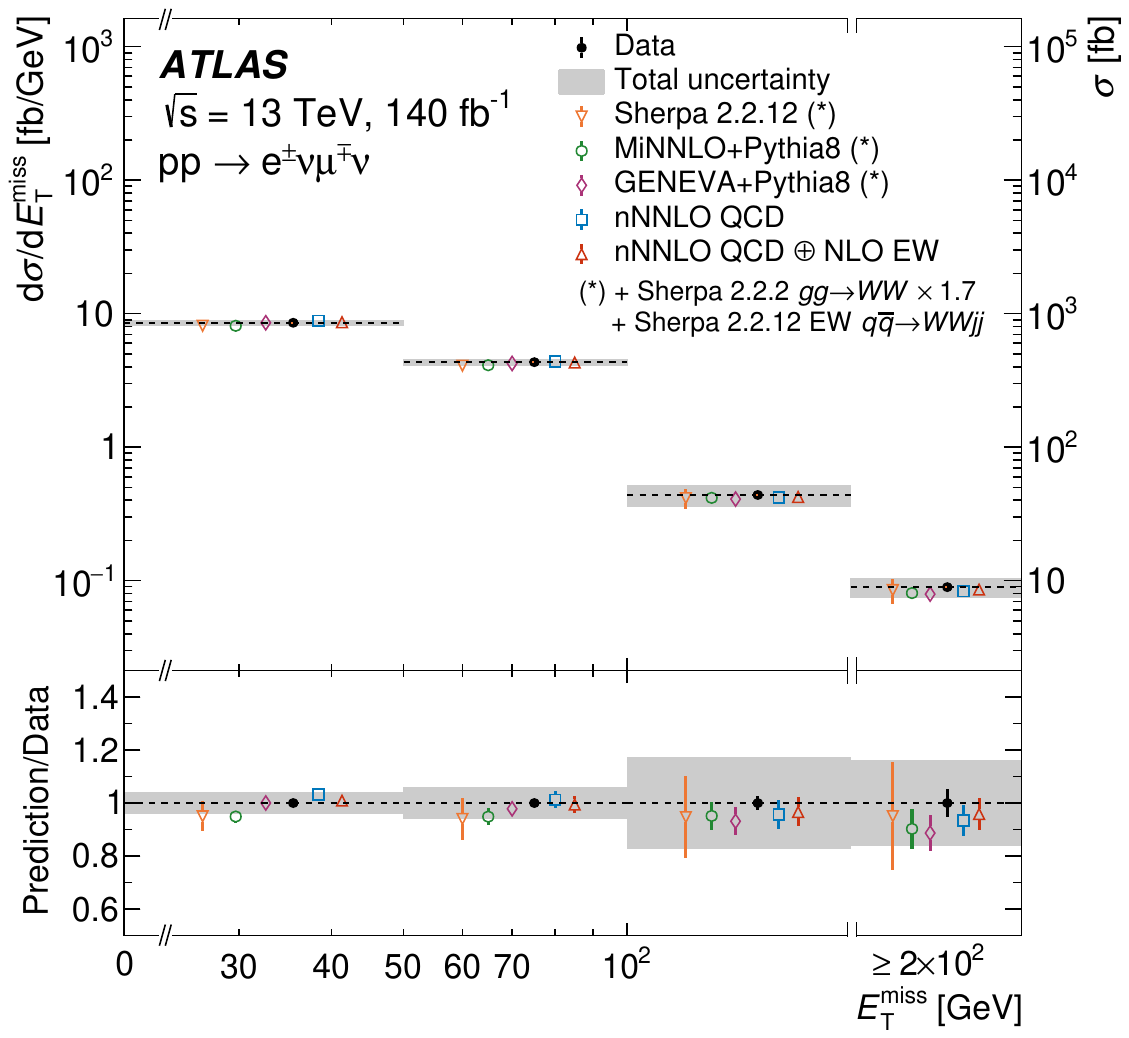}}
\caption{Fiducial differential cross-sections as a function of (a) \mll, (b) \dphill, (c) \costhetastar, and (d) \met.
The measured cross-section values are shown as points with error bars giving the statistical uncertainty and solid bands indicating the size of the total uncertainty. For distributions in which the rightmost bin is inclusive, the right-hand-side axis indicates the integrated cross-section of the rightmost bin. The results are compared with fixed-order nNNLO QCD (+ NLO EW) predictions from \textsc{Matrix\,2.1}, the NNLO+PS predictions from \textsc{Powheg MiNNLO}+\PYTHIA[8] and \textsc{Geneva}+\PYTHIA[8], and the \SHERPA[2.2.12]  NLO+PS predictions. The last three predictions are combined with the ones from \SHERPA[2.2.2] for the $gg$ initial state and \SHERPA[2.2.12] for electroweak $WWjj$ production. These contributions are modelled at LO but an NLO QCD $K$-factor of 1.7 is applied for gluon-induced production. Theoretical predictions are indicated as markers with vertical lines denoting PDF, scale and parton-shower uncertainties. Markers are staggered for better visibility.}
\end{figure}

\begin{figure}
\centering
\subfloat[]{\includegraphics[width=.5\linewidth]{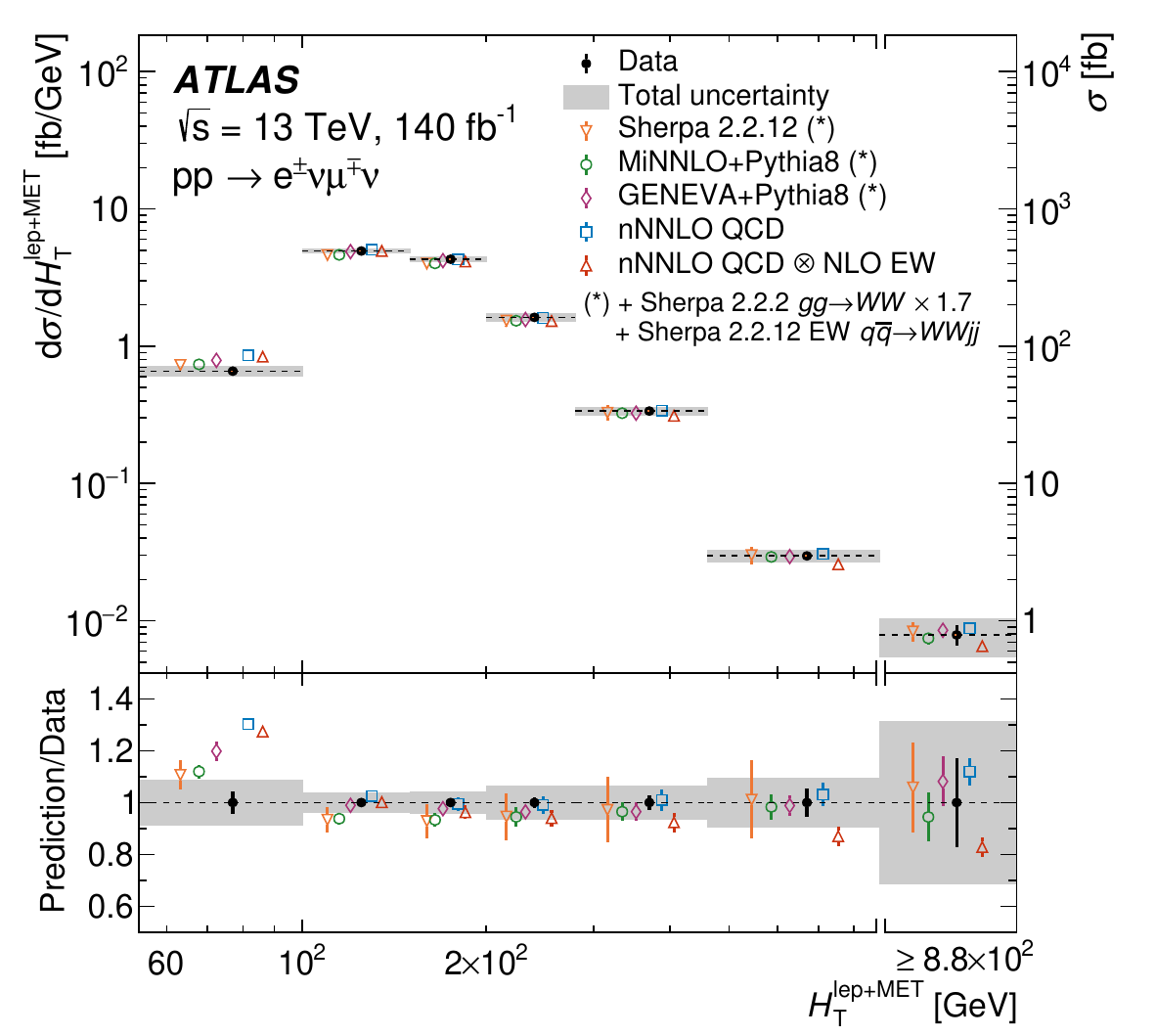}}%
\subfloat[]{\includegraphics[width=.5\linewidth]{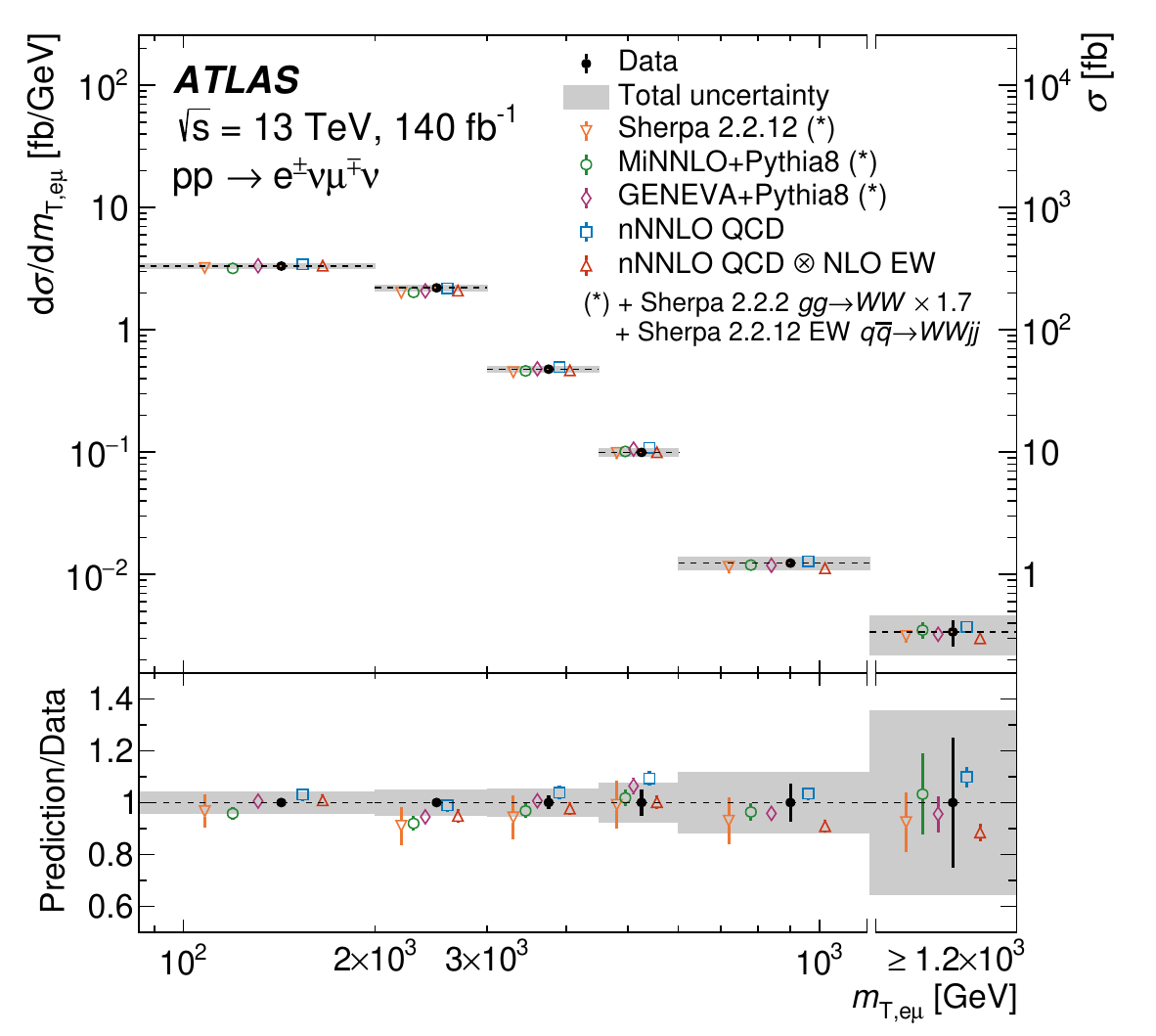}}\\
\subfloat[]{\includegraphics[width=.5\linewidth]{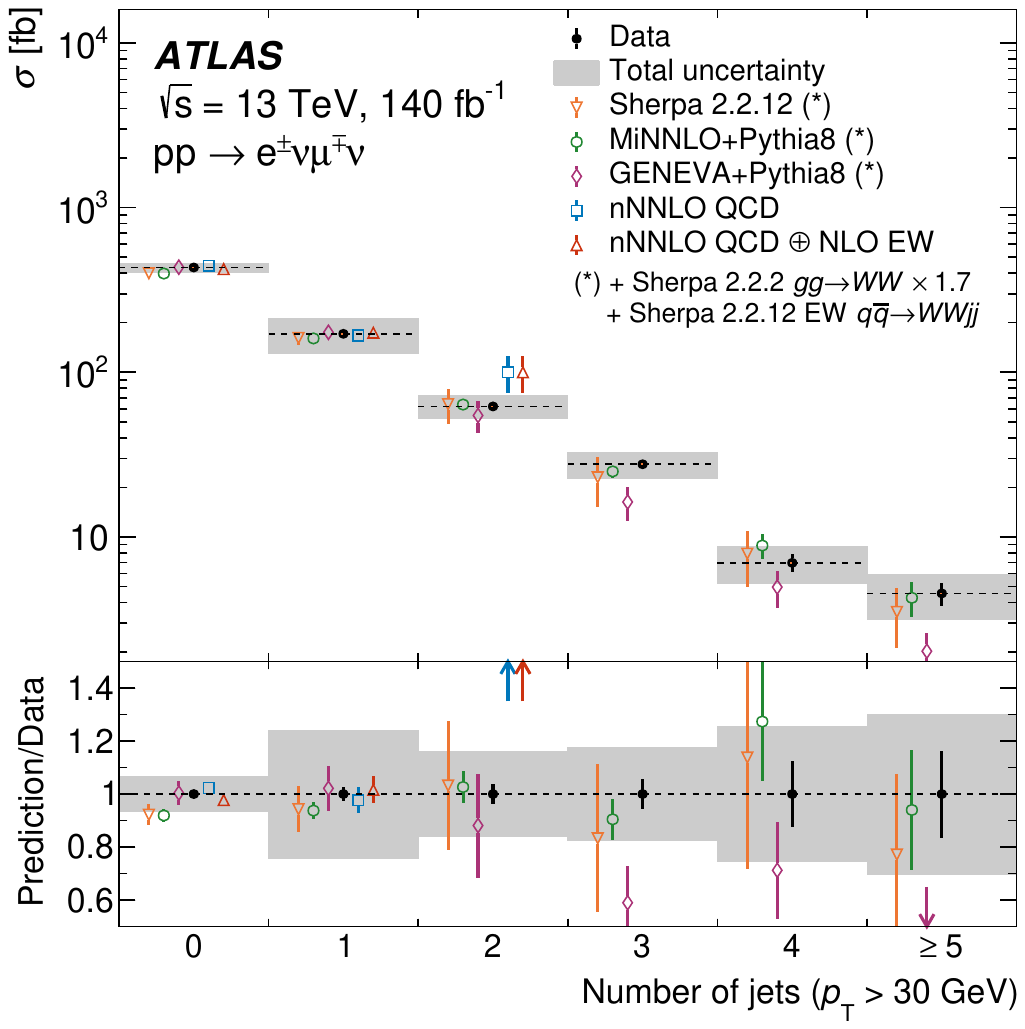}}%
\subfloat[]{\includegraphics[width=.5\linewidth]{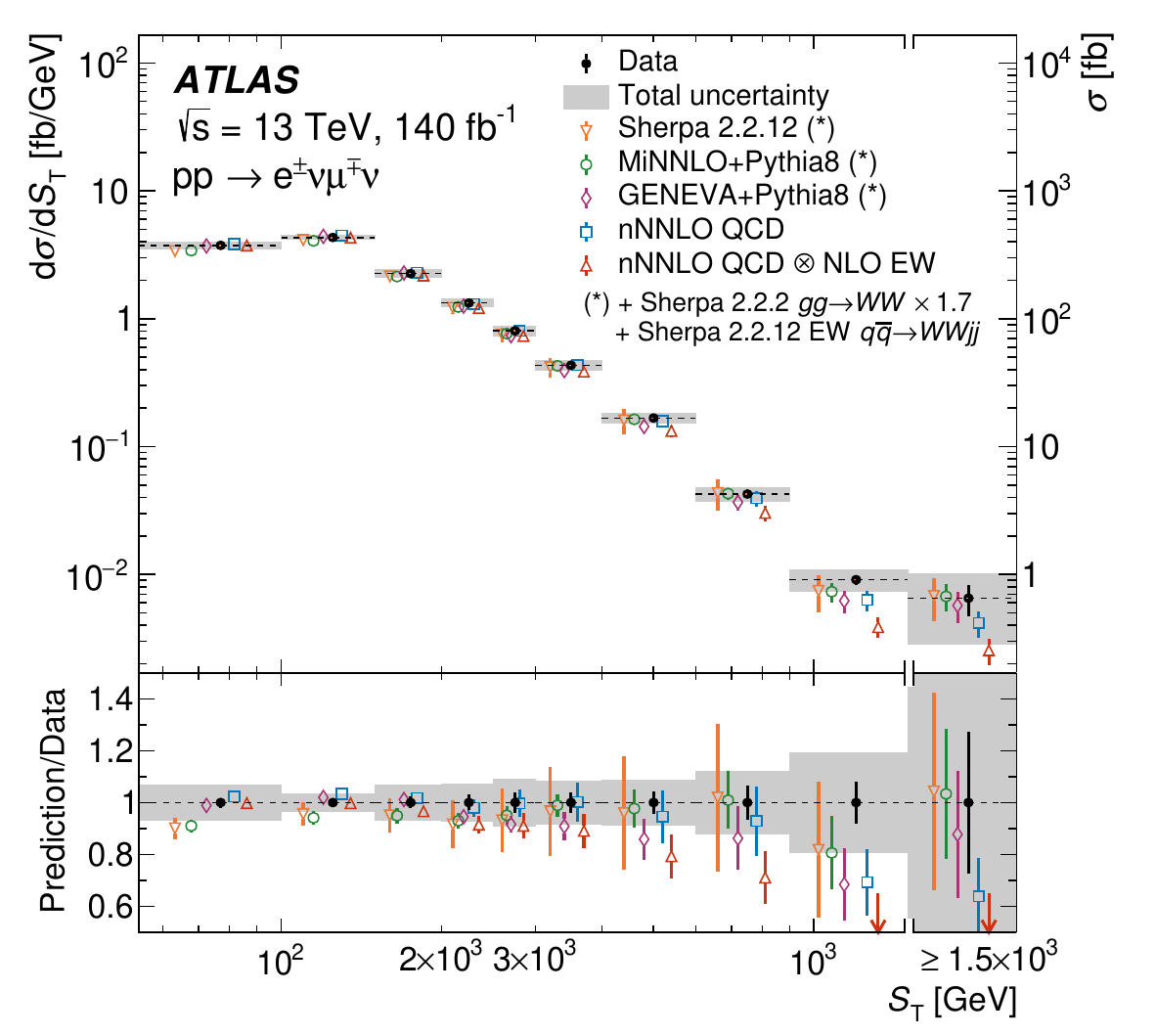}}\\
\caption{Fiducial differential cross-sections as a function of (a) \HTlep, (b) \mtll, (c) the jet multiplicity of the event, and (d) \ST.
The measured cross-section values are shown as points with error bars giving the statistical uncertainty and solid bands indicating the size of the total uncertainty. For distributions in which the rightmost bin is inclusive, the right-hand-side axis indicates the integrated cross-section of the rightmost bin. The results are compared with fixed-order nNNLO QCD (+ NLO EW) predictions from \textsc{Matrix\,2.1}, the NNLO+PS predictions from \textsc{Powheg MiNNLO}+\PYTHIA[8] and \textsc{Geneva}+\PYTHIA[8], and the \SHERPA[2.2.12]  NLO+PS predictions. The last three predictions are combined with the ones from \SHERPA[2.2.2] for the $gg$ initial state and \SHERPA[2.2.12] for electroweak $WWjj$ production. These contributions are modelled at LO but an NLO QCD $K$-factor of 1.7 is applied for gluon-induced production. Theoretical predictions are indicated as markers with vertical lines denoting PDF, scale and parton-shower uncertainties. Markers are staggered for better visibility.}
\label{fig:DiffResults-3}
\end{figure}

Figure~\ref{fig:DiffResultsTwoDpdf} displays the measured \ptll and |\yll{}| distributions in low ($85\,\GeV < \mll < 150\,\GeV$), intermediate ($150\,\GeV< \mll  < 250\,\GeV$), and high ($\mll  > 250\,\GeV$) \mll regions.
The two observables are proxies for the transverse and longitudinal boost of the \WW system and their measurement in different invariant mass regions offers sensitivity to parton distribution functions.
The measurement is in good agreement with modern four-flavour NNLO PDF sets, of which CT18~\cite{Hou:2019efy}, MSHT20~\cite{Bailey:2020ooq},  \NNPDF[3.0]~\cite{Ball:2014uwa}, \NNPDF[3.1]~\cite{NNPDF:2017mvq}, and \NNPDF[4.0]~\cite{NNPDF:2021njg} are shown.

\begin{figure}
\centering
\subfloat[]{\includegraphics[width=.4\linewidth]{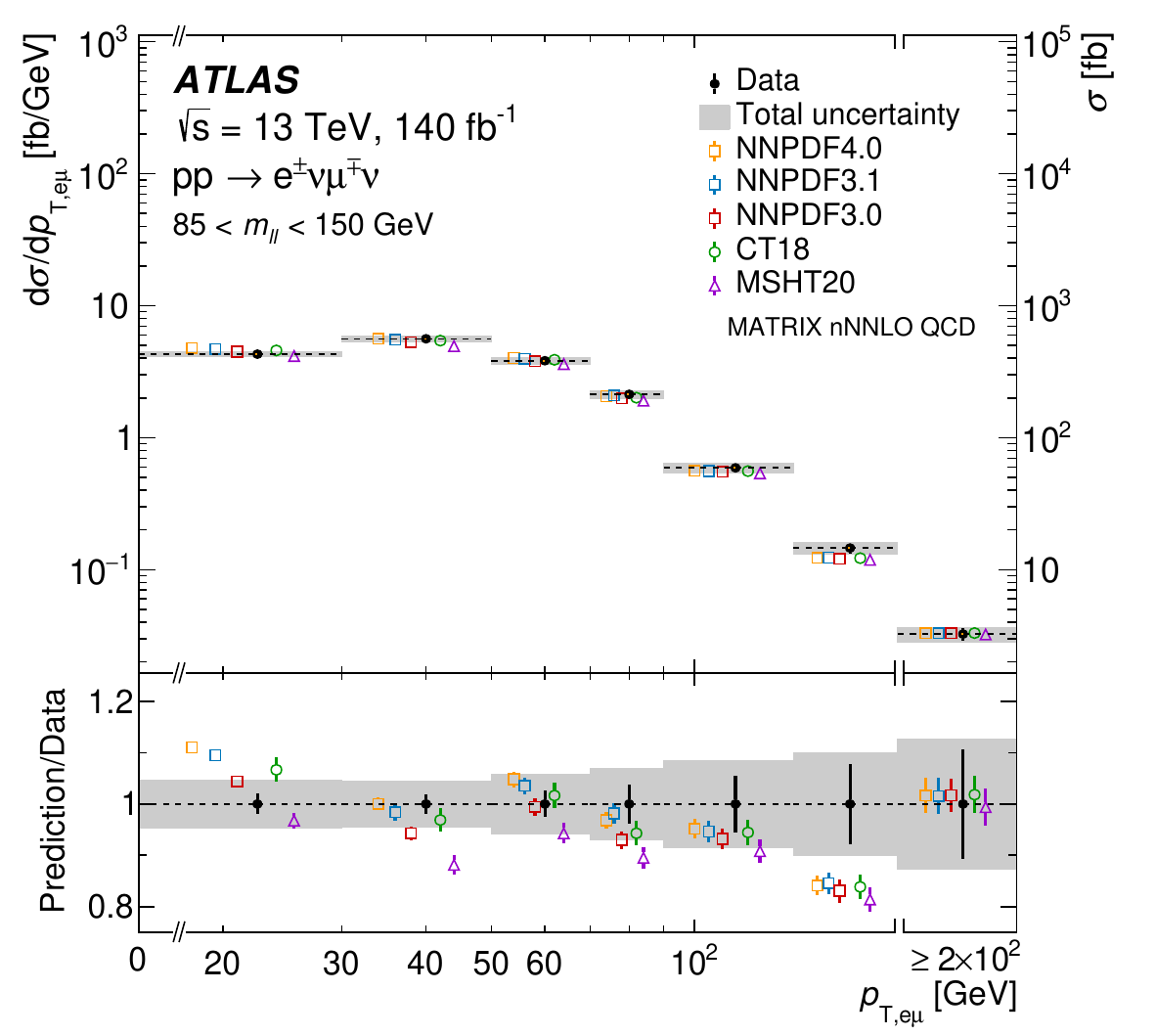}}%
\subfloat[]{\includegraphics[width=.4\linewidth]{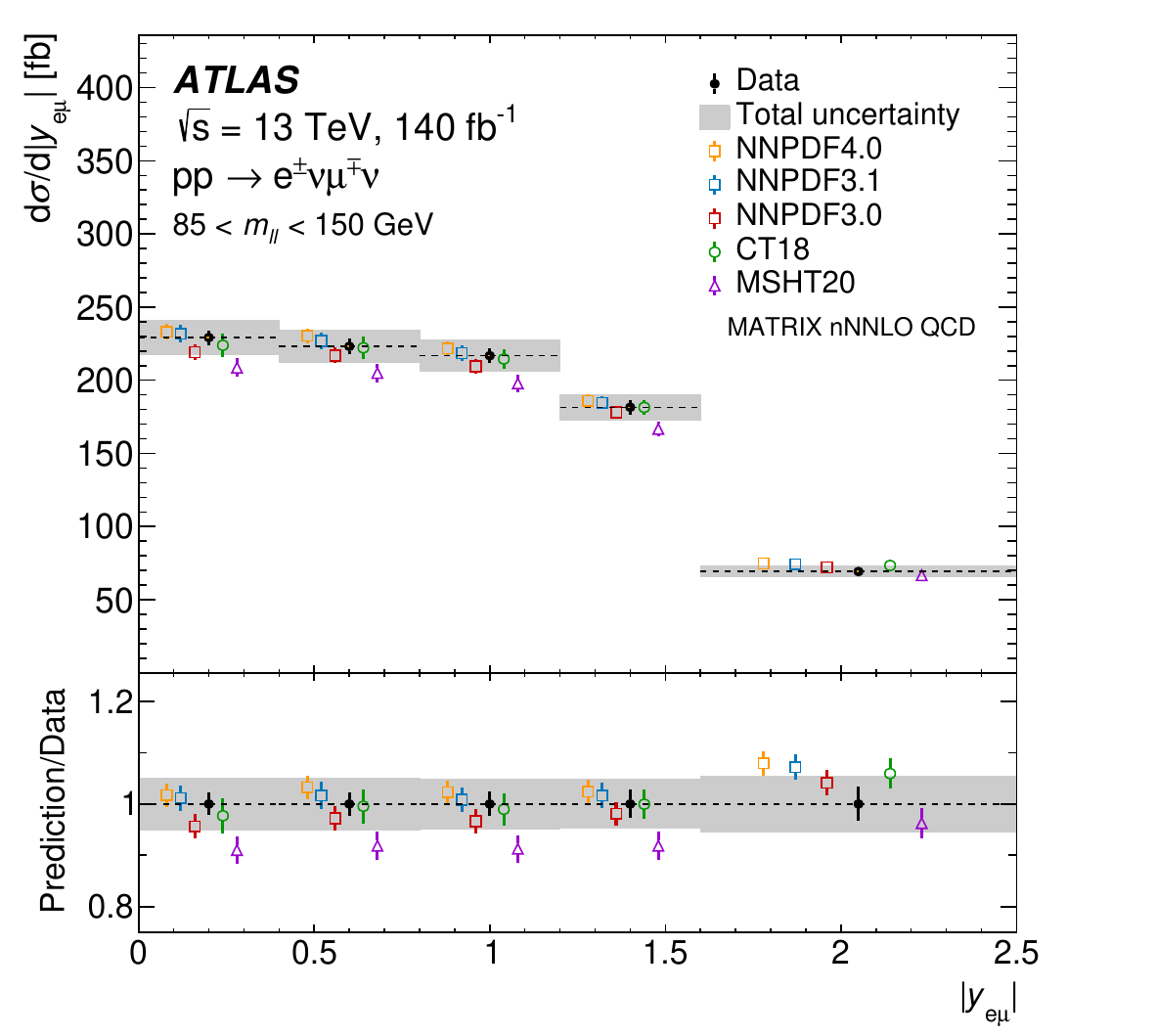}}\\
\subfloat[]{\includegraphics[width=.4\linewidth]{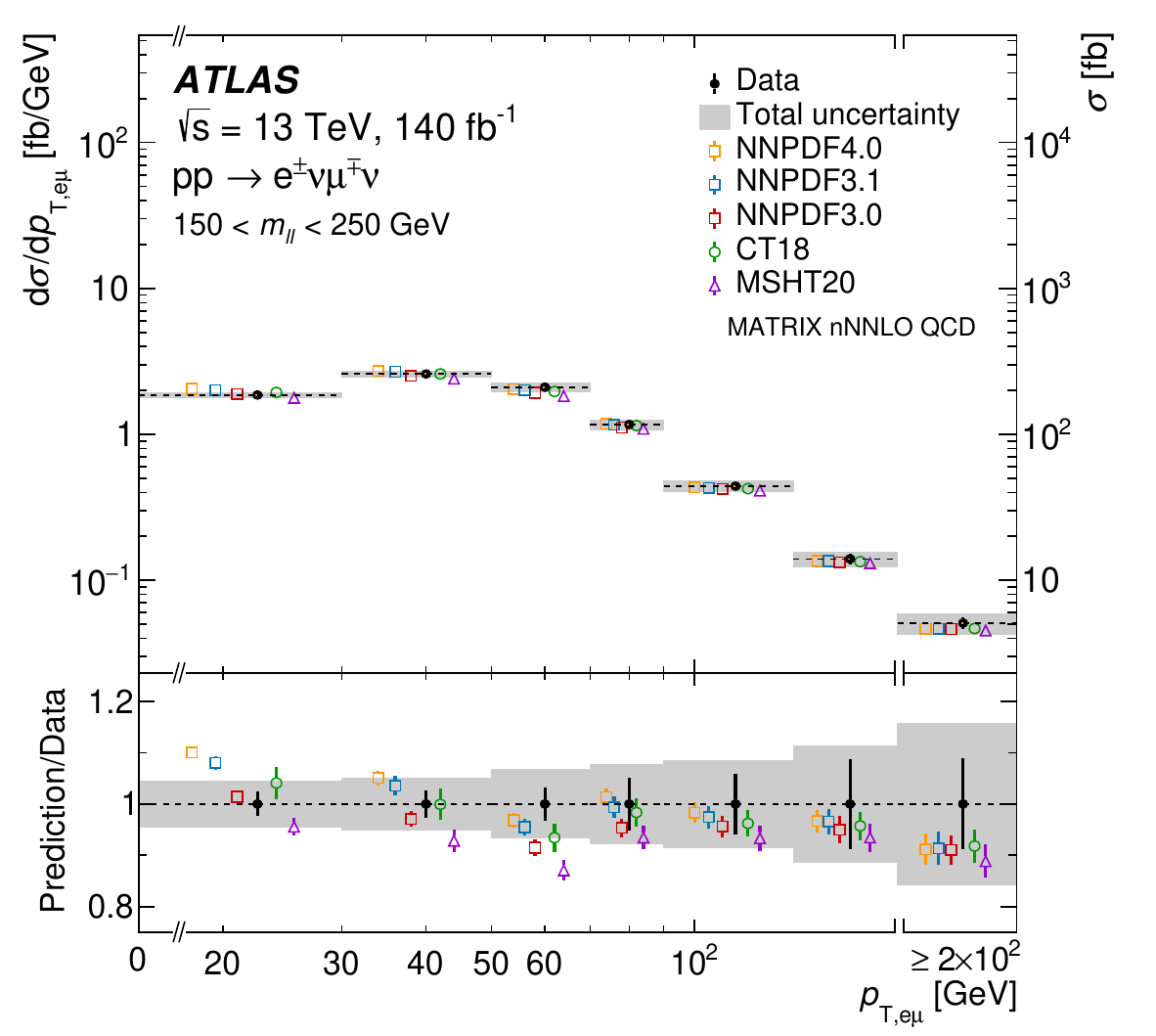}}%
\subfloat[]{\includegraphics[width=.4\linewidth]{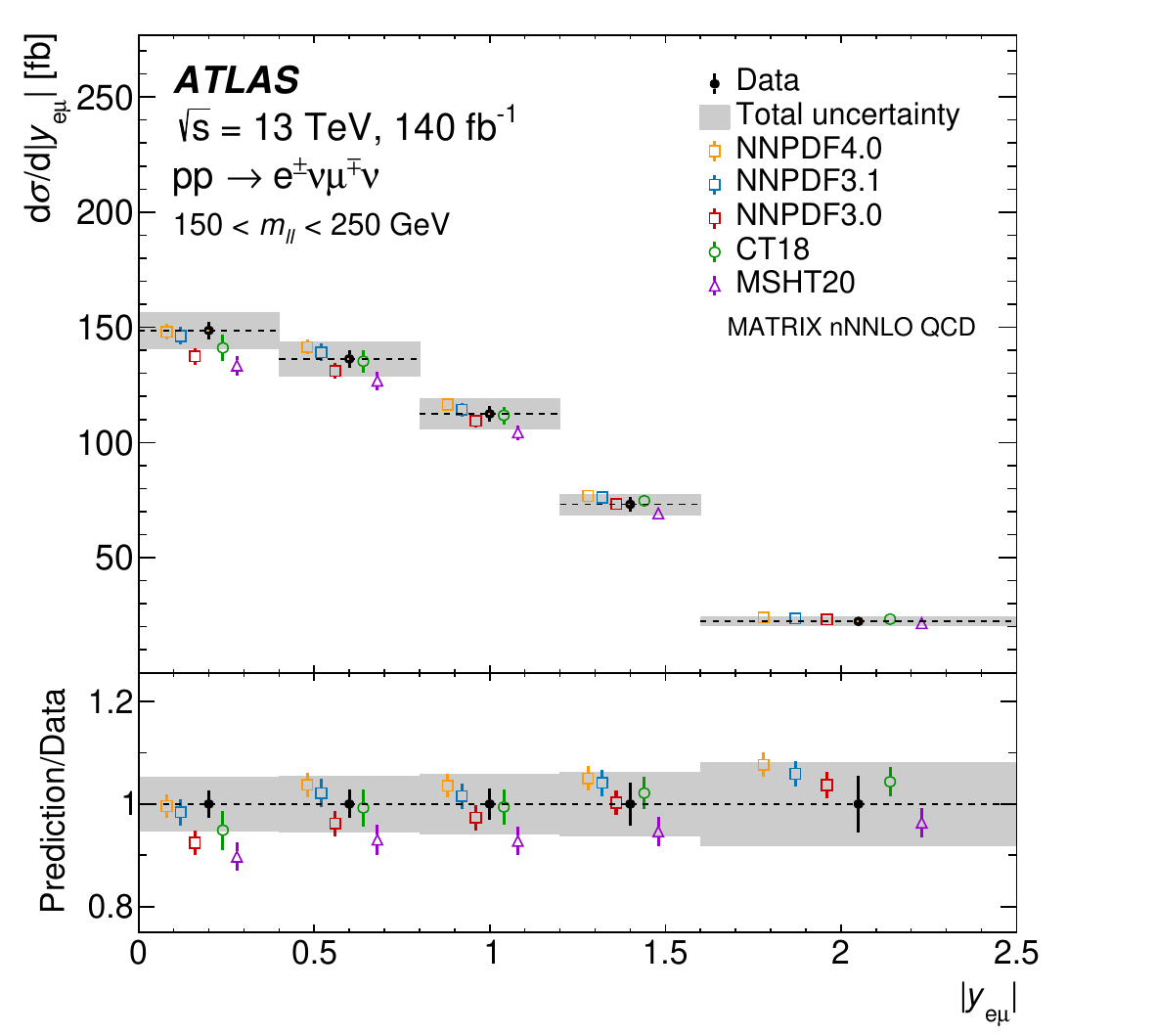}}\\
\subfloat[]{\includegraphics[width=.4\linewidth]{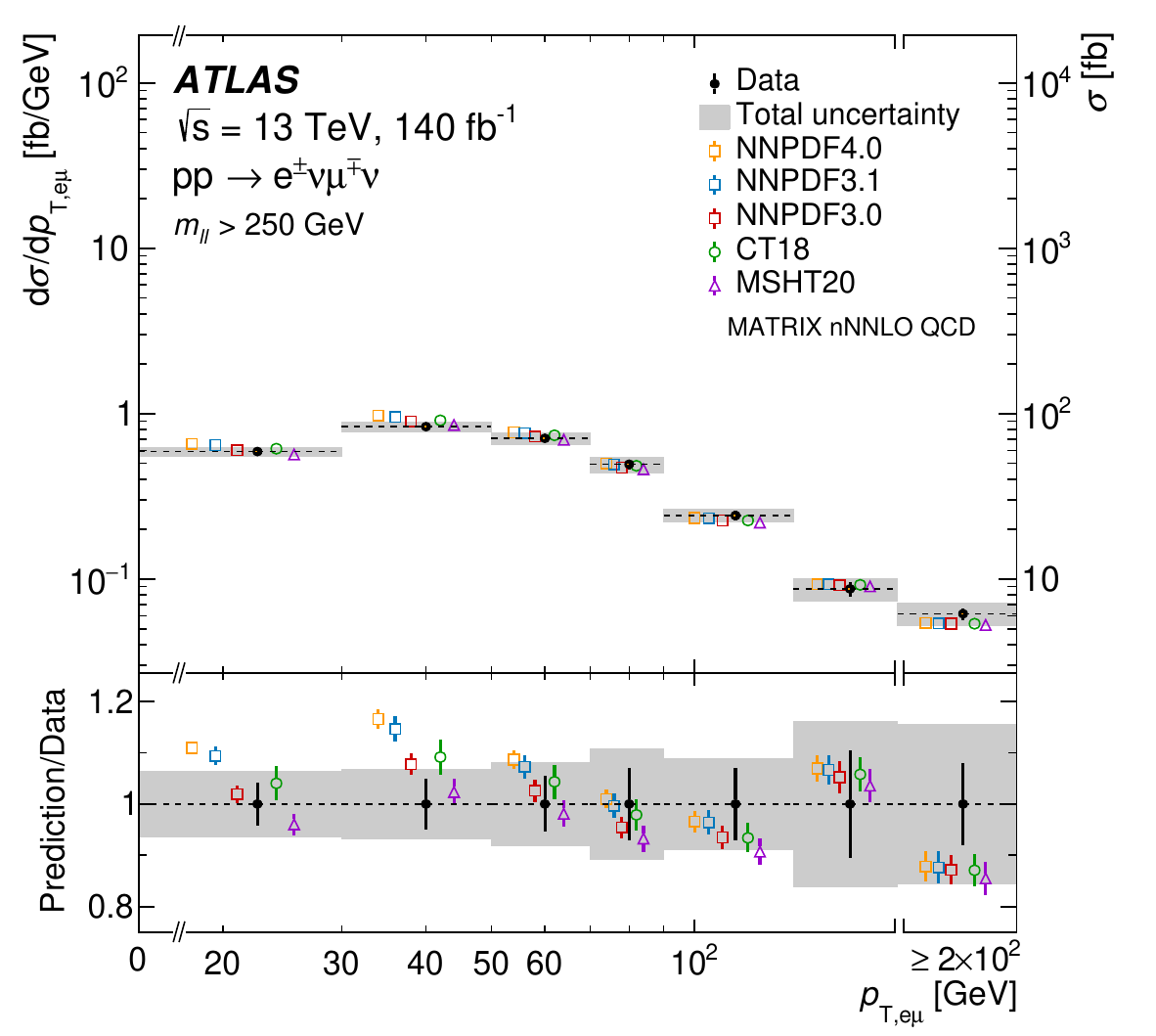}}%
\subfloat[]{\includegraphics[width=.4\linewidth]{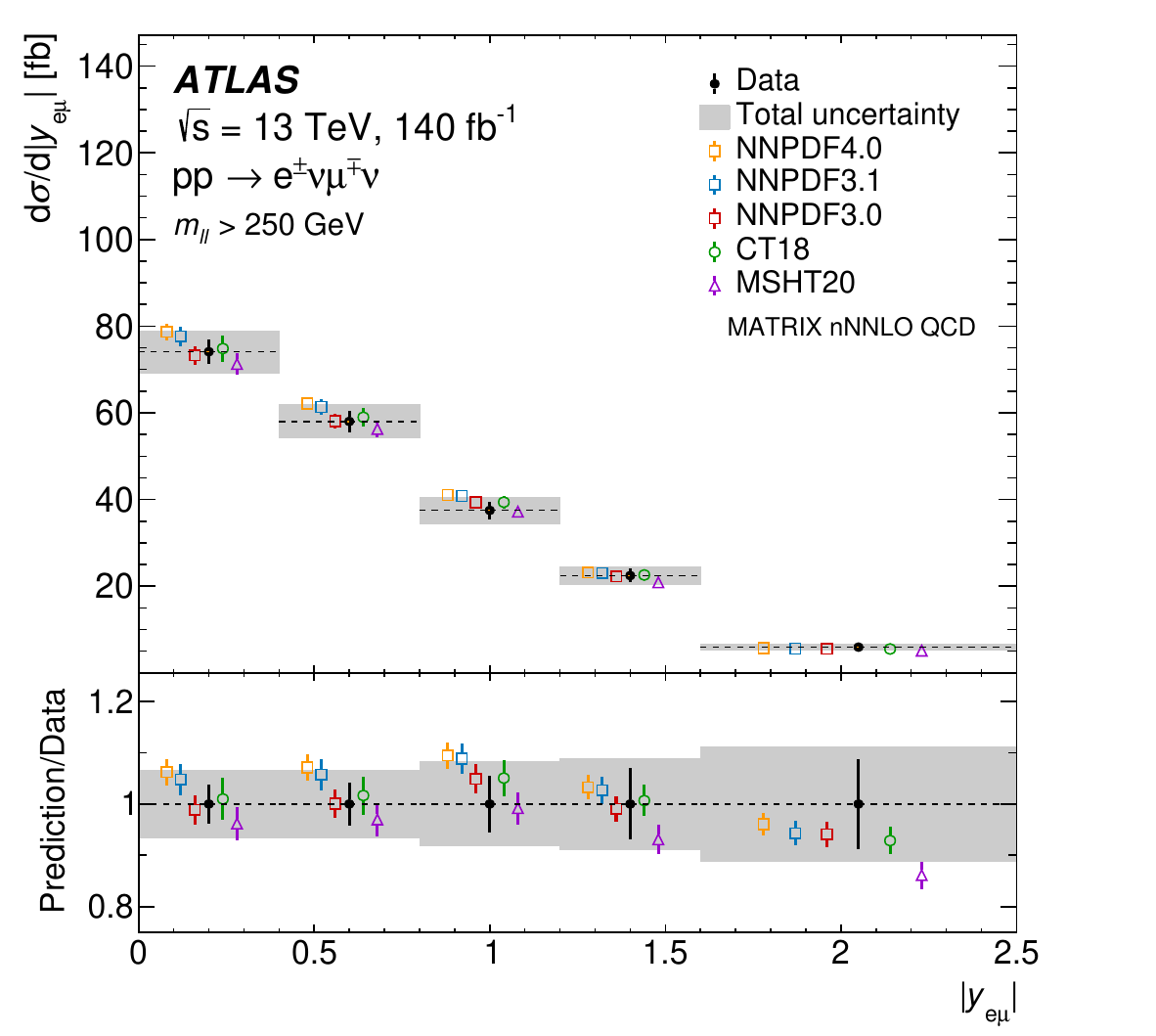}}\\
\caption{Fiducial differential cross-sections for \ptll and |\yll{}| for the nNNLO \textsc{Matrix} prediction and various four-flavour PDF sets in three \mll regions: ((a) and (b)) $85\,\GeV < \mll < 150\,\GeV$, ((c) and (d)) $150\,\GeV< \mll  <250\,\GeV$, and ((e) and (f)) $150\,\GeV< \mll  < 250\,\GeV$. The measured cross-section values are shown as points with error bars giving the statistical uncertainty and solid bands indicating the size of the total uncertainty. For \ptll, the right-hand-side axis indicates the integrated cross-section of the rightmost bin. The measurement is compared with fixed-order nNNLO QCD + NLO EW predictions from \textsc{Matrix\,2.1} using five different four-flavour NNLO PDF sets: \NNPDF[4.0], \NNPDF[3.1], \NNPDF[3.0], CT18, and MSHT20.}
\label{fig:DiffResultsTwoDpdf}
\end{figure}

\subsection{Asymmetries}
In this subsection the measurement of two asymmetries, $A_\text{C}$ and $A_\text{CP}$, is presented.
To determine these asymmetries free from detector effects, the distributions of the underlying observables, introduced in Section~\ref{sec:intro}, are background-subtracted and unfolded, analogously to the differential cross-section measurements described in the previous subsection.

In \WW production a rapidity-dependent charge asymmetry is expected~\cite{Re:2018vac}.
In the dominant $u\bar u$ production mode the $W^+$ boson tends to follow the more highly boosted up-quark direction, so the $W^+$ bosons tend to be more forward than $W^-$ bosons.
However, contributions from other production modes and the angular dependence of the boson decays lead to a much smaller observable charge asymmetry in the lepton rapidities, with the negatively charged lepton being more forward slightly more often.
Based on the \textsc{MiNNLO}+\PYTHIA[8] sample and including the \NNPDF[3.0nnlo] PDF uncertainty, the predicted lepton asymmetry $A_\text{C}$ (as introduced in Section~\ref{sec:intro}) is $-0.023\pm 0.003$.
Its measured value is
\[
A_\text{C}=-0.004 \pm 0.008~\mathrm{(stat.)} \pm 0.006~\mathrm{(syst.)} \,,
\]
which differs from the theoretical prediction by about two standard deviations.

The charge asymmetry is also measured as a function of the absolute difference of the absolute values of the lepton rapidities, $|\dabsetadef|$, and the invariant mass of the dilepton system, \mll.
An increase in the asymmetry is predicted for larger values of these quantities, as shown in Figure~\ref{fig:Asymmetries}.
Differences between predictions discussed in the previous section are small compared to the uncertainty of the measurement.
\begin{figure}
\centering
\subfloat[]{\includegraphics[width=.49\linewidth]{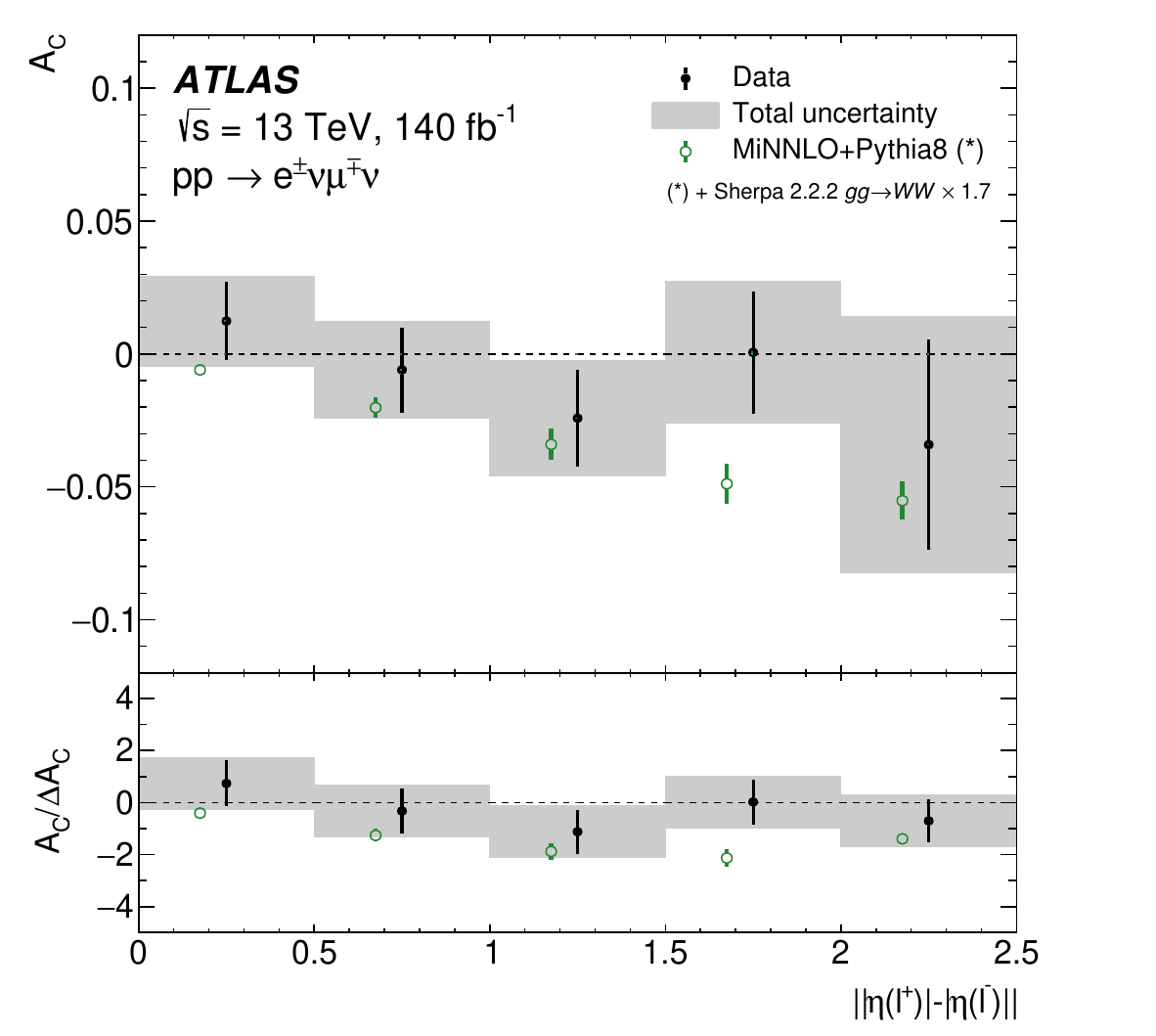}}
\subfloat[]{\includegraphics[width=.49\linewidth]{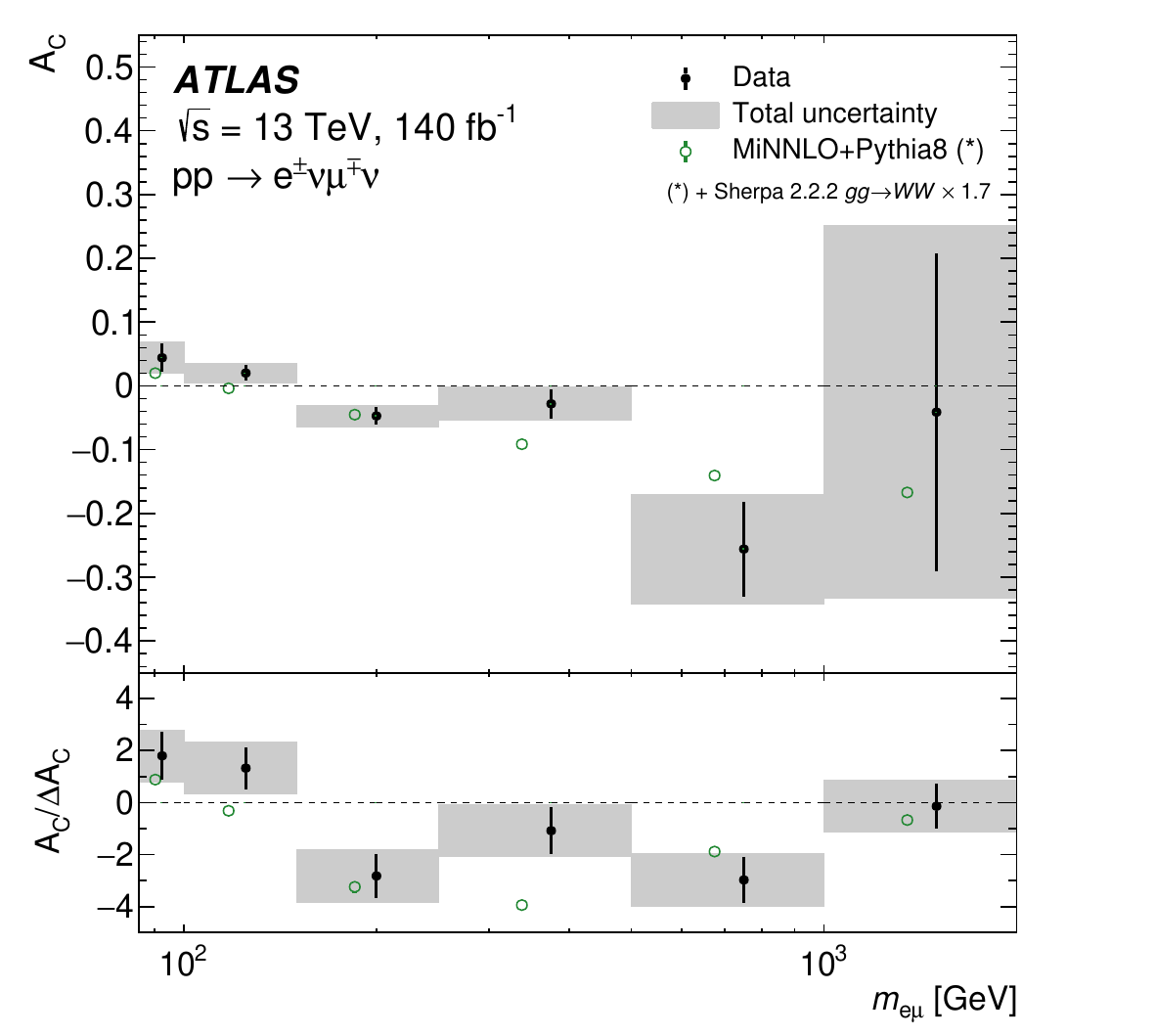}}
\caption{
Charge asymmetry $A_\text{C}$ measured as a function of (a) the absolute difference of the leptons' absolute rapidities $|\dabsetadef|$, and (b) the invariant mass of the dilepton system, \mll.
The measured values are shown as points, whose error bars represent the statistical uncertainty, while the solid bands indicate the size of the total uncertainty.
In the bottom panel, the central value of the charge asymmetry divided by the total uncertainty of the $A_\text{C}$ measurement is shown.
The measurement is compared with the NNLO+PS predictions from \textsc{Powheg MiNNLO}+\PYTHIA[8] with vertical lines denoting PDF uncertainties. For $A_\text{C}$ as a function of \mll, PDF uncertainties are smaller than the marker size and not visible.
}
\label{fig:Asymmetries}
\end{figure}
The significance of the measured differential asymmetry is evaluated using a hypothesis test based on the profile-likelihood-ratio test statistic. The measurement as a function of $|\dabsetadef|$ is compatible, within two standard deviations, with both the SM prediction and the hypothesis of no asymmetry. However, for the asymmetry distribution as a function of \mll, the absolute value of $A_\text{C}$ significantly increases for larger values of \mll, in agreement with the SM. In this case, the no-asymmetry hypothesis is rejected in favour of the SM with a significance of approximately three standard deviations.

A non-zero asymmetry in a CP-odd observable would signal CP violation. Since the SM does not predict any CP-violating effects in \WW production, such a finding would point to CP-violating new physics.
In this paper, the asymmetry $A_\text{CP}$ of the  $O_z$ observable, introduced in Section~\ref{sec:intro}, is measured for the first time.
It is found to be
\[
A_\text{CP} =0.014 \pm 0.008~\mathrm{(stat.)} \pm 0.008~\mathrm{(syst.)},
\]
which is compatible with the SM expectation of zero within about one standard deviation.


\subsection{Effective field theory interpretation}
The Standard Model effective field theory~\cite{Buchmuller:1985jz,Grzadkowski:2010es}
provides a framework for describing the effects of theories beyond the Standard Model (BSM theories) that introduce new physics at a scale $\Lambda$ much larger than the electroweak scale $v$.
The SMEFT allows observables to be expanded in $v/\Lambda$ and $E/\Lambda$, where $E$ is the typical energy exchanged in the process.
This is achieved by adding to the SM Lagrangian a series of operators ${\mathcal{O}_i^{(d)}}$, which consist of gauge invariant combinations of SM fields with an energy dimension $d$ greater than four:
\begin{equation*}
\mathcal{L}_\textrm{SMEFT} = \mathcal{L}_\textrm{SM} + \sum\limits_i \frac{c^{(5)}_i}{\Lambda} \mathcal{O}_i^{(5)} +  \sum\limits_i \frac{c^{(6)}_i}{\Lambda^2} \mathcal{O}_i^{(6)} + \dots.
\end{equation*}
Measurements of observables sensitive to the effect of SMEFT operators constrain $c_i^{(d)}/\Lambda^{d-4}$, where $c_i^{(d)}$ are the Wilson coefficients associated with the dimension-$d$ operator ${\mathcal{O}_i^{(d)}}$.
Operators of odd dimensionality introduce lepton and baryon number violation and are thus irrelevant for this analysis.
The leading effects of BSM physics are expected to manifest themselves as dimension-six operators, as higher-dimensional operators are suppressed by greater powers of $\Lambda$.

The Warsaw basis~\cite{Grzadkowski:2010es} provides a complete set of dimension-six operators consistent with the SM gauge symmetries.
For this analysis a $U(2)_q\times U(2)_u \times U(2)_d$  flavour symmetry is assumed, which requires identical couplings for quarks of the first two generations.
A quadratic parameterization of dimension-six SMEFT effects is obtained by reweighting~\cite{Mattelaer:2016gcx} a sample of simulated events generated  with \MADGRAPH[2.9.9]~\cite{Alwall:2014hca} and the \textsc{SMEFTsim}\,3.0~\cite{Brivio:2017btx,Brivio:2020onw} model.
Both the $e^\pm\nu\mu^\mp\nu$ and $e^\pm\nu\mu^\mp\nu$+1-jet final states are simulated at leading order at matrix-element level, then showered with \PYTHIA[8.307], and merged via the CKKW-L merging scheme~\cite{Lonnblad:2001iq,Lonnblad:2011xx} to avoid the double counting of jet emissions from the parton-shower and the matrix-element generators.
The merged sample shows good agreement with NLO QCD predictions generated with \textsc{SMEFT@NLO}~\cite{Degrande:2020evl} for the SM, and for effects linear and quadratic in the Wilson coefficients.
SMEFT effects are also simulated for the  $\gamma\gamma$ initial states; these contribute about 10\% of the predicted high-mass \WW modification due to anomalous triple-gauge-boson couplings.

Seven dimension-six Wilson coefficients that impact \WW production are analyzed.
Three of these coefficients -- \cW, \cHWB, and \cHDD{} -- modify triple-gauge-boson couplings (and in the case of \cHWB and \cHDD also $Z$-boson couplings to fermions), which strongly affects the high-mass \WW cross-section.
The remaining four coefficients -- \cHjone, \cHjthree, \cHu, and \cHd{} -- alter the couplings between quarks and weak bosons, resulting in significant effects at high invariant mass by disrupting the cancellations present in the SM amplitudes.
Other coefficients are not studied, either because their contributions do not increase with the \WW invariant mass (thus offering limited sensitivity) or because they correspond to four-fermion interactions that are more tightly constrained by high-mass Drell--Yan measurements.

The unfolded \mtll distribution in Figure~\ref{fig:DiffResults-3} is found to offer the best sensitivity to the targeted SMEFT effects.
It is verified, using simulated data, that the unfolding procedure accurately recovers the effects of the studied dimension-six operators.
Figure~\ref{fig:mTsmeft} shows the measured \mtll distribution again, now compared to the SM and SMEFT predictions described below
\begin{figure}
\centering
\includegraphics[width=0.6\textwidth]{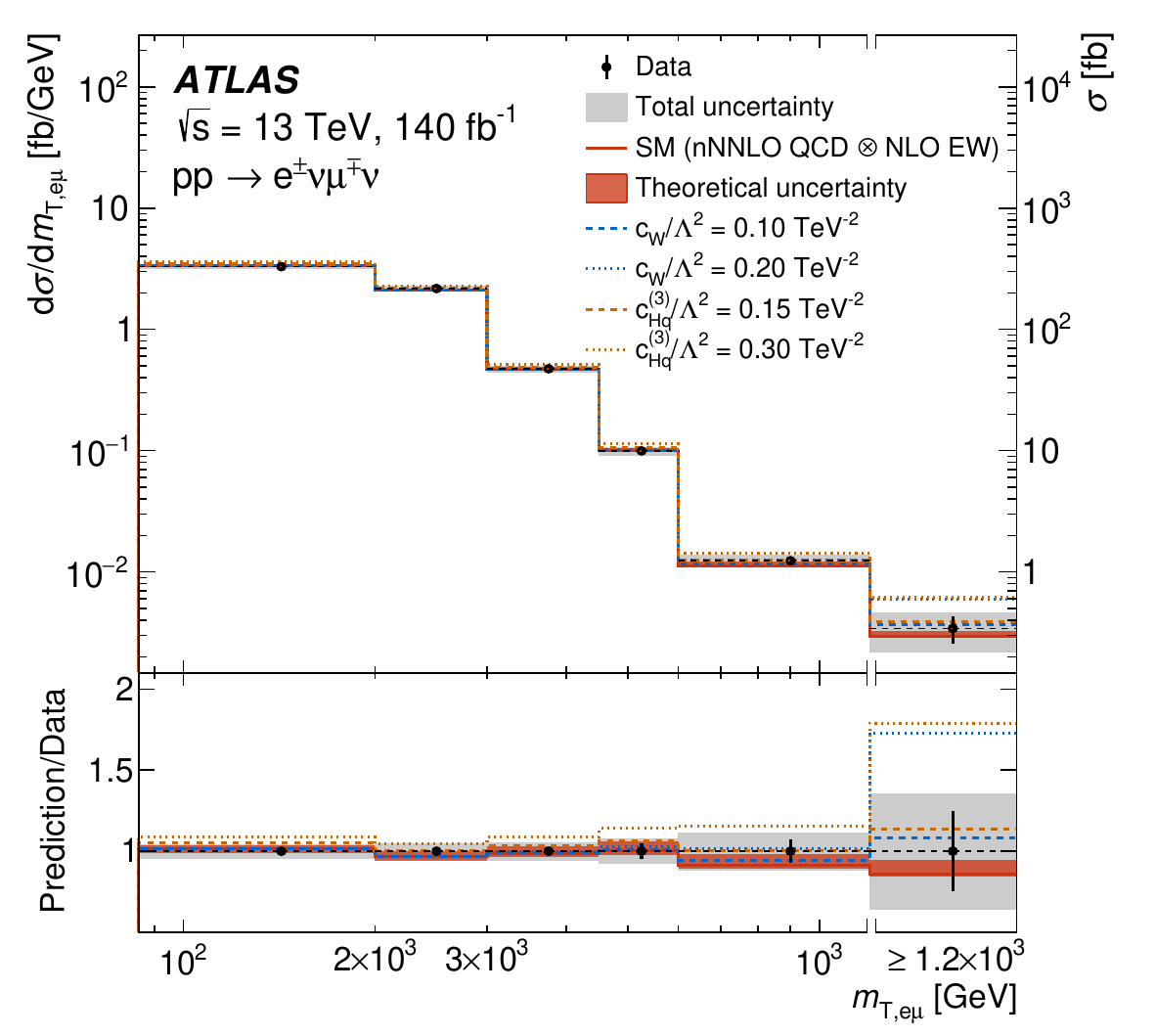}%
\caption{Measured \mtll distribution used in the SMEFT analysis, compared to the \textsc{Matrix} nNNLO QCD $\otimes$ NLO EW prediction obtained using \textsc{NNPDF3.1\,luxQED}, as well as SMEFT predictions for illustrative values of the Wilson coefficients \cW and \cHjthree.
Measured cross-section values are shown as points, with error bars representing statistical uncertainties and solid bands indicating the total experimental uncertainty. The right-hand vertical axis displays the integrated cross-section in the inclusive overflow bin.
The SM prediction is shown as a red line, with a shaded band denoting the total theoretical uncertainty used in the SMEFT fit, combined in quadrature. The uncertainty is asymmetric due to the dominant contribution from the two-point uncertainty defined by the difference between multiplicative and additive combinations of QCD and EW corrections.}
\label{fig:mTsmeft}
\end{figure}

The precise \textsc{Matrix} nNNLO QCD $\otimes$ NLO EW prediction introduced in Section~\ref{sec:predictions} serves as the basis for the SM prediction.
Theoretical uncertainties in this prediction arise from parton distribution functions, the QCD scales, and the unknown effect of mixed EW--QCD higher-order corrections.
The last is the dominant uncertainty, estimated as the difference of multiplicative and additive application of QCD and EW corrections, which affects the predicted yield in the last bin of the distribution by 10\%.

While the jet-merged SMEFT prediction agrees well with NLO QCD results from \textsc{SMEFT@NLO}, residual differences remain relative to the best SM prediction because NLO EW and NNLO QCD corrections to \WW production have not been computed within the SMEFT framework. As an approximation, it is assumed that the SM NLO EW and NNLO QCD corrections are applicable to the SMEFT prediction in each bin of the \mtll distribution. The QCD correction increases the SMEFT contribution in all bins by approximately 10\%, while the EW correction reduces it by up to 30\% in the last bin. This approach is conservative in the sense that not applying higher-order corrections would result in Wilson coefficient limits that are roughly 10\% more stringent.

To constrain the Wilson coefficients, a multivariate normal likelihood function is used. Statistical and systematic uncertainties are captured by the covariance of the multivariate distribution, while signal theory uncertainties are incorporated through Gaussian-constrained nuisance parameters.
Confidence intervals are derived using a profile-likelihood-ratio test statistic~\cite{Feldman:1997qc}, which is assumed to be $\chi^2$-distributed according to Wilks' theorem~\cite{Wilks:1938dza}.
The validity of the $\chi^2$-approximation is confirmed using pseudo-experiments.

\begin{table}
\centering
\caption{Expected and observed 95\% CL limits for various assumptions. For $O(\Lambda^{-2})$,  only the interference effect of dimension-six operators is considered, while for $O(\Lambda^{-4})$, both the interference and quadratic effects are included.  \enquote{Profiled'} means that all seven Wilson coefficients are fit simultaneously. Profiled limits cannot be obtained in the $O(\Lambda^{-2})$ fit, which is degenerate.
Confidence intervals are given for  $\Lambda=1\,\TeV$ and increase quadratically with the value of $\Lambda$.}
\begin{tabular}{@{}lcccccc@{}}
\toprule
& \multicolumn{2}{c}{$O(\Lambda^{-2})$, individual} & \multicolumn{2}{c}{$O(\Lambda^{-4})$, individual} & \multicolumn{2}{c}{$O(\Lambda^{-4})$, profiled}                                                          \\
& Expected                                          & Observed                                          & Expected                                        & Observed         & Expected         & Observed         \\
\midrule
\cW       & $ [-3.5,3.2] $                                    & $ [-3.5,3.4] $                                    & $ [-0.16,0.16] $                                & $ [-0.18,0.18] $ & $ [-0.17,0.16] $ & $ [-0.18,0.18] $ \\
\cHDD     & $ [-8.9,9.8] $                                    & $ [-11,8] $~~                                       & ~~$ [-7,21] $                                     & ~~$ [-8,21] $      & ~~$ [-7,21] $      & ~~$ [-8,21] $      \\
\cHWB     & $ [-8.4,9.2] $                                    & $ [-10,8] $~~                                       & $ [-1.5,1.7] $                                  & $ [-1.7,1.9] $   & $ [-1.7,1.7] $   & $ [-1.8,1.9] $   \\
\cHjone   & $ [-2.5,2.4] $                                    & $ [-2.2,2.8] $                                    & $ [-0.27,0.24] $                                & $ [-0.29,0.27] $ & $ [-0.29,0.29] $ & $ [-0.31,0.31] $ \\
\cHjthree & $ [-0.69,0.66] $                                  & ~~$ [-0.7,0.68] $                                   & $ [-0.28,0.22] $                                & $ [-0.31,0.24] $ & ~~$ [-0.3,0.27] $  & $ [-0.34,0.29] $ \\
\cHu      & $ [-3.2,3.0] $                                    & $ [-3.0,3.4] $                                    & $ [-0.31,0.29] $                                & $ [-0.35,0.32] $ & $ [-0.32,0.31] $ & $ [-0.35,0.33] $ \\
\cHd      & $ [-11,11] $                                      & $ [-11,11] $                                      & $ [-0.45,0.46] $                                & ~~$ [-0.5,0.51] $  & $ [-0.49,0.49] $ & $ [-0.52,0.53] $ \\
\bottomrule
\end{tabular}
\label{tab:eft_mt}
\end{table}

Table~\ref{tab:eft_mt} summarizes the expected and observed 95\% confidence level (CL) limits on Wilson coefficients, considering either only linear contributions, of $O(c_i/\Lambda^{2})$ -- arising from the interference between SM and BSM amplitudes -- or also quadratic terms of order $O(c_i^2/\Lambda^{4})$ -- arising purely from the BSM amplitudes.
Limits that account only for the linear contribution of the Wilson coefficients are much weaker, despite formally missing only a higher-order correction in $\Lambda^{-1}$.
This is largely due to helicity suppression of the  $O(\Lambda^{-2})$ interference effects~\cite{Azatov:2016sqh} and does not necessarily indicate a breakdown of the SMEFT expansion~\cite{Falkowski:2016cxu}.
In the case of the combined linear and quadratic parameterization, profiled limits can be derived by fitting all other parameters to their conditional best-fit values.
The linear fit suffers from degeneracies, as there are more degrees of freedom than data points, allowing constraints only on combinations of Wilson coefficients rather than individual parameters.

Limits driven by quadratic dimension-six contributions exhibit increased model dependence, as they are only valid for BSM models where linear contributions from dimension-eight operators are small relative to the quadratic dimension-six contributions -- despite both contributions being of the order $\Lambda^{-4}$.
To facilitate an \emph{a posteriori} assessment of the validity of limits for specific BSM models, a \enquote{clipping} scan is implemented following a recommendation of the LHC EFT Working Group~\cite{Brivio:2022pyi}.
In this approach, the effect of dimension-six operators is disregarded for a \WW invariant mass exceeding a cut-off, $M_\text{cut}$, allowing the extraction of limits sensitive only to data below a defined energy scale, enabling an estimation of unknown higher-order SMEFT contributions.
Profiled 95\% CL limits as a function of the cut-off value are shown in Figure~\ref{fig:mcutmain}.

\begin{figure}
\subfloat[]{\includegraphics[width=0.33\textwidth]{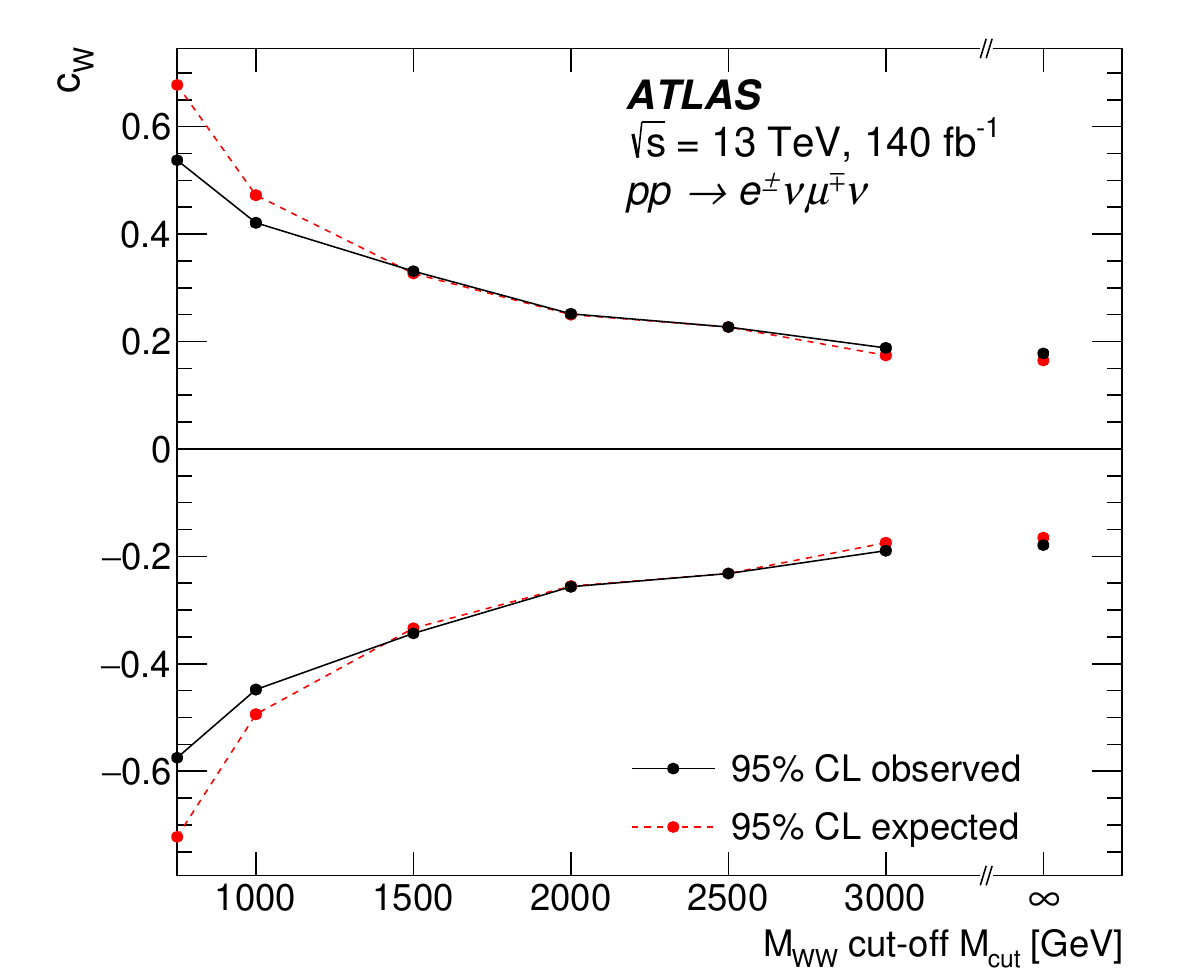}}%
\subfloat[]{\includegraphics[width=0.33\textwidth]{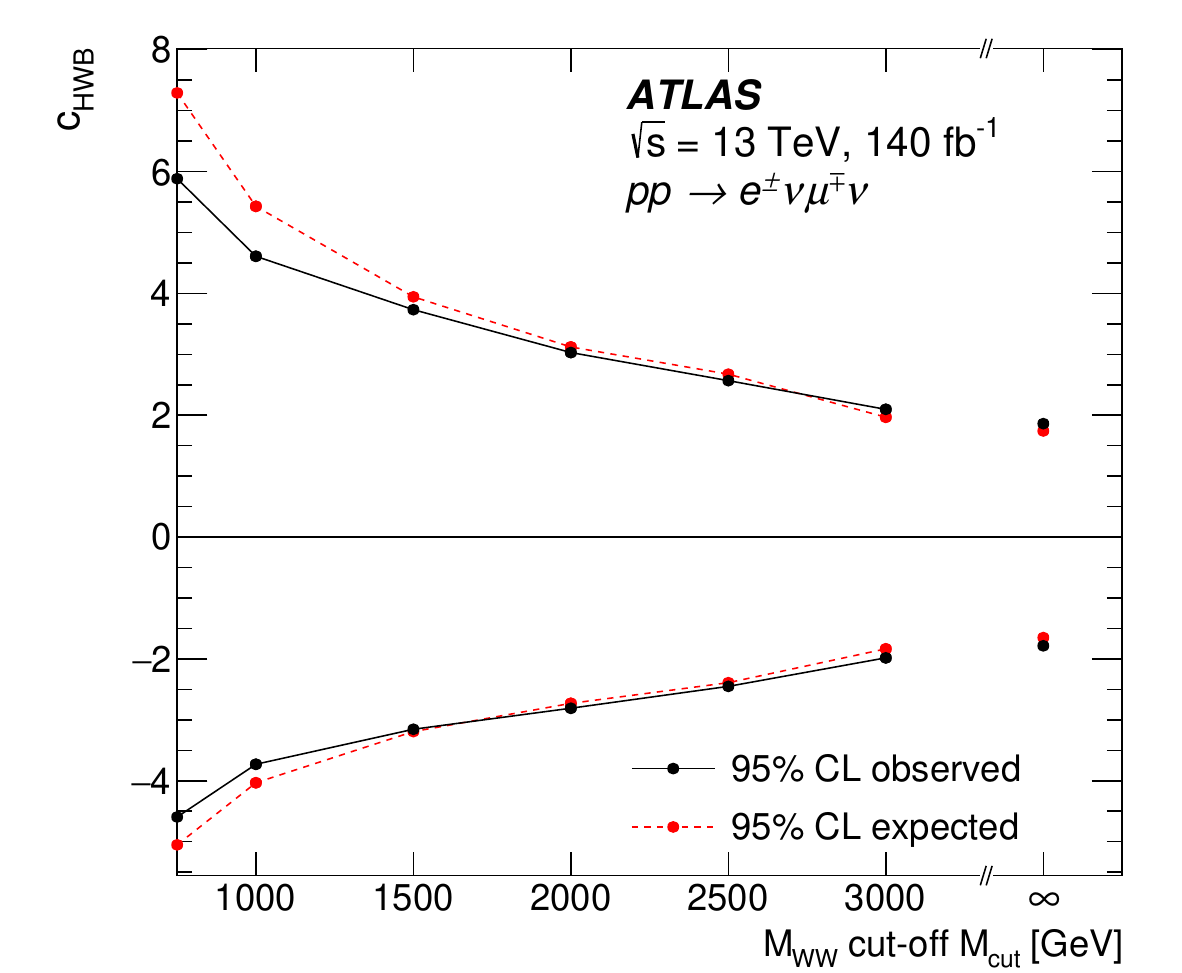}}%
\subfloat[]{\includegraphics[width=0.33\textwidth]{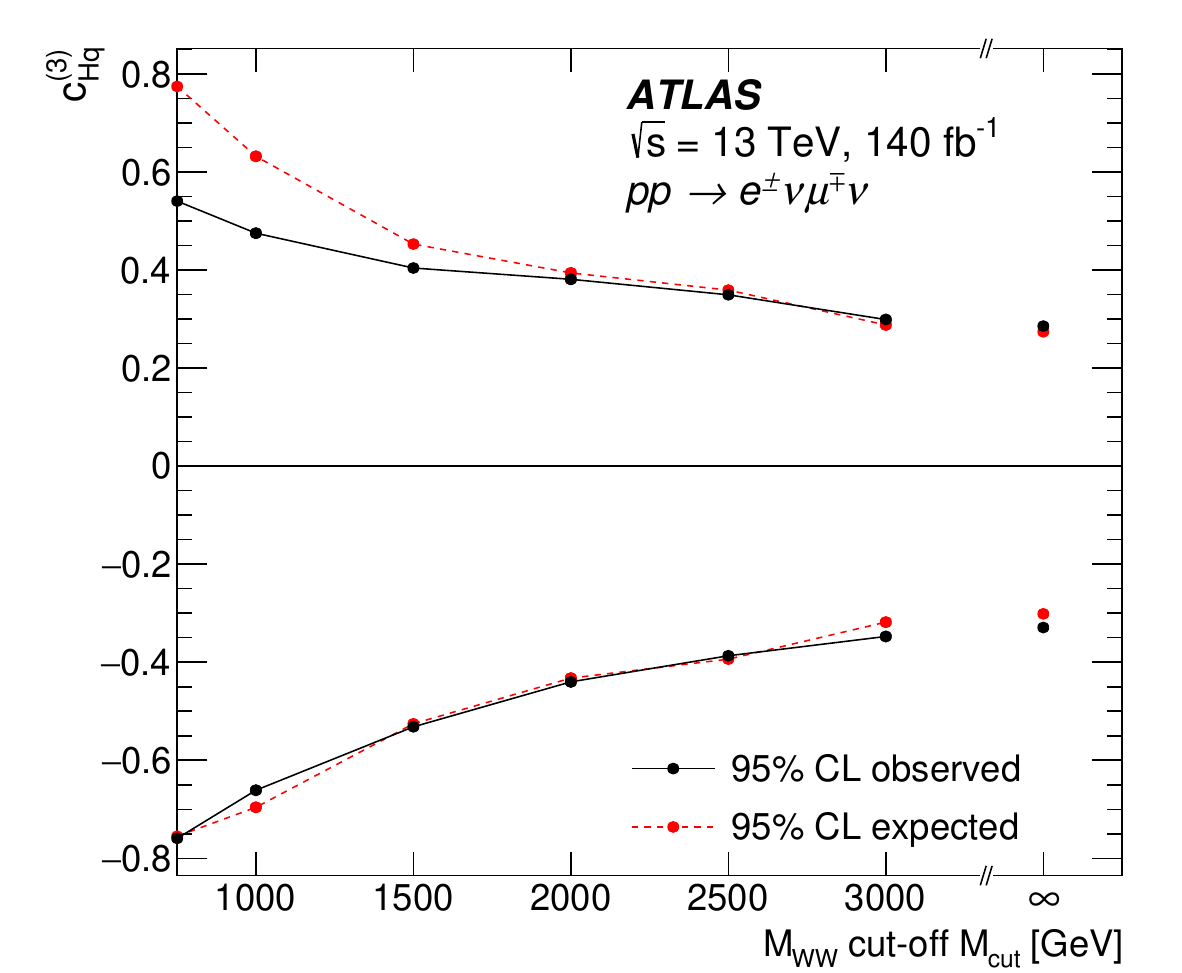}}\\
\subfloat[]{\includegraphics[width=0.33\textwidth]{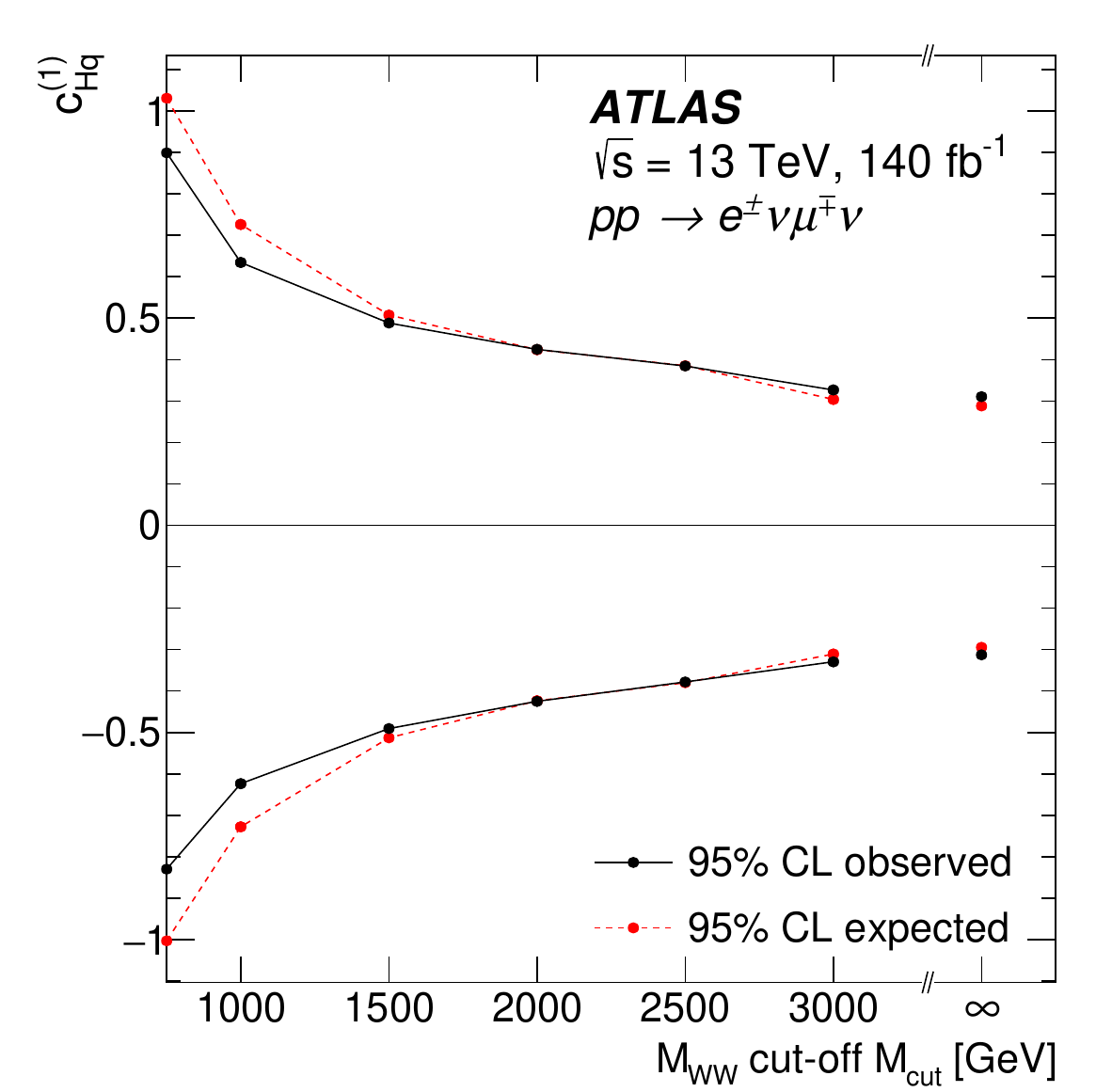}}%
\subfloat[]{\includegraphics[width=0.33\textwidth]{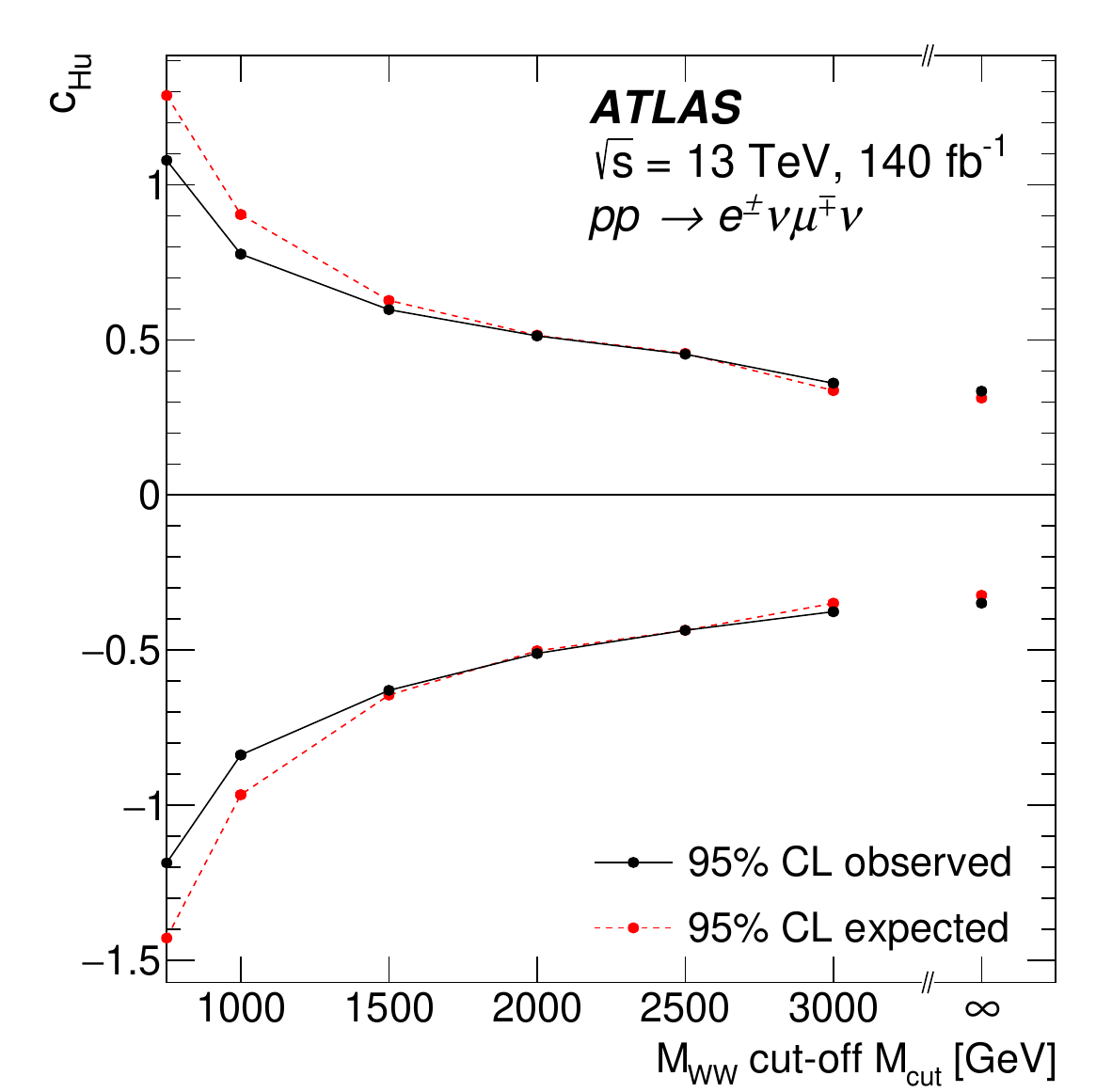}}%
\subfloat[]{\includegraphics[width=0.33\textwidth]{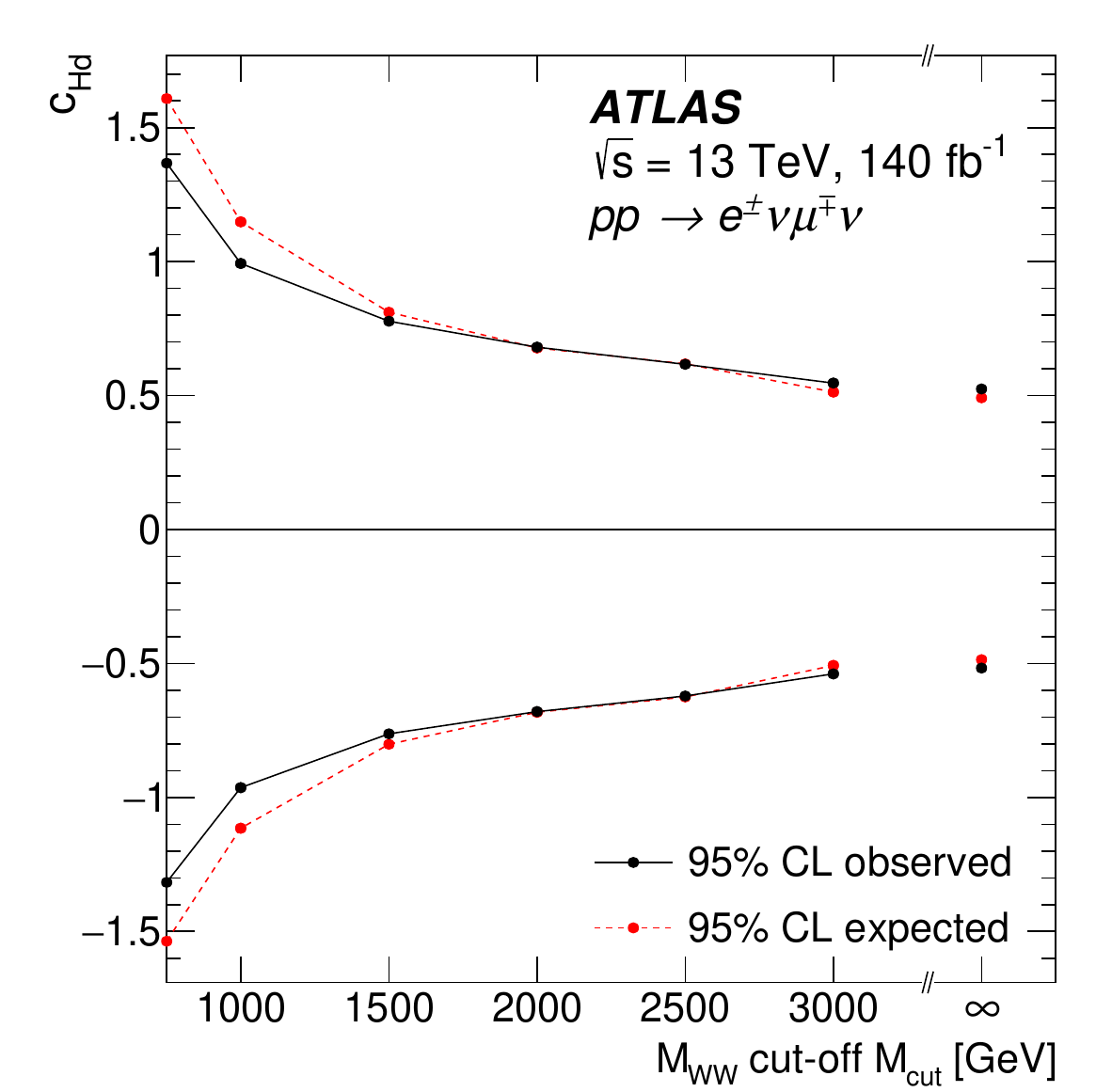}}%
\caption{Borders of the 95\% confidence intervals as a function of an upper mass cut-off $M_\text{cut}$, applied to the SMEFT effects only, for (a) \cW, (b) \cHWB, (c) \cHjthree, (d) \cHjone, (e) \cHu, and (f) \cHd. The six remaining Wilson coefficients are profiled in the fit. Not shown is \cHDD, which has no relevant $M_{WW}$ dependence.}
\label{fig:mcutmain}
\end{figure}

The limits  on the triple-gauge-boson coupling coefficient \cW, considering also quadratic contributions, represent an improvement on  the previous ATLAS \WW measurement~\cite{STDM-2017-24}, which provided constraints on $c_{WWW}$ in the HISZ basis~\cite{Hagiwara:1993ck,Degrande:2012wf} that translate into a 95\% confidence interval of $[-0.24,0.24]$ for the Warsaw basis coefficient \cW.
However, they are weaker than those obtained in $WZ$~\cite{CMS-SMP-20-014} and $W\gamma$~\cite{CMS-SMP-20-005,CMS-SMP-19-002} final states, which are the most stringent constraints from diboson data.
In the linear approximation, limits are much stronger than in the previous $WW$ measurement, owing to the inclusion of final states with jets, which partially mitigate the helicity surpression of the interference~\cite{Azatov:2017kzw}.
Nonetheless, in the linear case the sensitivity is still an order of magnitude weaker than in the ATLAS analysis of electroweak $Zjj$ production~\cite{STDM-2017-27} and the CMS analysis of $W\gamma$ production~\cite{CMS-SMP-20-005}, for which angular observables specifically sensitive to the interference can be constructed.

The individual constraints on the quark-coupling-modifying Wilson coefficients \cHjone, \cHjthree, \cHu, and \cHd are weaker than those obtained from global fits~\cite{Ellis:2020unq} that include electroweak precision data~\cite{ALEPH:2005ab}.
However, these global constraints are highly correlated, and profiled limits on these coefficients are considerably weaker and cannot effectively distinguish between anomalous couplings to left-handed light quarks (as introduced by \cHjone and \cHjthree) and right-handed light quarks (as introduced by \cHu and \cHd).
When profiling over other coefficients, this \WW measurement particularly improves constraints on the Wilson coefficients that modify right-handed quark couplings, as no sensitivity to these operators exists in $WZ$ or $W\gamma$ final states.
However, the improvements are mainly driven by quadratic contributions in the Wilson coefficients and thus carry larger model dependence than those from precision data.
While constraints on \cHjone, \cHjthree, \cHu, and \cHd from Higgs boson production in association with a vector boson ($VH$) are typically stronger by a factor of two~\cite{HIGG-2018-51,HIGG-2018-52}, the \WW measurement offers complementary sensitivity, as $VH$ is also affected by additional Wilson coefficients.


\FloatBarrier

\section{Conclusion}
\label{sec:conclusions}
Differential and integrated fiducial cross-sections for the production of $W$-boson pairs decaying into final states with one electron and one muon of opposite electric charge are measured using a 140\,fb$^{-1}$ dataset of $\sqrt{s}=13$\,TeV $pp$ collisions recorded by the ATLAS detector at the LHC.

A jet-inclusive event selection allows comparisons of the measured cross-sections with precise theoretical calculations.
Improved data-driven estimates of the top-quark and fake-lepton backgrounds reduce the uncertainty in the fiducial cross-section to 3.1\%.
The measurement is extrapolated to the full phase space, resulting in a total $WW$ cross-section of $127\pm4$\,pb.
Differential cross-sections are measured as a function of twelve observables.

The measurements are compared with several state-of-the-art theoretical predictions.
A fixed-order prediction at nNNLO in QCD using the \textsc{NNPDF3.1nnlo\,luxQED} set of PDFs is in excellent agreement with the measured total cross-section and gives a good description of differential cross-sections.
Its multiplicative combination with EW corrections improves the description of some observables, although it does not provide an adequate description in other cases because the combination cannot take into account non-factorizing EW--QCD effects.
Furthermore, parton-shower-matched predictions from \PowMiNNLO{}+\PYTHIA[8], \textsc{Geneva}+\PYTHIA[8],  and \SHERPA[2.2.12] are able to correctly describe observables sensitive to jet emission.
Given the experimental and theoretical uncertainties, excellent agreement between the measurements and predictions is observed.

The rapidity-dependent charge asymmetry in \WW production, $A_\text{C}$, is measured differentially in the dilepton invariant mass \mll and in the absolute difference of the leptons' absolute rapidities.
A significant asymmetry is observed as a function of \mll, in accordance with the SM expectation.
Furthermore, an asymmetry $A_\text{CP}$ in a CP-odd observable is found to be compatible with no CP violation.

Limits on Wilson coefficients that modify triple-gauge-boson couplings and the coupling of quarks to bosons are set in the SMEFT framework. When considering the effects of dimension-six operators at order $\Lambda^{-4}$, the limits on \cW, \cHu, and \cHd are competitive with those derived from electroweak precision data, as well as those from weak-boson pair and $ZH$ measurements at the LHC, although they do not provide the strongest constraints on any of these coefficients.


\FloatBarrier

\section*{Acknowledgements}

%
%

%
%

We thank CERN for the very successful operation of the LHC and its injectors, as well as the support staff at
CERN and at our institutions worldwide without whom ATLAS could not be operated efficiently.

The crucial computing support from all WLCG partners is acknowledged gratefully, in particular from CERN, the ATLAS Tier-1 facilities at TRIUMF/SFU (Canada), NDGF (Denmark, Norway, Sweden), CC-IN2P3 (France), KIT/GridKA (Germany), INFN-CNAF (Italy), NL-T1 (Netherlands), PIC (Spain), RAL (UK) and BNL (USA), the Tier-2 facilities worldwide and large non-WLCG resource providers. Major contributors of computing resources are listed in Ref.~\cite{ATL-SOFT-PUB-2025-001}.

We gratefully acknowledge the support of ANPCyT, Argentina; YerPhI, Armenia; ARC, Australia; BMWFW and FWF, Austria; ANAS, Azerbaijan; CNPq and FAPESP, Brazil; NSERC, NRC and CFI, Canada; CERN; ANID, Chile; CAS, MOST and NSFC, China; Minciencias, Colombia; MEYS CR, Czech Republic; DNRF and DNSRC, Denmark; IN2P3-CNRS and CEA-DRF/IRFU, France; SRNSFG, Georgia; BMFTR, HGF and MPG, Germany; GSRI, Greece; RGC and Hong Kong SAR, China; ISF and Benoziyo Center, Israel; INFN, Italy; MEXT and JSPS, Japan; CNRST, Morocco; NWO, Netherlands; RCN, Norway; MEiN, Poland; FCT, Portugal; MNE/IFA, Romania; MESTD, Serbia; MSSR, Slovakia; ARIS and MVZI, Slovenia; DSI/NRF, South Africa; MICINN, Spain; SRC and Wallenberg Foundation, Sweden; SERI, SNSF and Cantons of Bern and Geneva, Switzerland; NSTC, Taipei; TENMAK, T\"urkiye; STFC/UKRI, United Kingdom; DOE and NSF, United States of America.

Individual groups and members have received support from BCKDF, CANARIE, CRC and DRAC, Canada; CERN-CZ, PRIMUS 21/SCI/017 and UNCE SCI/013, Czech Republic; COST, ERC, ERDF, Horizon 2020, ICSC-NextGenerationEU and Marie Sk{\l}odowska-Curie Actions, European Union; Investissements d'Avenir Labex, Investissements d'Avenir Idex and ANR, France; DFG and AvH Foundation, Germany; Herakleitos, Thales and Aristeia programmes co-financed by EU-ESF and the Greek NSRF, Greece; BSF-NSF and MINERVA, Israel; Norwegian Financial Mechanism 2014-2021, Norway; NCN and NAWA, Poland; La Caixa Banking Foundation, CERCA Programme Generalitat de Catalunya and PROMETEO and GenT Programmes Generalitat Valenciana, Spain; G\"{o}ran Gustafssons Stiftelse, Sweden; The Royal Society and Leverhulme Trust, United Kingdom.

In addition, individual members wish to acknowledge support from Chile: Agencia Nacional de Investigaci\'on y Desarrollo (FONDECYT 1190886, FONDECYT 1210400); China: National Natural Science Foundation of China (NSFC - 12175119, NSFC 12275265, NSFC-12075060); Czech Republic: PRIMUS Research Programme (PRIMUS/21/SCI/017); EU: H2020 European Research Council (ERC - 101002463, H2020-MSCA-IF-2020: HPOFHIC - 10103); European Union: European Research Council (ERC - 948254), Horizon 2020 Framework Programme (MUCCA - CHIST-ERA-19-XAI-00), European Union, Future Artificial Intelligence Research (FAIR-NextGenerationEU PE00000013), Italian Center for High Performance Computing, Big Data and Quantum Computing (ICSC, NextGenerationEU), Marie Sklodowska-Curie Actions (EU H2020 MSC IF GRANT NO 101033496); France: Agence Nationale de la Recherche (ANR-20-CE31-0013, ANR-21-CE31-0013, ANR-21-CE31-0022), Investissements d'Avenir Idex (ANR-11-LABX-0012), Investissements d'Avenir Labex (ANR-11-LABX-0012); Germany: Baden-Württemberg Stiftung (BW Stiftung-Postdoc Eliteprogramme), Deutsche Forschungsgemeinschaft (DFG - CR 312/5-1); Italy: Istituto Nazionale di Fisica Nucleare (FELLINI G.A. n. 754496, ICSC, NextGenerationEU); Japan: Japan Society for the Promotion of Science (JSPS KAKENHI JP21H05085, JSPS KAKENHI JP22H01227, JSPS KAKENHI JP22H04944); Netherlands: Netherlands Organisation for Scientific Research (NWO Veni 2020 - VI.Veni.202.179); Norway: Research Council of Norway (RCN-314472); Poland: Polish National Agency for Academic Exchange (PPN/PPO/2020/1/00002/U/00001), Polish National Science Centre (NCN 2021/42/E/ST2/00350, NCN UMO-2019/34/E/ST2/00393, NCN \& H2020 MSCA 945339, UMO-2020/37/B/ST2/01043, UMO-2021/40/C/ST2/00187); Portugal: Foundation for Science and Technology (FCT); Slovenia: Slovenian Research Agency (ARIS grant J1-3010); Spain: BBVA Foundation (LEO22-1-603), Generalitat Valenciana (Artemisa, FEDER, IDIFEDER/2018/048), La Caixa Banking Foundation (LCF/BQ/PI20/11760025), Ministry of Science and Innovation (MCIN \& NextGenEU PCI2022-135018-2, MICIN \& FEDER PID2021-125273NB, RYC2019-028510-I, RYC2020-030254-I, RYC2021-031273-I, RYC2022-038164-I), PROMETEO and GenT Programmes Generalitat Valenciana (CIDEGENT/2019/023, CIDEGENT/2019/027); Sweden: Swedish Research Council (VR 2018-00482, VR 2021-03651, VR 2022-03845, VR 2022-04683), Knut and Alice Wallenberg Foundation (KAW 2017.0100, KAW 2018.0157, KAW 2018.0458, KAW 2019.0447); Switzerland: Swiss National Science Foundation (SNSF - PCEFP2\_194658); United Kingdom: Leverhulme Trust (Leverhulme Trust RPG-2020-004); United States of America: U.S. Department of Energy (ECA DE-AC02-76SF00515), Neubauer Family Foundation.

%
%


%

%

%
%
%
%
%
%
%

%
%
%
%

%
%

%
%
%

\printbibliography

@article{Degrande:2021zpv,
    author = "Degrande, C\'eline and Touch\`eque, Julien",
    title = "{A reduced basis for CP violation in SMEFT at colliders and its application to diboson production}",
    eprint = "2110.02993",
    archivePrefix = "arXiv",
    primaryClass = "hep-ph",
    doi = "10.1007/JHEP04(2022)032",
    journal = "JHEP",
    volume = "04",
    pages = "032",
    year = "2022"
}

@article{Grazzini:2019jkl,
    author = "Grazzini, Massimiliano and Kallweit, S. and Lindert, Jonas M. and Pozzorini, Stefano and Wiesemann, Marius",
    title = "{NNLO QCD + NLO EW with Matrix+OpenLoops: precise predictions for vector-boson pair production}",
    eprint = "1912.00068",
    archivePrefix = "arXiv",
    primaryClass = "hep-ph",
    reportNumber = "MPP-2019-175, ZU-TH 49/19",
    doi = "10.1007/JHEP02(2020)087",
    journal = "JHEP",
    volume = "02",
    pages = "087",
    year = "2020"
}

@article{Lombardi:2021rvg,
    author = "Lombardi, Daniele and Wiesemann, Marius and Zanderighi, Giulia",
    title = "{W$^+$W$^-$ production at NNLO+PS with MINNLO$_{PS}$}",
    eprint = "2103.12077",
    archivePrefix = "arXiv",
    primaryClass = "hep-ph",
    reportNumber = "MPP-2021-27",
    doi = "10.1007/JHEP11(2021)230",
    journal = "JHEP",
    volume = "11",
    pages = "230",
    year = "2021"
}

@article{Degrande:2020evl,
    author = "Degrande, C\'eline and Durieux, Gauthier and Maltoni, Fabio and Mimasu, Ken and Vryonidou, Eleni and Zhang, Cen",
    title = "{Automated one-loop computations in the standard model effective field theory}",
    eprint = "2008.11743",
    archivePrefix = "arXiv",
    primaryClass = "hep-ph",
    reportNumber = "CERN-TH-2020-140, CP3-20-42",
    doi = "10.1103/PhysRevD.103.096024",
    journal = "Phys. Rev. D",
    volume = "103",
    number = "9",
    pages = "096024",
    year = "2021"
}

@article{Re:2018vac,
    author = "Re, Emanuele and Wiesemann, Marius and Zanderighi, Giulia",
    title = "{NNLOPS accurate predictions for $W^+W^-$ production}",
    eprint = "1805.09857",
    archivePrefix = "arXiv",
    primaryClass = "hep-ph",
    reportNumber = "CERN-TH/2018-114, LAPTH-018/18, CERN-TH-2018-114, LAPTH-018-18",
    doi = "10.1007/JHEP12(2018)121",
    journal = "JHEP",
    volume = "12",
    pages = "121",
    year = "2018"
}

@article{Yang:2022bxv,
    author = "Yang, Siqi and Xie, Mingzhe and Fu, Yao and Zhao, Zihan and Liu, Minghui and Han, Liang and Hou, Tie-Jiun and Yuan, C. -P.",
    title = "{Boost asymmetry of the diboson productions in pp collisions}",
    eprint = "2207.02072",
    archivePrefix = "arXiv",
    primaryClass = "hep-ph",
    doi = "10.1103/PhysRevD.106.L051301",
    journal = "Phys. Rev. D",
    volume = "106",
    number = "5",
    pages = "L051301",
    year = "2022"
}

@article{Monni:2019whf,
    author = "Monni, Pier Francesco and Nason, Paolo and Re, Emanuele and Wiesemann, Marius and Zanderighi, Giulia",
    title = "{MiNNLO$_{PS}$: a new method to match NNLO QCD to parton showers}",
    eprint = "1908.06987",
    archivePrefix = "arXiv",
    primaryClass = "hep-ph",
    reportNumber = "CERN-TH-2019-117, LAPTH-042/19, MPP-2019-177",
    doi = "10.1007/JHEP05(2020)143",
    journal = "JHEP",
    volume = "05",
    pages = "143",
    year = "2020",
    related        = "Monni:2019whf-err",
    relatedstring  = "Erratum:",
}

@article{Grazzini:2020stb,
      author         = "Grazzini, Massimiliano and Kallweit, Stefan and
                        Wiesemann, Marius and Yook, Jeong Yeon",
      title          = "{$W^+W^-$ production at the LHC: NLO QCD corrections to
                        the loop-induced gluon fusion channel}",
      journal        = "Phys. Lett. B",
      volume         = "804",
      year           = "2020",
      pages          = "135399",
      doi            = "10.1016/j.physletb.2020.135399",
      eprint         = "2002.01877",
      archivePrefix  = "arXiv",
      primaryClass   = "hep-ph",
      reportNumber   = "MPP-2019-176, ZU-TH 04/20",
      SLACcitation   = "%%CITATION = ARXIV:2002.01877;%%"
}

@article{Grazzini:2016ctr,
      author         = "Grazzini, Massimiliano and Kallweit, Stefan and
                        Pozzorini, Stefano and Rathlev, Dirk and Wiesemann, Marius",
      title          = "{$W^{+}W^{-}$ production at the LHC: fiducial cross
                        sections and distributions in NNLO QCD}",
      journal        = "JHEP",
      volume         = "08",
      year           = "2016",
      pages          = "140",
      doi            = "10.1007/JHEP08(2016)140",
      eprint         = "1605.02716",
      archivePrefix  = "arXiv",
      primaryClass   = "hep-ph",
      reportNumber   = "ZU-TH-16-16, MITP-16-038, NSF-KITP-16-047, DESY-16-075",
      SLACcitation   = "%%CITATION = ARXIV:1605.02716;%%"
}

@article{Gehrmann:2014fva,
      author         = "{Gehrmann, T. and Grazzini, M. and Kallweit, S. and
                        Maierh\"ofer, P. and von Manteuffel, A. and Pozzorini, S.
                        and Rathlev, D. and Tancredi, L.}",
      title          = "{$W^+W^-$ Production at Hadron Colliders in Next to Next
                        to Leading Order QCD}",
      journal        = "Phys. Rev. Lett.",
      volume         = "113",
      year           = "2014",
      number         = "21",
      pages          = "212001",
      doi            = "10.1103/PhysRevLett.113.212001",
      eprint         = "1408.5243",
      archivePrefix  = "arXiv",
      primaryClass   = "hep-ph",
      reportNumber   = "ZU-TH-29-14, MITP-14-053",
      SLACcitation   = "%%CITATION = ARXIV:1408.5243;%%"
}

@article{Grazzini:2017mhc,
    author = "Grazzini, Massimiliano and Kallweit, Stefan and Wiesemann, Marius",
    title = "{Fully differential NNLO computations with MATRIX}",
    eprint = "1711.06631",
    archivePrefix = "arXiv",
    primaryClass = "hep-ph",
    reportNumber = "ZU-TH-30-17, CERN-TH-2017-232, ZU-TH 30/17",
    doi = "10.1140/epjc/s10052-018-5771-7",
    journal = "Eur. Phys. J. C",
    volume = "78",
    number = "7",
    pages = "537",
    year = "2018"
}

@article{vonManteuffel:2015msa,
      author         = "von Manteuffel, Andreas and Tancredi, Lorenzo",
      title          = "{The two-loop helicity amplitudes for $gg \to V_1 V_2 \to 4~\mathrm{leptons}$}",
      eprint         = "1503.08835",
      archivePrefix  = "arXiv",
      primaryClass   = "hep-ph",
      reportNumber   = "MITP-15-012, TTP15-014",
      doi            = "10.1007/JHEP06(2015)197",
      journal        = "JHEP",
      volume         = "06",
      pages          = "197",
      year           = "2015"
}

@article{Gehrmann:2015ora,
      author         = "Gehrmann, Thomas and von Manteuffel, Andreas and
                        Tancredi, Lorenzo",
      title          = "{The two-loop helicity amplitudes for $q\overline{q}^{\prime} \to {V}_1{V}_2 \to 4~\mathrm{leptons}$}",
      journal        = "JHEP",
      volume         = "09",
      year           = "2015",
      pages          = "128",
      doi            = "10.1007/JHEP09(2015)128",
      eprint         = "1503.04812",
      archivePrefix  = "arXiv",
      primaryClass   = "hep-ph",
      reportNumber   = "ZU-TH-03-15, MITP-15-011, TTP15-011",
      SLACcitation   = "%%CITATION = ARXIV:1503.04812;%%"
}

@article{Bertone:2017bme,
    author = "Bertone, Valerio and Carrazza, Stefano and Hartland, Nathan P. and Rojo, Juan",
    collaboration = "NNPDF",
    title = "{Illuminating the photon content of the proton within a global PDF analysis}",
    eprint = "1712.07053",
    archivePrefix = "arXiv",
    primaryClass = "hep-ph",
    reportNumber = "Nikhef/2017-064, CERN-TH-2017-235, NIKHEF-2017-064",
    doi = "10.21468/SciPostPhys.5.1.008",
    journal = "SciPost Phys.",
    volume = "5",
    number = "1",
    pages = "008",
    year = "2018"
}

@article{NNPDF:2017mvq,
    author = "Ball, Richard D. and others",
    collaboration = "NNPDF",
    title = "{Parton distributions from high-precision collider data}",
    eprint = "1706.00428",
    archivePrefix = "arXiv",
    primaryClass = "hep-ph",
    reportNumber = "EDINBURGH-2017-08, NIKHEF-2017-006, OUTP-17-04P, TIF-UNIMI-2017-3, CAVENDISH-HEP-17-06, CERN-TH-2017-077, Edinburgh 2017/08,
  Nikhef/2017-006, OUTP-17-04P,TIF-UNIMI-2017-3",
    doi = "10.1140/epjc/s10052-017-5199-5",
    journal = "Eur. Phys. J. C",
    volume = "77",
    number = "10",
    pages = "663",
    year = "2017"
}

@article{Brivio:2017btx,
    author = "Brivio, Ilaria and Jiang, Yun and Trott, Michael",
    title = "{The SMEFTsim package, theory and tools}",
    eprint = "1709.06492",
    archivePrefix = "arXiv",
    primaryClass = "hep-ph",
    doi = "10.1007/JHEP12(2017)070",
    journal = "JHEP",
    volume = "12",
    pages = "070",
    year = "2017"
}

@article{Brivio:2020onw,
    author = "Brivio, Ilaria",
    title = "{SMEFTsim 3.0 \textemdash{} a practical guide}",
    eprint = "2012.11343",
    archivePrefix = "arXiv",
    primaryClass = "hep-ph",
    doi = "10.1007/JHEP04(2021)073",
    journal = "JHEP",
    volume = "04",
    pages = "073",
    year = "2021"
}

@article{Han:2009ra,
    author = "Han, Tao and Li, Yingchuan",
    title = "{Genuine CP-odd Observables at the LHC}",
    eprint = "0911.2933",
    archivePrefix = "arXiv",
    primaryClass = "hep-ph",
    reportNumber = "MADPH-09-1549, NPAC-09-15",
    doi = "10.1016/j.physletb.2009.12.047",
    journal = "Phys. Lett. B",
    volume = "683",
    pages = "278--281",
    year = "2010"
}

@article{Azatov:2017kzw,
    author = "Azatov, A. and Elias-Mir{\'o}, J. and Reyimuaji, Y. and Venturini, E.",
    title = "{Novel measurements of anomalous triple gauge couplings for the LHC}",
    eprint = "1707.08060",
    archivePrefix = "arXiv",
    primaryClass = "hep-ph",
    reportNumber = "SISSA-35-2017-FISI",
    doi = "10.1007/JHEP10(2017)027",
    journal = "JHEP",
    volume = "10",
    pages = "027",
    year = "2017"
}

@article{Brivio:2022pyi,
    author = "Brivio, Ilaria and others",
    title = "{Truncation, validity, uncertainties}",
    eprint = "2201.04974",
    archivePrefix = "arXiv",
    primaryClass = "hep-ph",
    reportNumber = "CERN-LHCEFTWG-2021-002, CERN-LPCC-2022-01",
    month = "1",
    year = "2022"
}

@article{Degrande:2012wf,
    author = "Degrande, Celine and Greiner, Nicolas and Kilian, Wolfgang and Mattelaer, Olivier and Mebane, Harrison and Stelzer, Tim and Willenbrock, Scott and Zhang, Cen",
    title = "{Effective Field Theory: A Modern Approach to Anomalous Couplings}",
    eprint = "1205.4231",
    archivePrefix = "arXiv",
    primaryClass = "hep-ph",
    reportNumber = "MPP-2011-149, SI-HEP-2011-17, CP3-12-25",
    doi = "10.1016/j.aop.2013.04.016",
    journal = "Annals Phys.",
    volume = "335",
    pages = "21--32",
    year = "2013"
}

@article{Ellis:2020unq,
    author = "Ellis, John and Madigan, Maeve and Mimasu, Ken and Sanz, Veronica and You, Tevong",
    title = "{Top, Higgs, Diboson and Electroweak Fit to the Standard Model Effective Field Theory}",
    eprint = "2012.02779",
    archivePrefix = "arXiv",
    primaryClass = "hep-ph",
    reportNumber = "KCL-PH-TH/2020-73, CERN-TH-2020-202",
    doi = "10.1007/JHEP04(2021)279",
    journal = "JHEP",
    volume = "04",
    pages = "279",
    year = "2021"
}

@article{Mattelaer:2016gcx,
    author = "Mattelaer, Olivier",
    title = "{On the maximal use of Monte Carlo samples: re-weighting events at NLO accuracy}",
    eprint = "1607.00763",
    archivePrefix = "arXiv",
    primaryClass = "hep-ph",
    reportNumber = "IPPP-16-55, MCNET-16-25",
    doi = "10.1140/epjc/s10052-016-4533-7",
    journal = "Eur. Phys. J. C",
    volume = "76",
    number = "12",
    pages = "674",
    year = "2016"
}

@article{Azatov:2016sqh,
    author = "Azatov, Aleksandr and Contino, Roberto and Machado, Camila S. and Riva, Francesco",
    title = "{Helicity selection rules and noninterference for BSM amplitudes}",
    eprint = "1607.05236",
    archivePrefix = "arXiv",
    primaryClass = "hep-ph",
    reportNumber = "CERN-TH-2016-165",
    doi = "10.1103/PhysRevD.95.065014",
    journal = "Phys. Rev. D",
    volume = "95",
    number = "6",
    pages = "065014",
    year = "2017"
}

@article{Hagiwara:1993ck,
    author = "Hagiwara, Kaoru and Ishihara, S. and Szalapski, R. and Zeppenfeld, D.",
    title = "{Low energy effects of new interactions in the electroweak boson sector}",
    reportNumber = "MAD-PH-737, UT-635, KEK-TH-356, KEK-PREPRINT-92-214",
    doi = "10.1103/PhysRevD.48.2182",
    journal = "Phys. Rev. D",
    volume = "48",
    pages = "2182--2203",
    year = "1993"
}

@article{Buchmuller:1985jz,
    author = "{Buchm\"uller, W. and Wyler, D.}",
    title = "{ Effective lagrangian analysis of new interactions and flavour conservation}",
    reportNumber = "CERN-TH-4254/85",
    doi = "10.1016/0550-3213(86)90262-2",
    journal = "Nucl. Phys. B",
    volume = "268",
    pages = "621--653",
    year = "1986"
}

@article{ALEPH:2005ab,
    author = "Schael, S. and others",
    collaboration = "ALEPH, DELPHI, L3, OPAL, SLD, LEP Electroweak Working Group, SLD Electroweak Group, SLD Heavy Flavour Group",
    title = "{Precision electroweak measurements on the $Z$ resonance}",
    eprint = "hep-ex/0509008",
    archivePrefix = "arXiv",
    reportNumber = "SLAC-R-774",
    doi = "10.1016/j.physrep.2005.12.006",
    journal = "Phys. Rept.",
    volume = "427",
    pages = "257--454",
    year = "2006"
}

@article{Feldman:1997qc,
    author = "Feldman, Gary J. and Cousins, Robert D.",
    title = "{Unified approach to the classical statistical analysis of small signals}",
    eprint = "physics/9711021",
    archivePrefix = "arXiv",
    reportNumber = "HUTP-97-A096",
    doi = "10.1103/PhysRevD.57.3873",
    journal = "Phys. Rev. D",
    volume = "57",
    pages = "3873--3889",
    year = "1998"
}

@article{Falkowski:2016cxu,
    author = "Falkowski, Adam and Gonz{\'a}lez-Alonso, Martin and Greljo, Admir and Marzocca, David and Son, Minho",
    title = "{Anomalous Triple Gauge Couplings in the Effective Field Theory Approach at the LHC}",
    eprint = "1609.06312",
    archivePrefix = "arXiv",
    primaryClass = "hep-ph",
    reportNumber = "ZU-TH-34-16",
    doi = "10.1007/JHEP02(2017)115",
    journal = "JHEP",
    volume = "02",
    pages = "115",
    year = "2017"
}

@article{Meade:2014fca,
    author = "Meade, Patrick and Ramani, Harikrishnan and Zeng, Mao",
    title = "{Transverse momentum resummation effects in $W^+W^-$ measurements}",
    eprint = "1407.4481",
    archivePrefix = "arXiv",
    primaryClass = "hep-ph",
    doi = "10.1103/PhysRevD.90.114006",
    journal = "Phys. Rev. D",
    volume = "90",
    number = "11",
    pages = "114006",
    year = "2014"
}

@article{Hou:2019efy,
    author = "Hou, Tie-Jiun and others",
    title = "{New CTEQ global analysis of quantum chromodynamics with high-precision data from the LHC}",
    eprint = "1912.10053",
    archivePrefix = "arXiv",
    primaryClass = "hep-ph",
    reportNumber = "MSUHEP-19-025, PITT-PACC-1911, SMU-HEP-19-03",
    doi = "10.1103/PhysRevD.103.014013",
    journal = "Phys. Rev. D",
    volume = "103",
    number = "1",
    pages = "014013",
    year = "2021"
}

@article{Bailey:2020ooq,
    author = "Bailey, S. and Cridge, T. and Harland-Lang, L. A. and Martin, A. D. and Thorne, R. S.",
    title = "{Parton distributions from LHC, HERA, Tevatron and fixed target data: MSHT20 PDFs}",
    eprint = "2012.04684",
    archivePrefix = "arXiv",
    primaryClass = "hep-ph",
    reportNumber = "IPPP/20/58",
    doi = "10.1140/epjc/s10052-021-09057-0",
    journal = "Eur. Phys. J. C",
    volume = "81",
    number = "4",
    pages = "341",
    year = "2021"
}

@article{ParticleDataGroup:2024cfk,
    author = "{Particle Data Group}",
    collaboration = "Particle Data Group",
    title = "{Review of particle physics}",
    doi = "10.1103/PhysRevD.110.030001",
    journal = "Phys. Rev. D",
    volume = "110",
    number = "3",
    pages = "030001",
    year = "2024"
}

@article{Cascioli:2013gfa,
    author = {Cascioli, F. and H\"oche, S. and Krauss, F. and Maierh\"ofer, P. and Pozzorini, S. and Siegert, F.},
    title = "{Precise Higgs-background predictions: merging NLO QCD and squared quark-loop corrections to four-lepton + 0,1 jet production}",
    eprint = "1309.0500",
    archivePrefix = "arXiv",
    primaryClass = "hep-ph",
    reportNumber = "IPPP-13-66, DCPT-13-132, MCNET-13-12, SLAC-PUB-15714, ZU-TH-15-13, LPN13-056, FR-PHENO-2013-007",
    doi = "10.1007/JHEP01(2014)046",
    journal = "JHEP",
    volume = "01",
    pages = "046",
    year = "2014"
}

@article{NNPDF:2021njg,
    author = "{NNPDF Collaboration}",
    collaboration = "NNPDF",
    title = "{The path to proton structure at 1\% accuracy}",
    eprint = "2109.02653",
    archivePrefix = "arXiv",
    primaryClass = "hep-ph",
    reportNumber = "Edinburgh 2021/12, Nikhef-2021-013, TIF-UNIMI-2021-11",
    doi = "10.1140/epjc/s10052-022-10328-7",
    journal = "Eur. Phys. J. C",
    volume = "82",
    number = "5",
    pages = "428",
    year = "2022"
}

@article{Grazzini:2016swo,
	author = "Grazzini, Massimiliano and Kallweit, Stefan and Rathlev, Dirk and Wiesemann, Marius",
	title = "{$W^{\pm}Z$ production at hadron colliders in NNLO QCD}",
	eprint = "1604.08576",
	archivePrefix = "arXiv",
	primaryClass = "hep-ph",
	reportNumber = "ZU-TH-15-16, MITP-16-037, NSF-KITP-16-046, DESY-16-074",
	doi = "10.1016/j.physletb.2016.08.017",
	journal = "Phys. Lett. B",
	volume = "761",
	pages = "179--183",
	year = "2016"
}

@article{Grazzini:2017ckn,
	author = "Grazzini, Massimiliano and Kallweit, Stefan and Rathlev, Dirk and Wiesemann, Marius",
	title = "{$W^\pm Z$ production at the LHC: fiducial cross sections and distributions in NNLO QCD}",
	eprint = "1703.09065",
	archivePrefix = "arXiv",
	primaryClass = "hep-ph",
	reportNumber = "ZU-TH-06-17, CERN-TH-2017-065",
	doi = "10.1007/JHEP05(2017)139",
	journal = "JHEP",
	volume = "05",
	pages = "139",
	year = "2017"
}

@article{Grazzini:2015hta,
	author = "Grazzini, Massimiliano and Kallweit, Stefan and Rathlev, Dirk",
	title = "{ZZ production at the LHC: Fiducial cross sections and distributions in NNLO QCD}",
	eprint = "1507.06257",
	archivePrefix = "arXiv",
	primaryClass = "hep-ph",
	doi = "10.1016/j.physletb.2015.09.055",
	journal = "Phys. Lett. B",
	volume = "750",
	pages = "407--410",
	year = "2015"
}

@article{Cascioli:2014yka,
	author = "Cascioli, F. and Gehrmann, T. and Grazzini, M. and Kallweit, S. and Maierhöfer, P. and von Manteuffel, A. and Pozzorini, S. and Rathlev, D. and Tancredi, L. and Weihs, E.",
	title = "{ZZ production at hadron colliders in NNLO QCD}",
	eprint = "1405.2219",
	archivePrefix = "arXiv",
	primaryClass = "hep-ph",
	reportNumber = "ZU-TH-19-14, MITP-14-033",
	doi = "10.1016/j.physletb.2014.06.056",
	journal = "Phys. Lett. B",
	volume = "735",
	pages = "311--313",
	year = "2014"
}

@article{BayesUnfold2,
	Archiveprefix = {arXiv},
	Author = {D'Agostini, G.},
	Eprint = {1010.0632},
	Primaryclass = {physics.data-an},
	Title = {{Improved iterative Bayesian unfolding}},
	Year = {2010}}

@article{Grazzini:2015nwa,
	author = "Grazzini, Massimiliano and Kallweit, Stefan and Rathlev, Dirk",
	title = "{$W\gamma$ and $Z\gamma$ production at the LHC in NNLO QCD}",
	eprint = "1504.01330",
	archivePrefix = "arXiv",
	primaryClass = "hep-ph",
	reportNumber = "ZU-TH-04-15, MITP-15-021",
	doi = "10.1007/JHEP07(2015)085",
	journal = "JHEP",
	volume = "07",
	pages = "085",
	year = "2015"
}

@article{Barr:2005dz,
	author = "Barr, A.J.",
	title = "{Measuring slepton spin at the LHC}",
	eprint = "hep-ph/0511115",
	archivePrefix = "arXiv",
	reportNumber = "ATL-PHYS-PUB-2005-023",
	doi = "10.1088/1126-6708/2006/02/042",
	journal = "JHEP",
	volume = "02",
	pages = "042",
	year = "2006"
}

@article{Grzadkowski:2010es,
    author = "Grzadkowski, B. and Iskrzy{\'{n}}ski, M. and Misiak, M. and Rosiek, J.",
    title = "{Dimension-Six Terms in the Standard Model Lagrangian}",
    eprint = "1008.4884",
    archivePrefix = "arXiv",
    primaryClass = "hep-ph",
    reportNumber = "IFT-9-2010, TTP10-35",
    doi = "10.1007/JHEP10(2010)085",
    journal = "JHEP",
    volume = "10",
    pages = "085",
    year = "2010"
}

@article{schael:2013ita,
	Archiveprefix = {arXiv},
	Author = {{ALEPH, DELPHI, L3 and OPAL Collaborations and the LEP Electroweak Working Group}},
	Collaboration = {DELPHI, OPAL, LEP Electroweak, ALEPH, L3},
	Doi = {10.1016/j.physrep.2013.07.004},
	Eprint = {1302.3415},
	Journal = {Phys. Rept.},
	Pages = {119-244},
	Primaryclass = {hep-ex},
	Reportnumber = {CERN-PH-EP-2013-022},
	Slaccitation = {%%CITATION = ARXIV:1302.3415;%%},
	Title = {{Electroweak measurements in electron--positron collisions at W-boson-pair energies at LEP}},
	Volume = {532},
	Year = {2013}
}

@article{Abe:1996dw,
	author = {{CDF Collaboration}},
	collaboration = {CDF},
    title = "{Evidence of $W^+ W^-$ Production in $\bar{p}p$ Collisions at $\sqrt{s} = 1.8$ TeV}",
    reportNumber = "FERMILAB-PUB-96-311-E, CDF-PUB-ELECTROWEAK-PUBLIC-3725, FERMILAB-PUB-98-173-EH",
    doi = "10.1103/PhysRevLett.78.4536",
    journal = "Phys. Rev. Lett.",
    volume = "78",
    pages = "4536--4540",
    year = "1997"
}

@article{Aaltonen:2009aa,
	Archiveprefix = {arXiv},
	Author = {{CDF Collaboration}},
	Collaboration = {CDF},
	Doi = {10.1103/PhysRevLett.104.201801},
	Eprint = {0912.4500},
	Journal = {Phys. Rev. Lett.},
	Pages = {201801},
	Primaryclass = {hep-ex},
	Reportnumber = {FERMILAB-PUB-09-636},
	Slaccitation = {%%CITATION = ARXIV:0912.4500;%%},
	Title = {{Measurement of the $W^+ W^-$ Production Cross Section and Search for Anomalous $WW \gamma$ and $WWZ$ Couplings in $p \bar{p}$ Collisions at $\sqrt{s}=1.96$ \TeV}},
	Volume = {104},
	Year = {2010},
        related        = "Aaltonen:2009aa-err",
        relatedstring  = "Erratum:",
}

@article{Abazov:2009ys,
	Archiveprefix = {arXiv},
	Author = {{D\O{} Collaboration}},
	Doi = {10.1103/PhysRevLett.103.191801},
	Eprint = {0904.0673},
	Journal = {Phys. Rev. Lett.},
	Pages = {191801},
	Primaryclass = {hep-ex},
	Reportnumber = {FERMILAB-PUB-09-094-E},
	Slaccitation = {%%CITATION = ARXIV:0904.0673;%%},
	Title = {{Measurement of the WW Production Cross Section with Dilepton Final States in $p\bar{p}$ Collisions at $\sqrt{s}$ = 1.96 \TeV\ and Limits on Anomalous Trilinear Gauge Couplings}},
	Volume = {103},
	Year = {2009}
}

@article{Wilks:1938dza,
	author = "Wilks, S.S.",
	title = "{The Large-Sample Distribution of the Likelihood Ratio for Testing Composite Hypotheses}",
	doi = "10.1214/aoms/1177732360",
	journal = "Annals Math. Statist.",
	volume = "9",
	number = "1",
	pages = "60--62",
	year = "1938"
}

@article{Cacciari:2007fd,
    author = "Cacciari, Matteo and Salam, Gavin P.",
    title = "{Pileup subtraction using jet areas}",
    eprint = "0707.1378",
    archivePrefix = "arXiv",
    primaryClass = "hep-ph",
    reportNumber = "LPTHE-07-01",
    doi = "10.1016/j.physletb.2007.09.077",
    journal = "Phys. Lett. B",
    volume = "659",
    pages = "119--126",
    year = "2008"
}

@article{Catani:2012qa,
    author = "Catani, Stefano and Cieri, Leandro and de Florian, Daniel and Ferrera, Giancarlo and Grazzini, Massimiliano",
    title = "{Vector boson production at hadron colliders: hard-collinear coefficients at the NNLO}",
    eprint = "1209.0158",
    archivePrefix = "arXiv",
    primaryClass = "hep-ph",
    reportNumber = "ZU-TH-16-12",
    doi = "10.1140/epjc/s10052-012-2195-7",
    journal = "Eur. Phys. J. C",
    volume = "72",
    pages = "2195",
    year = "2012"
}

@article{Gavardi:2023aco,
    author = "Gavardi, Alessandro and Lim, Matthew A. and Alioli, Simone and Tackmann, Frank J.",
    title = "{NNLO+PS W$^{+}$W$^{-}$ production using jet veto resummation at NNLL'}",
    eprint = "2308.11577",
    archivePrefix = "arXiv",
    primaryClass = "hep-ph",
    doi = "10.1007/JHEP12(2023)069",
    journal = "JHEP",
    volume = "12",
    pages = "069",
    year = "2023"
}

@Article{Evans:2008zzb,
      author         = "Evans, Lyndon and Bryant, Philip",
      title          = "{LHC Machine}",
      journal        = "JINST",
      volume         = "3",
      pages          = "S08001",
      doi            = "10.1088/1748-0221/3/08/S08001",
      year           = "2008",
      SLACcitation   = "%%CITATION = JINST,3,S08001;%%",
}

@Article{Cacciari:2008gp,
     author    = "Cacciari, Matteo and Salam, Gavin P. and Soyez, Gregory",
     title     = "{The anti-\(k_{t}\) jet clustering algorithm}",
     journal   = "JHEP",
     volume    = "04",
     year      = "2008",
     pages     = "063",
     eprint    = "0802.1189",
     archivePrefix = "arXiv",
     primaryClass  =  "hep-ph",
     doi       = "10.1088/1126-6708/2008/04/063",
     SLACcitation  = "%%CITATION = 0802.1189;%%"
}

@Article{Fastjet,
      author         = "Cacciari, Matteo and Salam, Gavin P. and Soyez, Gregory",
      title          = "{FastJet user manual}",
      journal        = "Eur. Phys. J. C",
      volume         = "72",
      year           = "2012",
      pages          = "1896",
      doi            = "10.1140/epjc/s10052-012-1896-2",
      eprint         = "1111.6097",
      archivePrefix  = "arXiv",
      primaryClass   = "hep-ph",
      reportNumber   = "CERN-PH-TH-2011-297",
      SLACcitation   = "%%CITATION = ARXIV:1111.6097;%%"
}

@article{Harland-Lang:2014zoa,
      author         = "Harland-Lang, L. A. and Martin, A. D. and Motylinski, P.
                        and Thorne, R. S.",
      title          = "{Parton distributions in the LHC era: MMHT 2014 PDFs}",
      journal        = "Eur. Phys. J. C",
      number         = "5",
      volume         = "75",
      pages          = "204",
      doi            = "10.1140/epjc/s10052-015-3397-6",
      year           = "2015",
      eprint         = "1412.3989",
      archivePrefix  = "arXiv",
      primaryClass   = "hep-ph",
      reportNumber   = "LCTS-2014-47, IPPP-14-97, DCPT-14-194",
      SLACcitation   = "%%CITATION = ARXIV:1412.3989;%%",
}

@article{Ball:2012cx,
      author         = "{NNPDF Collaboration} and Ball, Richard D. and others",
      title          = "{Parton distributions with LHC data}",
      journal        = "Nucl. Phys. B",
      volume         = "867",
      year           = "2013",
      pages          = "244",
      doi            = "10.1016/j.nuclphysb.2012.10.003",
      eprint         = "1207.1303",
      archivePrefix  = "arXiv",
      primaryClass   = "hep-ph",
      reportNumber   = "EDINBURGH-2012-08, IFUM-FT-997, FR-PHENO-2012-014,
                        RWTH-TTK-12-25, CERN-PH-TH-2012-037, SFB-CPP-12-47\,
                        --CERN-PH-TH-2012-037",
      SLACcitation   = "%%CITATION = ARXIV:1207.1303;%%"
}

@article{Ball:2014uwa,
      author         = "{NNPDF Collaboration} and Ball, Richard D. and others",
      title          = "{Parton distributions for the LHC run II}",
      journal        = "JHEP",
      volume         = "04",
      year           = "2015",
      pages          = "040",
      doi            = "10.1007/JHEP04(2015)040",
      eprint         = "1410.8849",
      archivePrefix  = "arXiv",
      primaryClass   = "hep-ph",
      reportNumber   = "EDINBURGH-2014-15, IFUM-1034-FT, CERN-PH-TH-2013-253,
                        OUTP-14-11P, CAVENDISH-HEP-14-11",
      SLACcitation   = "%%CITATION = ARXIV:1410.8849;%%"
}

@article{Czakon:2012pz,
      author         = "Czakon, Michal and Mitov, Alexander",
      title          = "{NNLO corrections to top pair production at hadron
                        colliders: the quark-gluon reaction}",
      journal        = "JHEP",
      volume         = "01",
      pages          = "080",
      doi            = "10.1007/JHEP01(2013)080",
      year           = "2013",
      eprint         = "1210.6832",
      archivePrefix  = "arXiv",
      primaryClass   = "hep-ph",
      SLACcitation   = "%%CITATION = ARXIV:1210.6832;%%",
}

@article{Skands:2014pea,
      author         = "Skands, Peter and Carrazza, Stefano and Rojo, Juan",
      title          = "{Tuning PYTHIA 8.1: the Monash 2013 Tune}",
      journal        = "Eur. Phys. J. C",
      volume         = "74",
      year           = "2014",
      number         = "8",
      pages          = "3024",
      doi            = "10.1140/epjc/s10052-014-3024-y",
      eprint         = "1404.5630",
      archivePrefix  = "arXiv",
      primaryClass   = "hep-ph",
      reportNumber   = "CERN-PH-TH-2014-069, MCNET-14-08, OUTP-14-05P",
      SLACcitation   = "%%CITATION = ARXIV:1404.5630;%%"
}

@Article{Sjostrand:2007gs,
	author    = "Sj{\"o}strand, T. and Mrenna, S. and Skands, P.",
	title     = "{A brief introduction to PYTHIA 8.1}",
	journal   = "Comput. Phys. Commun.",
	volume    = "178",
	year      = "2008",
	pages     = "852-867",
	eprint    = "0710.3820",
	archivePrefix = "arXiv",
	primaryClass  =  "hep-ph",
	doi       = "10.1016/j.cpc.2008.01.036",
	SLACcitation  = "%%CITATION = 0710.3820;%%"
}

@article{Sjostrand:2014zea,
      author         = "Sj{\"o}strand, Torbj{\"o}rn and Ask, Stefan and Christiansen,
                        Jesper R. and Corke, Richard and Desai, Nishita and Ilten,
                        Philip and Mrenna, Stephen and Prestel, Stefan and
                        Rasmussen, Christine O. and Skands, Peter Z.",
      title          = "{An introduction to PYTHIA 8.2}",
      journal        = "Comput. Phys. Commun.",
      volume         = "191",
      year           = "2015",
      pages          = "159",
      doi            = "10.1016/j.cpc.2015.01.024",
      eprint         = "1410.3012",
      archivePrefix  = "arXiv",
      primaryClass   = "hep-ph",
      reportNumber   = "LU-TP-14-36, MCNET-14-22, CERN-PH-TH-2014-190,
                        FERMILAB-PUB-14-316-CD, DESY-14-178, SLAC-PUB-16122,
                        --FERMILAB-PUB-14-316-CD",
      SLACcitation   = "%%CITATION = ARXIV:1410.3012;%%"
}

@article{Lonnblad:2001iq,
      author         = "L{\"o}nnblad, Leif",
      title          = "{Correcting the Colour-Dipole Cascade Model with Fixed Order Matrix Elements}",
      journal        = "JHEP",
      volume         = "05",
      year           = "2002",
      pages          = "046",
      doi            = "10.1088/1126-6708/2002/05/046",
      eprint         = "hep-ph/0112284",
      archivePrefix  = "arXiv",
      XprimaryClass   = "hep-ph",
      reportNumber   = "LU-TP-01-38",
      SLACcitation   = "%%CITATION = HEP-PH/0112284;%%"
}

@article{Lonnblad:2011xx,
      author         = "L{\"o}nnblad, Leif and Prestel, Stefan",
      title          = "{Matching tree-level matrix elements with interleaved
                        showers}",
      journal        = "JHEP",
      volume         = "03",
      year           = "2012",
      pages          = "019",
      doi            = "10.1007/JHEP03(2012)019",
      eprint         = "1109.4829",
      archivePrefix  = "arXiv",
      primaryClass   = "hep-ph",
      SLACcitation   = "%%CITATION = ARXIV:1109.4829;%%"
}

@Article{Lange:2001uf,
      author         = "Lange, D. J.",
      title          = "{The EvtGen particle decay simulation package}",
      booktitle      = "{Proceedings, 7th International Conference on B physics
                        at hadron machines (BEAUTY 2000)}",
      journal        = "Nucl. Instrum. Meth. A",
      volume         = "462",
      year           = "2001",
      pages          = "152",
      doi            = "10.1016/S0168-9002(01)00089-4",
      SLACcitation   = "%%CITATION = NUIMA,A462,152;%%"
}

@Article{Frixione:2008yi,
     author    = "Frixione, Stefano and Laenen, Eric and Motylinski, Patrick and White, Chris and Webber, Bryan R.",
     title     = "{Single-top hadroproduction in association with a \(W\) boson}",
     journal   = "JHEP",
     volume    = "07",
     year      = "2008",
     pages     = "029",
     eprint    = "0805.3067",
     archivePrefix = "arXiv",
     primaryClass  =  "hep-ph",
     doi       = "10.1088/1126-6708/2008/07/029"
}

@Article{Alwall:2014hca,
      author         = "Alwall, J. and Frederix, R. and Frixione, S. and Hirschi,
                        V. and Maltoni, F. and Mattelaer, O. and Shao, H. -S. and
                        Stelzer, T. and Torrielli, P. and Zaro, M.",
      title          = "{The automated computation of tree-level and
                        next-to-leading order differential cross sections, and
                        their matching to parton shower simulations}",
      journal        = "JHEP",
      volume         = "07",
      year           = "2014",
      pages          = "079",
      doi            = "10.1007/JHEP07(2014)079",
      eprint         = "1405.0301",
      archivePrefix  = "arXiv",
      primaryClass   = "hep-ph",
      reportNumber   = "CERN-PH-TH-2014-064, CP3-14-18, LPN14-066, MCNET-14-09,
                        ZU-TH-14-14",
      SLACcitation   = "%%CITATION = ARXIV:1405.0301;%%"
}

@Article{Bahr:2008pv,
      author         = "B{\"a}hr, M. and others",
      title          = "{Herwig++ physics and manual}",
      journal        = "Eur. Phys. J. C",
      volume         = "58",
      year           = "2008",
      pages          = "639",
      doi            = "10.1140/epjc/s10052-008-0798-9",
      eprint         = "0803.0883",
      archivePrefix  = "arXiv",
      primaryClass   = "hep-ph",
      reportNumber   = "CERN-PH-TH-2008-038, CAVENDISH-HEP-08-03, KA-TP-05-2008,
                        DCPT-08-22, IPPP-08-11, CP3-08-05",
      SLACcitation   = "%%CITATION = ARXIV:0803.0883;%%"
}

@Article{Bellm:2015jjp,
      author         = "Bellm, Johannes and others",
      title          = "{Herwig 7.0/Herwig++ 3.0 release note}",
      journal        = "Eur. Phys. J. C",
      volume         = "76",
      year           = "2016",
      number         = "4",
      pages          = "196",
      doi            = "10.1140/epjc/s10052-016-4018-8",
      eprint         = "1512.01178",
      archivePrefix  = "arXiv",
      primaryClass   = "hep-ph",
      reportNumber   = "CERN-PH-TH-2015-289, MAN-HEP-2015-15, IFJPAN-IV-2015-13,
                        HERWIG-2015-01, KA-TP-18-2015, DCPT-15-142, MCNET-15-28,
                        IPPP-15-71, --HERWIG-2015-01",
      SLACcitation   = "%%CITATION = ARXIV:1512.01178;%%"
}

@Article{Nason:2004rx,
      author         = "Nason, Paolo",
      title          = "{A new method for combining NLO QCD with shower Monte Carlo algorithms}",
      journal        = "JHEP",
      volume         = "11",
      pages          = "040",
      doi            = "10.1088/1126-6708/2004/11/040",
      year           = "2004",
      eprint         = "hep-ph/0409146",
      archivePrefix  = "arXiv",
}

@Article{Frixione:2007vw,
      author         = "Frixione, Stefano and Nason, Paolo and Oleari, Carlo",
      title          = "{Matching NLO QCD computations with parton shower
                        simulations: the POWHEG method}",
      journal        = "JHEP",
      volume         = "11",
      pages          = "070",
      doi            = "10.1088/1126-6708/2007/11/070",
      year           = "2007",
      eprint         = "0709.2092",
      archivePrefix  = "arXiv",
      primaryClass   = "hep-ph",
}

@Article{Alioli:2010xd,
      author         = "Alioli, Simone and Nason, Paolo and Oleari, Carlo and Re,
                        Emanuele",
      title          = "{A general framework for implementing NLO calculations in
                        shower Monte Carlo programs: the POWHEG BOX}",
      journal        = "JHEP",
      volume         = "06",
      pages          = "043",
      doi            = "10.1007/JHEP06(2010)043",
      year           = "2010",
      eprint         = "1002.2581",
      archivePrefix  = "arXiv",
      primaryClass   = "hep-ph",
}

@article{Re:2010bp,
      author         = "Re, Emanuele",
      title          = "{Single-top \(Wt\)-channel production matched with parton
                        showers using the POWHEG method}",
      journal        = "Eur. Phys. J. C",
      volume         = "71",
      year           = "2011",
      pages          = "1547",
      doi            = "10.1140/epjc/s10052-011-1547-z",
      eprint         = "1009.2450",
      archivePrefix  = "arXiv",
      primaryClass   = "hep-ph",
      reportNumber   = "IPPP-10-74, DCPT-10-148",
      SLACcitation   = "%%CITATION = ARXIV:1009.2450;%%"
}

@article{Kidonakis:2010ux,
      author         = "Kidonakis, Nikolaos",
      title          = "{Two-loop soft anomalous dimensions for single top quark
                        associated production with a \(W^{-}\) or \(H^{-}\)}",
      journal        = "Phys. Rev. D",
      volume         = "82",
      year           = "2010",
      pages          = "054018",
      doi            = "10.1103/PhysRevD.82.054018",
      eprint         = "1005.4451",
      archivePrefix  = "arXiv",
      primaryClass   = "hep-ph",
      SLACcitation   = "%%CITATION = ARXIV:1005.4451;%%"
}

@inproceedings{Kidonakis:2013zqa,
      author         = "Kidonakis, Nikolaos",
      title          = "{Top Quark Production}",
      booktitle      = "{Proceedings, Helmholtz International Summer School on Physics of Heavy Quarks and Hadrons (HQ2013)}",
      eventdate      = "2013-07-15/2013-07-28",
      venue          = "JINR, Dubna, Russia",
      pages          = "139-168",
      doi            = "10.3204/DESY-PROC-2013-03/Kidonakis",
      eprint         = "1311.0283",
      archivePrefix  = "arXiv",
      primaryClass   = "hep-ph",
      reportNumber   = "KSU-HEP-110113",
      SLACcitation   = "%%CITATION = ARXIV:1311.0283;%%"
}

@article{Beneke:2011mq,
      author         = "Beneke, M. and Falgari, P. and Klein, S. and Schwinn, C.",
      title          = "{Hadronic top-quark pair production with NNLL threshold
                        resummation}",
      journal        = "Nucl. Phys. B",
      volume         = "855",
      year           = "2012",
      pages          = "695-741",
      doi            = "10.1016/j.nuclphysb.2011.10.021",
      eprint         = "1109.1536",
      archivePrefix  = "arXiv",
      primaryClass   = "hep-ph",
      reportNumber   = "TTK-11-38, ITP-UU-11-26, SPIN-11-19, FR-PHENO-2011-015,
                        SFB-CPP-11-49",
      SLACcitation   = "%%CITATION = ARXIV:1109.1536;%%"
}

@article{Anastasiou:2003ds,
      author         = "Anastasiou, Charalampos and Dixon, Lance and Melnikov,
                        Kirill and Petriello, Frank",
      title          = "{High-precision QCD at hadron colliders: Electroweak
                        gauge boson rapidity distributions at next-to-next-to leading order}",
      journal        = "Phys. Rev. D",
      volume         = "69",
      year           = "2004",
      pages          = "094008",
      doi            = "10.1103/PhysRevD.69.094008",
      eprint         = "hep-ph/0312266",
      archivePrefix  = "arXiv",
      reportNumber   = "SLAC-PUB-10288, UH-511-1042-03",
      SLACcitation   = "%%CITATION = HEP-PH/0312266;%%"
}

@article{Cacciari:2011hy,
      author         = "Cacciari, Matteo and Czakon, Michal and Mangano,
                        Michelangelo and Mitov, Alexander and Nason, Paolo",
      title          = "{Top-pair production at hadron colliders with
                        next-to-next-to-leading logarithmic soft-gluon
                        resummation}",
      journal        = "Phys. Lett. B",
      volume         = "710",
      pages          = "612-622",
      doi            = "10.1016/j.physletb.2012.03.013",
      year           = "2012",
      eprint         = "1111.5869",
      primaryClass   = "hep-ph",
      archivePrefix  = "arXiv",
      reportNumber   = "CERN-PH-TH-2011-277, TTK-11-54",
      SLACcitation   = "%%CITATION = ARXIV:1111.5869;%%",
}

@article{Czakon:2012zr,
      author         = "Czakon, Michal and Mitov, Alexander",
      title          = "{NNLO corrections to top-pair production at hadron
                        colliders: the all-fermionic scattering channels}",
      journal        = "JHEP",
      volume         = "12",
      pages          = "054",
      doi            = "10.1007/JHEP12(2012)054",
      year           = "2012",
      eprint         = "1207.0236",
      archivePrefix  = "arXiv",
      primaryClass   = "hep-ph",
      SLACcitation   = "%%CITATION = ARXIV:1207.0236;%%",
}

@article{Czakon:2013goa,
      author         = "Czakon, Michal and Fiedler, Paul and Mitov, Alexander",
      title          = "{Total Top-Quark Pair-Production Cross Section at
                        Hadron Colliders Through \(\mathcal{O}(\alpha_S^4)\)}",
      journal        = "Phys. Rev. Lett.",
      volume         = "110",
      pages          = "252004",
      doi            = "10.1103/PhysRevLett.110.252004",
      year           = "2013",
      eprint         = "1303.6254",
      archivePrefix  = "arXiv",
      primaryClass   = "hep-ph",
      reportNumber   = "CERN-PH-TH-2013-056, TTK-13-08",
      SLACcitation   = "%%CITATION = ARXIV:1303.6254;%%",
}

@article{Baernreuther:2012ws,
      author         = "B{\"a}rnreuther, Peter and Czakon, Michal and Mitov,
                        Alexander",
      title          = "{Percent-Level-Precision Physics at the
                  Tevatron: Next-to-Next-to-Leading Order QCD
                  Corrections to \(q \bar{q} \to t \bar{t} + X\)}",
      journal        = "Phys. Rev. Lett.",
      volume         = "109",
      pages          = "132001",
      doi            = "10.1103/PhysRevLett.109.132001",
      year           = "2012",
      eprint         = "1204.5201",
      archivePrefix  = "arXiv",
      primaryClass   = "hep-ph",
      SLACcitation   = "%%CITATION = ARXIV:1204.5201;%%",
}

@article{Frixione:2007nw,
      author         = "Frixione, Stefano and Ridolfi, Giovanni and Nason, Paolo",
      title          = "{A positive-weight next-to-leading-order Monte Carlo for
                        heavy flavour hadroproduction}",
      journal        = "JHEP",
      volume         = "09",
      pages          = "126",
      doi            = "10.1088/1126-6708/2007/09/126",
      year           = "2007",
      eprint         = "0707.3088",
      archivePrefix  = "arXiv",
      primaryClass   = "hep-ph",
      SLACcitation   = "%%CITATION = ARXIV:0707.3088;%%",
}

@Article{Hoeche:2009rj,
      author         = "H{\"o}che, Stefan and Krauss, Frank and Schumann, Steffen and Siegert, Frank",
      title          = "{QCD matrix elements and truncated showers}",
      journal        = "JHEP",
      volume         = "05",
      pages          = "053",
      doi            = "10.1088/1126-6708/2009/05/053",
      year           = "2009",
      eprint         = "0903.1219",
      archivePrefix  = "arXiv",
      primaryClass   = "hep-ph",
}

@Article{Gleisberg:2008fv,
      author         = "Gleisberg, Tanju and H{\"o}che, Stefan",
      title          = "{Comix, a new matrix element generator}",
      journal        = "JHEP",
      volume         = "12",
      pages          = "039",
      doi            = "10.1088/1126-6708/2008/12/039",
      year           = "2008",
      eprint         = "0808.3674",
      archivePrefix  = "arXiv",
      primaryClass   = "hep-ph",
}

@Article{Schumann:2007mg,
      author         = "Schumann, Steffen and Krauss, Frank",
      title          = "{A parton shower algorithm based on Catani--Seymour dipole factorisation}",
      journal        = "JHEP",
      volume         = "03",
      pages          = "038",
      doi            = "10.1088/1126-6708/2008/03/038",
      year           = "2008",
      eprint         = "0709.1027",
      archivePrefix  = "arXiv",
      primaryClass   = "hep-ph",
}

@article{Hoeche:2012yf,
      author         = "H{\"o}che, Stefan and Krauss, Frank and Sch{\"o}nherr, Marek and
                        Siegert, Frank",
      title          = "{QCD matrix elements + parton showers. The NLO case}",
      journal        = "JHEP",
      volume         = "04",
      year           = "2013",
      pages          = "027",
      doi            = "10.1007/JHEP04(2013)027",
      eprint         = "1207.5030",
      archivePrefix  = "arXiv",
      primaryClass   = "hep-ph",
      reportNumber   = "SLAC-PUB-15191, IPPP-12-52, DCPT-12-104, LPN12-081,
                        FR-PHENO-2012-017, MCNET-12-09, --FR-PHENO-2012-017",
      SLACcitation   = "%%CITATION = ARXIV:1207.5030;%%"
}

@article{Bothmann:2019yzt,
      author = "Bothmann, Enrico and others",
      title = "{Event generation with Sherpa 2.2}",
      journal = "SciPost Phys.",
      volume = "7",
      year = "2019",
      number = "3",
      pages = "034",
      doi = "10.21468/SciPostPhys.7.3.034",
      reportNumber = "FERMILAB-PUB-19-218-T, SLAC-PUB-17433, IPPP/19/42, MCNET-19-11",
      eprint = "1905.09127",
      archivePrefix = "arXiv",
      primaryClass = "hep-ph",
}

@article{Hoeche:2011fd,
      author         = "H{\"o}che, Stefan and Krauss, Frank and Sch{\"o}nherr, Marek and
                        Siegert, Frank",
      title          = "{A critical appraisal of NLO+PS matching methods}",
      journal        = "JHEP",
      volume         = "09",
      year           = "2012",
      pages          = "049",
      doi            = "10.1007/JHEP09(2012)049",
      eprint         = "1111.1220",
      archivePrefix  = "arXiv",
      primaryClass   = "hep-ph",
      reportNumber   = "SLAC-PUB-14661, IPPP-11-67, DCPT-11-134, LPN11-58,
                        FR-PHENO-2011-019, MCNET-11-24",
      SLACcitation   = "%%CITATION = ARXIV:1111.1220;%%"
}

@article{Catani:2001cc,
      author         = "Catani, S. and Krauss, F. and Webber, B. R. and Kuhn, R.",
      title          = "{QCD Matrix Elements + Parton Showers}",
      journal        = "JHEP",
      volume         = "11",
      year           = "2001",
      pages          = "063",
      doi            = "10.1088/1126-6708/2001/11/063",
      eprint         = "hep-ph/0109231",
      archivePrefix  = "arXiv",
      reportNumber   = "CERN-TH-2000-367, CAVENDISH-HEP-00-03",
      SLACcitation   = "%%CITATION = HEP-PH/0109231;%%"
}

@article{Cascioli:2011va,
      author         = "Cascioli, Fabio and Maierh{\"o}fer, Philipp and Pozzorini, Stefano",
      title          = "{Scattering Amplitudes with Open Loops}",
      journal        = "Phys. Rev. Lett.",
      volume         = "108",
      year           = "2012",
      pages          = "111601",
      doi            = "10.1103/PhysRevLett.108.111601",
      eprint         = "1111.5206",
      archivePrefix  = "arXiv",
      primaryClass   = "hep-ph",
      reportNumber   = "ZU-TH-23-11, LPN11-66",
      SLACcitation   = "%%CITATION = ARXIV:1111.5206;%%"
}

@article{Buccioni:2017yxi,
    author = "Buccioni, Federico and Pozzorini, Stefano and Zoller, Max",
    title = "{On-the-fly reduction of open loops}",
    eprint = "1710.11452",
    archivePrefix = "arXiv",
    primaryClass = "hep-ph",
    reportNumber = "ZU-TH-29-17",
    doi = "10.1140/epjc/s10052-018-5562-1",
    journal = "Eur. Phys. J. C",
    volume = "78",
    number = "1",
    pages = "70",
    year = "2018"
}

@article{Buccioni:2019sur,
    author = "Buccioni, Federico and Lang, Jean-Nicolas and Lindert, Jonas M. and Maierh{\"o}fer, Philipp and Pozzorini, Stefano and Zhang, Hantian and Zoller, Max F.",
    title = "{OpenLoops 2}",
    eprint = "1907.13071",
    archivePrefix = "arXiv",
    primaryClass = "hep-ph",
    reportNumber = "IPPP/19/62, FR-PHENO-2019-12, PSI-PR-19-15, ZU-TH 37/19",
    doi = "10.1140/epjc/s10052-019-7306-2",
    journal = "Eur. Phys. J. C",
    volume = "79",
    number = "10",
    pages = "866",
    year = "2019"
}

@article{Denner:2016kdg,
      author         = "Denner, Ansgar and Dittmaier, Stefan and Hofer, Lars",
      title          = "{\textsc{Collier}: A fortran-based complex one-loop library in
                        extended regularizations}",
      journal        = "Comput. Phys. Commun.",
      volume         = "212",
      year           = "2017",
      pages          = "220-238",
      doi            = "10.1016/j.cpc.2016.10.013",
      eprint         = "1604.06792",
      archivePrefix  = "arXiv",
      primaryClass   = "hep-ph",
      reportNumber   = "FR-PHENO-2016-003, ICCUB-16-016",
      SLACcitation   = "%%CITATION = ARXIV:1604.06792;%%"
}

@article{Czakon:2011xx,
  author         = "Czakon, Michal and Mitov, Alexander",
  title          = "{Top++: A program for the calculation of the top-pair cross-section at hadron colliders}",
  journal        = "Comput. Phys. Commun.",
  volume         = "185",
  year           = "2014",
  pages          = "2930",
  doi            = "10.1016/j.cpc.2014.06.021",
  eprint         = "1112.5675",
  archivePrefix  = "arXiv",
  primaryClass   = "hep-ph",
  reportNumber   = "CERN-PH-TH-2011-303, TTK-11-58",
}

@article{Catani:2007vq,
      author         = "Catani, Stefano and Grazzini, Massimiliano",
      title          = "{Next-to-Next-to-Leading-Order Subtraction Formalism in Hadron Collisions and
                        its Application to Higgs-Boson Production at the Large Hadron Collider}",
      journal        = "Phys. Rev. Lett.",
      volume         = "98",
      year           = "2007",
      pages          = "222002",
      doi            = "10.1103/PhysRevLett.98.222002",
      eprint         = "hep-ph/0703012",
      archivePrefix  = "arXiv",
      SLACcitation   = "%%CITATION = HEP-PH/0703012;%%"
}

@article{Agostinelli:2002hh,
      author         = "Agostinelli, S. and others",
      title          = "{\textsc{Geant4} -- a simulation toolkit}",
      journal        = "Nucl. Instrum. Meth. A",
      volume         = "506",
      year           = "2003",
      pages          = "250",
      doi            = "10.1016/S0168-9002(03)01368-8",
      reportNumber   = "SLAC-PUB-9350, FERMILAB-PUB-03-339",
      SLACcitation   = "%%CITATION = NUIMA,A506,250;%%"
}

@Booklet{RooFit,
      author         = "Verkerke, Wouter and Kirkby, David",
      title          = "{The RooFit toolkit for data modeling}",
      year           = "2003",
      eprint         = "physics/0306116",
      archivePrefix  = "arXiv",
      primaryClass   = "physics.data-an",
}

@article{DAgostini:1994fjx,
      author         = "D'Agostini, G.",
      title          = "{A multidimensional unfolding method based on Bayes' theorem}",
      journal        = "Nucl. Instrum. Meth. A",
      volume         = "362",
      year           = "1995",
      pages          = "487-498",
      doi            = "10.1016/0168-9002(95)00274-X",
      reportNumber   = "DESY-94-099",
      SLACcitation   = "%%CITATION = NUIMA,A362,487;%%"
}

@Article{PERF-2007-01,
    author         = "{ATLAS Collaboration}",
    title          = "{The ATLAS Experiment at the CERN Large Hadron Collider}",
    journal        = "JINST",
    volume         = "3",
    year           = "2008",
    pages          = "S08003",
    doi            = "10.1088/1748-0221/3/08/S08003",
    primaryClass   = "hep-ex",
}

@Article{SOFT-2010-01,
    author         = "{ATLAS Collaboration}",
    title          = "{The ATLAS Simulation Infrastructure}",
    journal        = "Eur. Phys. J. C",
    volume         = "70",
    year           = "2010",
    pages          = "823",
    doi            = "10.1140/epjc/s10052-010-1429-9",
    eprint         = "1005.4568",
    archivePrefix  = "arXiv",
    primaryClass   = "physics.ins-det",
}

@Article{STDM-2012-01,
    author         = "{ATLAS Collaboration}",
    title          = "{Measurement of \(W^+W^-\) production in \(pp\) collisions at \(\sqrt{s} = 7\,\text{TeV}\) with the ATLAS detector and limits on anomalous \(WWZ\) and \(WW\gamma\) couplings}",
    journal        = "Phys. Rev. D",
    volume         = "87",
    year           = "2013",
    pages          = "112001",
    doi            = "10.1103/PhysRevD.87.112001",
    reportNumber   = "CERN-PH-EP-2012-242",
    eprint         = "1210.2979",
    archivePrefix  = "arXiv",
    primaryClass   = "hep-ex",
    related        = "STDM-2012-01-err",
    relatedstring  = "Erratum:",
}

@Article{STDM-2013-07,
    author         = "{ATLAS Collaboration}",
    title          = "{Measurement of total and differential \(W^+W^-\) production cross sections in proton--proton collisions at \(\sqrt{s} = 8\,\text{TeV}\) with the ATLAS detector and limits on anomalous triple-gauge-boson couplings}",
    journal        = "JHEP",
    volume         = "09",
    year           = "2016",
    pages          = "029",
    doi            = "10.1007/JHEP09(2016)029",
    reportNumber   = "CERN-PH-EP-2015-323",
    eprint         = "1603.01702",
    archivePrefix  = "arXiv",
    primaryClass   = "hep-ex",
}

@Article{PERF-2014-03,
    author         = "{ATLAS Collaboration}",
    title          = "{Performance of pile-up mitigation techniques for jets in \(pp\) collisions at \(\sqrt{s} = 8\,\text{TeV}\) using the ATLAS detector}",
    journal        = "Eur. Phys. J. C",
    volume         = "76",
    year           = "2016",
    pages          = "581",
    doi            = "10.1140/epjc/s10052-016-4395-z",
    reportNumber   = "CERN-PH-EP-2015-206",
    eprint         = "1510.03823",
    archivePrefix  = "arXiv",
    primaryClass   = "hep-ex",
}

@Article{PERF-2015-09,
    author         = "{ATLAS Collaboration}",
    title          = "{Jet reconstruction and performance using particle flow with the ATLAS Detector}",
    journal        = "Eur. Phys. J. C",
    volume         = "77",
    year           = "2017",
    pages          = "466",
    doi            = "10.1140/epjc/s10052-017-5031-2",
    reportNumber   = "CERN-EP-2017-024",
    eprint         = "1703.10485",
    archivePrefix  = "arXiv",
    primaryClass   = "hep-ex",
}

@Article{PERF-2015-10,
    author         = "{ATLAS Collaboration}",
    title          = "{Muon reconstruction performance of the ATLAS detector in proton--proton collision data at \(\sqrt{s} = 13\,\text{TeV}\)}",
    journal        = "Eur. Phys. J. C",
    volume         = "76",
    year           = "2016",
    pages          = "292",
    doi            = "10.1140/epjc/s10052-016-4120-y",
    reportNumber   = "CERN-EP-2016-033",
    eprint         = "1603.05598",
    archivePrefix  = "arXiv",
    primaryClass   = "hep-ex",
}

@Article{STDM-2015-03,
    author         = "{ATLAS Collaboration}",
    title          = "{Measurement of \(W^{\pm}\) and \(Z\)-boson production cross sections in \(pp\) collisions at \(\sqrt{s} = 13\,\text{TeV}\) with the ATLAS detector}",
    journal        = "Phys. Lett. B",
    volume         = "759",
    year           = "2016",
    pages          = "601",
    doi            = "10.1016/j.physletb.2016.06.023",
    reportNumber   = "CERN-EP-2016-069",
    eprint         = "1603.09222",
    archivePrefix  = "arXiv",
    primaryClass   = "hep-ex",
}

@Article{STDM-2015-20,
    author         = "{ATLAS Collaboration}",
    title          = "{Measurement of the \(W^+W^-\) production cross section in \(pp\) collisions at a centre-of-mass energy of \(\sqrt{s} = 13\,\text{TeV}\) with the ATLAS experiment}",
    journal        = "Phys. Lett. B",
    volume         = "773",
    year           = "2017",
    pages          = "354",
    doi            = "10.1016/j.physletb.2017.08.047",
    reportNumber   = "CERN-EP-2016-267",
    eprint         = "1702.04519",
    archivePrefix  = "arXiv",
    primaryClass   = "hep-ex",
}

@Article{STDM-2015-24,
    author         = "{ATLAS Collaboration}",
    title          = "{Measurement of \(W^+W^-\) production in association with one jet in proton--proton collisions at \(\sqrt{s} = 8\,\text{TeV}\) with the ATLAS detector}",
    journal        = "Phys. Lett. B",
    volume         = "763",
    year           = "2016",
    pages          = "114",
    doi            = "10.1016/j.physletb.2016.10.014",
    reportNumber   = "CERN-EP-2016-186",
    eprint         = "1608.03086",
    archivePrefix  = "arXiv",
    primaryClass   = "hep-ex",
}

@Article{STDM-2017-24,
    author         = "{ATLAS Collaboration}",
    title          = "{Measurement of fiducial and differential \(W^+W^-\) production cross-sections at \(\sqrt{s} = 13\,\text{TeV}\) with the ATLAS detector}",
    journal        = "Eur. Phys. J. C",
    volume         = "79",
    year           = "2019",
    pages          = "884",
    doi            = "10.1140/epjc/s10052-019-7371-6",
    reportNumber   = "CERN-EP-2019-055",
    eprint         = "1905.04242",
    archivePrefix  = "arXiv",
    primaryClass   = "hep-ex",
}

@Article{STDM-2017-27,
    author         = "{ATLAS Collaboration}",
    title          = "{Differential cross-section measurements for the electroweak production of dijets in association with a \(Z\) boson in proton--proton collisions at ATLAS}",
    journal        = "Eur. Phys. J. C",
    volume         = "81",
    year           = "2021",
    pages          = "163",
    doi            = "10.1140/epjc/s10052-020-08734-w",
    reportNumber   = "CERN-EP-2020-045",
    eprint         = "2006.15458",
    archivePrefix  = "arXiv",
    primaryClass   = "hep-ex",
}

@Article{DAPR-2018-01,
    author         = "{ATLAS Collaboration}",
    title          = "{ATLAS data quality operations and performance for 2015--2018 data-taking}",
    journal        = "JINST",
    volume         = "15",
    year           = "2020",
    pages          = "P04003",
    doi            = "10.1088/1748-0221/15/04/P04003",
    reportNumber   = "CERN-EP-2019-207",
    eprint         = "1911.04632",
    archivePrefix  = "arXiv",
    primaryClass   = "physics.ins-det",
}

@Article{EGAM-2018-01,
    author         = "{ATLAS Collaboration}",
    title          = "{Electron and photon performance measurements with the ATLAS detector using the 2015--2017 LHC proton--proton collision data}",
    journal        = "JINST",
    volume         = "14",
    year           = "2019",
    pages          = "P12006",
    doi            = "10.1088/1748-0221/14/12/P12006",
    reportNumber   = "CERN-EP-2019-145",
    eprint         = "1908.00005",
    archivePrefix  = "arXiv",
    primaryClass   = "hep-ex",
}

@Article{FTAG-2018-01,
    author         = "{ATLAS Collaboration}",
    title          = "{ATLAS \(b\)-jet identification performance and efficiency measurement with \(t\bar{t}\) events in \(pp\) collisions at \(\sqrt{s} = 13\,\text{TeV}\)}",
    journal        = "Eur. Phys. J. C",
    volume         = "79",
    year           = "2019",
    pages          = "970",
    doi            = "10.1140/epjc/s10052-019-7450-8",
    reportNumber   = "CERN-EP-2019-132",
    eprint         = "1907.05120",
    archivePrefix  = "arXiv",
    primaryClass   = "hep-ex",
}

@Article{HIGG-2018-51,
    author         = "{ATLAS Collaboration}",
    title          = "{Measurements of \(WH\) and \(ZH\) production in the \(H \to b\bar{b}\) decay channel in \(pp\) collisions at \(13\,\text{TeV}\) with the ATLAS detector}",
    journal        = "Eur. Phys. J. C",
    volume         = "81",
    year           = "2021",
    pages          = "178",
    doi            = "10.1140/epjc/s10052-020-08677-2",
    reportNumber   = "CERN-EP-2020-087",
    eprint         = "2007.02873",
    archivePrefix  = "arXiv",
    primaryClass   = "hep-ex",
}

@Article{HIGG-2018-52,
    author         = "{ATLAS Collaboration}",
    title          = "{Measurement of the associated production of a Higgs boson decaying into \(b\)-quarks with a vector boson at high transverse momentum in \(pp\) collisions at \(\sqrt{s} = 13\,\text{TeV}\) with the ATLAS detector}",
    journal        = "Phys. Lett. B",
    volume         = "816",
    year           = "2021",
    pages          = "136204",
    doi            = "10.1016/j.physletb.2021.136204",
    reportNumber   = "CERN-EP-2020-093",
    eprint         = "2008.02508",
    archivePrefix  = "arXiv",
    primaryClass   = "hep-ex",
}

@Article{JETM-2018-05,
    author         = "{ATLAS Collaboration}",
    title          = "{Jet energy scale and resolution measured in proton--proton collisions at \(\sqrt{s} = 13\,\text{TeV}\) with the ATLAS detector}",
    journal        = "Eur. Phys. J. C",
    volume         = "81",
    year           = "2021",
    pages          = "689",
    doi            = "10.1140/epjc/s10052-021-09402-3",
    reportNumber   = "CERN-EP-2020-083",
    eprint         = "2007.02645",
    archivePrefix  = "arXiv",
    primaryClass   = "hep-ex",
}

@Article{MUON-2018-03,
    author         = "{ATLAS Collaboration}",
    title          = "{Muon reconstruction and identification efficiency in ATLAS using the full Run~2 \(pp\) collision data set at \(\sqrt{s} = 13\,\text{TeV}\)}",
    journal        = "Eur. Phys. J. C",
    volume         = "81",
    year           = "2021",
    pages          = "578",
    doi            = "10.1140/epjc/s10052-021-09233-2",
    reportNumber   = "CERN-EP-2020-199",
    eprint         = "2012.00578",
    archivePrefix  = "arXiv",
    primaryClass   = "hep-ex",
}

@Article{PIX-2018-001,
    author         = "Abbott, B. and others",
    title          = "{Production and integration of the ATLAS Insertable B-Layer}",
    journal        = "JINST",
    volume         = "13",
    year           = "2018",
    pages          = "T05008",
    doi            = "10.1088/1748-0221/13/05/T05008",
    eprint         = "1803.00844",
    archivePrefix  = "arXiv",
    primaryClass   = "physics.ins-det",
}

@Article{STDM-2018-03,
    author         = "{ATLAS Collaboration}",
    title          = "{Measurement of \(W^{\pm}Z\) production cross sections and gauge boson polarisation in \(pp\) collisions at \(\sqrt{s} = 13\,\text{TeV}\) with the ATLAS detector}",
    journal        = "Eur. Phys. J. C",
    volume         = "79",
    year           = "2019",
    pages          = "535",
    doi            = "10.1140/epjc/s10052-019-7027-6",
    reportNumber   = "CERN-EP-2018-327",
    eprint         = "1902.05759",
    archivePrefix  = "arXiv",
    primaryClass   = "hep-ex",
}

@Article{STDM-2018-34,
    author         = "{ATLAS Collaboration}",
    title          = "{Measurements of \(W^{+}W^{-} + \ge 1\) jet production cross-sections in \(pp\) collisions at \(\sqrt{s} = 13\,\text{TeV}\) with the ATLAS detector}",
    journal        = "JHEP",
    volume         = "06",
    year           = "2021",
    pages          = "003",
    doi            = "10.1007/JHEP06(2021)003",
    reportNumber   = "CERN-EP-2021-030",
    eprint         = "2103.10319",
    archivePrefix  = "arXiv",
    primaryClass   = "hep-ex",
}

@Article{TOPQ-2018-17,
    author         = "{ATLAS Collaboration}",
    title          = "{Measurement of the \(t\bar{t}\) production cross-section and lepton differential distributions in \(e\mu\) dilepton events from \(pp\) collisions at \(\sqrt{s} = 13\,\text{TeV}\) with the ATLAS detector}",
    journal        = "Eur. Phys. J. C",
    volume         = "80",
    year           = "2020",
    pages          = "528",
    doi            = "10.1140/epjc/s10052-020-7907-9",
    reportNumber   = "CERN-EP-2019-203",
    eprint         = "1910.08819",
    archivePrefix  = "arXiv",
    primaryClass   = "hep-ex",
}

@Article{TRIG-2018-01,
    author         = "{ATLAS Collaboration}",
    title          = "{Performance of the ATLAS muon triggers in Run~2}",
    journal        = "JINST",
    volume         = "15",
    year           = "2020",
    pages          = "P09015",
    doi            = "10.1088/1748-0221/15/09/p09015",
    reportNumber   = "CERN-EP-2020-031",
    eprint         = "2004.13447",
    archivePrefix  = "arXiv",
    primaryClass   = "physics.ins-det",
}

@Article{TRIG-2018-05,
    author         = "{ATLAS Collaboration}",
    title          = "{Performance of electron and photon triggers in ATLAS during LHC Run~2}",
    journal        = "Eur. Phys. J. C",
    volume         = "80",
    year           = "2020",
    pages          = "47",
    doi            = "10.1140/epjc/s10052-019-7500-2",
    reportNumber   = "CERN-EP-2019-169",
    eprint         = "1909.00761",
    archivePrefix  = "arXiv",
    primaryClass   = "hep-ex",
}

@Article{FTAG-2019-07,
    author         = "{ATLAS Collaboration}",
    title          = "{ATLAS flavour-tagging algorithms for the LHC Run~2 \(pp\) collision dataset}",
    journal        = "Eur. Phys. J. C",
    volume         = "83",
    year           = "2023",
    pages          = "681",
    doi            = "10.1140/epjc/s10052-023-11699-1",
    reportNumber   = "CERN-EP-2022-226",
    eprint         = "2211.16345",
    archivePrefix  = "arXiv",
    primaryClass   = "physics.data-an",
}

@Article{JETM-2020-03,
    author         = "{ATLAS Collaboration}",
    title          = "{The performance of missing transverse momentum reconstruction and its significance with the ATLAS detector using \(140\,\text{fb}^{-1}\) of \(\sqrt{s} = 13\,\text{TeV}\) \(pp\) collisions}",
    journal        = "Eur. Phys. J. C",
    volume         = "85",
    year           = "2025",
    pages          = "606",
    doi            = "10.1140/epjc/s10052-025-14062-8",
    reportNumber   = "CERN-EP-2024-023",
    eprint         = "2402.05858",
    archivePrefix  = "arXiv",
    primaryClass   = "hep-ex",
}

@Article{STDM-2020-16,
    author         = "{ATLAS Collaboration}",
    title          = "{Measurements of \(W^+W^-\) production cross-sections in \(pp\) collisions at \(\sqrt{s} = 13\,\text{TeV}\) with the ATLAS detector}",
    year           = "2025",
    reportNumber   = "CERN-EP-2025-091",
    eprint         = "2505.11310",
    archivePrefix  = "arXiv",
    primaryClass   = "hep-ex",
}

@Article{DAPR-2021-01,
    author         = "{ATLAS Collaboration}",
    title          = "{Luminosity determination in \(pp\) collisions at \(\sqrt{s} = 13\,\text{TeV}\) using the ATLAS detector at the LHC}",
    journal        = "Eur. Phys. J. C",
    volume         = "83",
    year           = "2023",
    pages          = "982",
    doi            = "10.1140/epjc/s10052-023-11747-w",
    reportNumber   = "CERN-EP-2022-281",
    eprint         = "2212.09379",
    archivePrefix  = "arXiv",
    primaryClass   = "hep-ex",
}

@Article{SUSY-2021-06,
    author         = "{ATLAS Collaboration}",
    title          = "{Measurements of \(W^{+}W^{-}\) production in decay topologies inspired by searches for electroweak supersymmetry}",
    journal        = "Eur. Phys. J. C",
    volume         = "83",
    year           = "2023",
    pages          = "718",
    doi            = "10.1140/epjc/s10052-023-11508-9",
    reportNumber   = "CERN-EP-2022-087",
    eprint         = "2206.15231",
    archivePrefix  = "arXiv",
    primaryClass   = "hep-ex",
}

@Article{MUON-2022-01,
    author         = "{ATLAS Collaboration}",
    title          = "{Studies of the muon momentum calibration and performance of the ATLAS detector with \(pp\) collisions at \(\sqrt{s} = 13\,\text{TeV}\)}",
    journal        = "Eur. Phys. J. C",
    volume         = "83",
    year           = "2023",
    pages          = "686",
    doi            = "10.1140/epjc/s10052-023-11584-x",
    reportNumber   = "CERN-EP-2022-274",
    eprint         = "2212.07338",
    archivePrefix  = "arXiv",
    primaryClass   = "hep-ex",
}

@Article{SOFT-2022-02,
    author         = "{ATLAS Collaboration}",
    title          = "{Software and computing for Run~3 of the ATLAS experiment at the LHC}",
    journal        = "Eur. Phys. J. C",
    volume         = "85",
    year           = "2025",
    pages          = "234",
    doi            = "10.1140/epjc/s10052-024-13701-w",
    reportNumber   = "CERN-EP-2024-100",
    eprint         = "2404.06335",
    archivePrefix  = "arXiv",
    primaryClass   = "hep-ex",
}

@Booklet{ATL-SOFT-PUB-2025-001,
    author         = "{ATLAS Collaboration}",
    title          = "{ATLAS Computing Acknowledgements}",
    howpublished   = "{ATL-SOFT-PUB-2025-001}",
    url            = "https://cds.cern.ch/record/2922210",
    year           = "2025",
}

@Report{ATLAS-TDR-2010-19,
    author         = "{ATLAS Collaboration}",
    title          = "{ATLAS Insertable B-Layer Technical Design Report}",
    type           = "ATLAS-TDR-19; CERN-LHCC-2010-013",
    year           = "2010",
    url            = "https://cds.cern.ch/record/1291633",
    related        = "ATLAS-TDR-2010-19-addm",
    relatedstring  = "Addendum:",
}

@Article{LUCID2,
    author         = "Avoni, G. and others",
    title          = "{The new LUCID-2 detector for luminosity measurement and monitoring in ATLAS}",
    journal        = "JINST",
    volume         = "13",
    year           = "2018",
    pages          = "P07017",
    doi            = "10.1088/1748-0221/13/07/P07017",
}

@Article{CMS-SMP-12-005,
    author         = "{CMS Collaboration}",
    title          = "{Measurement of the \(W^+W^-\) cross section in \(pp\) collisions at \(\sqrt{s} = 7\,\text{TeV}\) and limits on anomalous \(WW\gamma\) and \(WWZ\) couplings}",
    journal        = "Eur. Phys. J. C",
    volume         = "73",
    year           = "2013",
    pages          = "2610",
    doi            = "10.1140/epjc/s10052-013-2610-8",
    reportNumber   = "CERN-PH-EP-2013-075",
    eprint         = "1306.1126",
    archivePrefix  = "arXiv",
    primaryClass   = "hep-ex",
}

@Article{CMS-SMP-14-016,
    author         = "{CMS Collaboration}",
    title          = "{Measurement of the \(W^+W^-\) cross section in \(pp\) collisions at \(\sqrt{s} = 8\,\text{TeV}\) and limits on anomalous gauge couplings}",
    journal        = "Eur. Phys. J. C",
    volume         = "76",
    year           = "2016",
    pages          = "401",
    doi            = "10.1140/epjc/s10052-016-4219-1",
    reportNumber   = "CERN-PH-EP-2015-122",
    eprint         = "1507.03268",
    archivePrefix  = "arXiv",
    primaryClass   = "hep-ex",
}

@Article{CMS-SMP-18-004,
    author         = "{CMS Collaboration}",
    title          = "{\(W^+W^-\) boson pair production in proton--proton collisions at \(\sqrt{s} = 13\,\text{TeV}\)}",
    journal        = "Phys. Rev. D",
    volume         = "102",
    year           = "2020",
    pages          = "092001",
    doi            = "10.1103/PhysRevD.102.092001",
    reportNumber   = "CERN-EP-2020-144",
    eprint         = "2009.00119",
    archivePrefix  = "arXiv",
    primaryClass   = "hep-ex",
}

@Article{CMS-SMP-19-002,
    author         = "{CMS Collaboration}",
    title          = "{Measurement of the \(W\gamma\) production cross section in proton--proton collisions at \(\sqrt{s} = 13\,\text{TeV}\) and constraints on effective field theory coefficients}",
    journal        = "Phys. Rev. Lett.",
    volume         = "126",
    year           = "2021",
    pages          = "252002",
    doi            = "10.1103/PhysRevLett.126.252002",
    reportNumber   = "CERN-EP-2021-008",
    eprint         = "2102.02283",
    archivePrefix  = "arXiv",
    primaryClass   = "hep-ex",
}

@Article{CMS-SMP-20-005,
    author         = "{CMS Collaboration}",
    title          = "{Measurement of \(W^{\pm}\gamma\) differential cross sections in proton--proton collisions at \(\sqrt{s} = 13\,\text{TeV}\) and effective field theory constraints}",
    journal        = "Phys. Rev. D",
    volume         = "105",
    year           = "2022",
    pages          = "052003",
    doi            = "10.1103/PhysRevD.105.052003",
    reportNumber   = "CERN-EP-2021-219",
    eprint         = "2111.13948",
    archivePrefix  = "arXiv",
    primaryClass   = "hep-ex",
}

@Article{CMS-SMP-20-012,
    author         = "{CMS Collaboration}",
    title          = "{Measurements of the electroweak diboson production cross sections in proton--proton collisions at \(\sqrt{s} = 5.02\,\text{TeV}\) using leptonic decays}",
    journal        = "Phys. Rev. Lett.",
    volume         = "127",
    year           = "2021",
    pages          = "191801",
    doi            = "10.1103/PhysRevLett.127.191801",
    reportNumber   = "CERN-EP-2021-111",
    eprint         = "2107.01137",
    archivePrefix  = "arXiv",
    primaryClass   = "hep-ex",
}

@Article{CMS-SMP-20-014,
    author         = "{CMS Collaboration}",
    title          = "{Measurement of the inclusive and differential \(WZ\) production cross sections, polarization angles, and triple gauge couplings in \(pp\) collisions at \(\sqrt{s} = 13\,\text{TeV}\)}",
    journal        = "JHEP",
    volume         = "07",
    year           = "2022",
    pages          = "032",
    doi            = "10.1007/JHEP07(2022)032",
    reportNumber   = "CERN-EP-2021-163",
    eprint         = "2110.11231",
    archivePrefix  = "arXiv",
    primaryClass   = "hep-ex",
}

@Article{CMS-SMP-24-001,
    author         = "{CMS Collaboration}",
    title          = "{Measurement of inclusive and differential cross sections for \(W^{+}W^{-}\) production in proton--proton collisions at \(\sqrt{s} = 13.6\,\text{TeV}\)}",
    journal        = "Phys. Lett. B",
    volume         = "861",
    year           = "2025",
    pages          = "139231",
    doi            = "10.1016/j.physletb.2024.139231",
    reportNumber   = "CERN-EP-2024-129",
    eprint         = "2406.05101",
    archivePrefix  = "arXiv",
    primaryClass   = "hep-ex",
}

@Booklet{ATL-PHYS-PUB-2014-021,
    author         = "{ATLAS Collaboration}",
    title          = "{ATLAS Pythia~8 tunes to \(7~\text{TeV}\) data}",
    howpublished   = "{ATL-PHYS-PUB-2014-021}",
    url            = "https://cds.cern.ch/record/1966419",
    year           = "2014",
}

@Booklet{ATL-PHYS-PUB-2016-002,
    author         = "{ATLAS Collaboration}",
    title          = "{Multi-boson simulation for \(13~\text{TeV}\) ATLAS analyses}",
    howpublished   = "{ATL-PHYS-PUB-2016-002}",
    url            = "https://cds.cern.ch/record/2119986",
    year           = "2016",
}

@Booklet{ATL-PHYS-PUB-2016-017,
    author         = "{ATLAS Collaboration}",
    title          = "{The Pythia~8 A3 tune description of ATLAS minimum bias and inelastic measurements incorporating the Donnachie--Landshoff diffractive model}",
    howpublished   = "{ATL-PHYS-PUB-2016-017}",
    url            = "https://cds.cern.ch/record/2206965",
    year           = "2016",
}

@Booklet{ATL-PHYS-PUB-2016-020,
    author         = "{ATLAS Collaboration}",
    title          = "{Studies on top-quark Monte Carlo modelling for Top2016}",
    howpublished   = "{ATL-PHYS-PUB-2016-020}",
    url            = "https://cds.cern.ch/record/2216168",
    year           = "2016",
}

@Booklet{ATL-PHYS-PUB-2017-007,
    author         = "{ATLAS Collaboration}",
    title          = "{Studies on top-quark Monte Carlo modelling with Sherpa and MG5\_aMC@NLO}",
    howpublished   = "{ATL-PHYS-PUB-2017-007}",
    url            = "https://cds.cern.ch/record/2261938",
    year           = "2017",
}

@Booklet{ATL-PHYS-PUB-2020-023,
    author         = "{ATLAS Collaboration}",
    title          = "{Study of top-quark pair modelling and uncertainties using ATLAS measurements at \(\sqrt{s} = 13\,\text{TeV}\)}",
    howpublished   = "{ATL-PHYS-PUB-2020-023}",
    url            = "https://cds.cern.ch/record/2730443",
    year           = "2020",
}
\clearpage
 
\begin{flushleft}
\hypersetup{urlcolor=black}
{\Large The ATLAS Collaboration}

\bigskip

\AtlasOrcid[0000-0002-6665-4934]{G.~Aad}$^\textrm{\scriptsize 102}$,
\AtlasOrcid[0000-0002-5888-2734]{B.~Abbott}$^\textrm{\scriptsize 120}$,
\AtlasOrcid[0000-0002-1002-1652]{K.~Abeling}$^\textrm{\scriptsize 55}$,
\AtlasOrcid[0000-0001-5763-2760]{N.J.~Abicht}$^\textrm{\scriptsize 49}$,
\AtlasOrcid[0000-0002-8496-9294]{S.H.~Abidi}$^\textrm{\scriptsize 29}$,
\AtlasOrcid[0000-0002-9987-2292]{A.~Aboulhorma}$^\textrm{\scriptsize 35e}$,
\AtlasOrcid[0000-0001-5329-6640]{H.~Abramowicz}$^\textrm{\scriptsize 152}$,
\AtlasOrcid[0000-0002-1599-2896]{H.~Abreu}$^\textrm{\scriptsize 151}$,
\AtlasOrcid[0000-0003-0403-3697]{Y.~Abulaiti}$^\textrm{\scriptsize 117}$,
\AtlasOrcid[0000-0003-0762-7204]{A.C.~Abusleme~Hoffman}$^\textrm{\scriptsize 137a}$
% ...
\bigskip
\\

% ...
$^{29}$Physics Department, Brookhaven National Laboratory, Upton NY; United States of America.\\
% ...
$^{35}$$^{(a)}$Facult\'e des Sciences Ain Chock, Universit\'e Hassan II de Casablanca;$^{(b)}$Facult\'{e} des Sciences, Universit\'{e} Ibn-Tofail, K\'{e}nitra;$^{(c)}$Facult\'e des Sciences Semlalia, Universit\'e Cadi Ayyad, LPHEA-Marrakech;$^{(d)}$LPMR, Facult\'e des Sciences, Universit\'e Mohamed Premier, Oujda;$^{(e)}$Facult\'e des sciences, Universit\'e Mohammed V, Rabat;$^{(f)}$Institute of Applied Physics, Mohammed VI Polytechnic University, Ben Guerir; Morocco.\\
% ...
$^{49}$Fakult\"{a}t Physik , Technische Universit{\"a}t Dortmund, Dortmund; Germany.\\
% ...
$^{55}$II. Physikalisches Institut, Georg-August-Universit\"{a}t G\"ottingen, G\"ottingen; Germany.\\
% ...
$^{102}$CPPM, Aix-Marseille Universit\'e, CNRS/IN2P3, Marseille; France.\\
% ...
$^{117}$Department of Physics, New York University, New York NY; United States of America.\\
% ...
$^{120}$Homer L. Dodge Department of Physics and Astronomy, University of Oklahoma, Norman OK; United States of America.\\
% ...
$^{137}$$^{(a)}$Departamento de F\'isica, Pontificia Universidad Cat\'olica de Chile, Santiago;$^{(b)}$Millennium Institute for Subatomic physics at high energy frontier (SAPHIR), Santiago;$^{(c)}$Instituto de Investigaci\'on Multidisciplinario en Ciencia y Tecnolog\'ia, y Departamento de F\'isica, Universidad de La Serena;$^{(d)}$Universidad Andres Bello, Department of Physics, Santiago;$^{(e)}$Instituto de Alta Investigaci\'on, Universidad de Tarapac\'a, Arica;$^{(f)}$Departamento de F\'isica, Universidad T\'ecnica Federico Santa Mar\'ia, Valpara\'iso; Chile.\\
% ...
$^{151}$Department of Physics, Technion, Israel Institute of Technology, Haifa; Israel.\\
$^{152}$Raymond and Beverly Sackler School of Physics and Astronomy, Tel Aviv University, Tel Aviv; Israel.\\
% ...
$^{a}$ Also Affiliated with an institute covered by a cooperation agreement with CERN.\\
$^{b}$ Also at An-Najah National University, Nablus; Palestine.\\
$^{c}$ Also at Borough of Manhattan Community College, City University of New York, New York NY; United States of America.\\
$^{d}$ Also at Center for High Energy Physics, Peking University; China.\\
$^{e}$ Also at Center for Interdisciplinary Research and Innovation (CIRI-AUTH), Thessaloniki; Greece.\\
$^{f}$ Also at Centro Studi e Ricerche Enrico Fermi; Italy.\\
% ...

\end{flushleft}


%

%
%

%

%
%
%
%

\end{document}